\documentclass[journal abbreviation]{copernicus}

\usepackage{amssymb}
\usepackage{commath}
\usepackage{amsmath}
\usepackage{amsthm}
\usepackage{multirow}
\usepackage{bigdelim}

\begin{document}

\linenumbers

\title{Using the UM dynamical cores to reproduce idealised 3D flows.}

\author[1]{N. J. Mayne}
\author[1]{I. Baraffe}
\author[1]{David M. Acreman}
\author[2]{Chris Smith}
\author[2]{Nigel Wood}
\author[1]{David Sk\aa lid Amundsen}
\author[3]{John Thuburn}
\author[2]{David R. Jackson}

\affil[1]{Physics and Astronomy, University of Exeter, Exeter, EX4 4QL, United Kingdom.}
\affil[2]{Met Office, Exeter, EX1 3PB, United Kingdom}
\affil[3]{Mathematics, University of Exeter, Exeter, EX4 4QL, United Kingdom.}


\runningtitle{Dynamical Cores}

\runningauthor{N. J. Mayne et al}

\correspondence{N. J. Mayne\\ (nathan@astro.ex.ac.uk)}

\received{}
\pubdiscuss{} 
\revised{}
\accepted{}
\published{}


\firstpage{1}

\maketitle  

\begin{abstract}
  We demonstrate that both the current (New Dynamics), and next
  generation (ENDGame) dynamical cores of the UK Met Office global
  circulation model, the UM, reproduce consistently, the long--term,
  large--scale flows found in several published idealised tests. The
  cases presented are the Held--Suarez test, a simplified model of
  Earth (including a stratosphere), and a hypothetical tidally locked
  Earth. Furthermore, we show that using simplifications to the
  dynamical equations, which are expected to be justified for the
  physical domains and flow regimes we have studied, and which are
  supported by the ENDGame dynamical core, also produces matching
  long--term, large--scale flows. Finally, we present evidence for
  differences in the detail of the planetary flows and circulations
  resulting from improvements in the ENDGame formulation over New
  Dynamics.
\end{abstract}

\introduction 
Global circulation models (GCMs) are used for both numerical weather
and climate prediction. The accuracy of predictions made by GCMs of
the Earth system are constantly being improved, driven by the
requirement to understand our changing climate, improve severe weather
warnings for the public, and inform weather sensitive businesses and
industries.

The UK Met Office Unified Model (UM) incorporates both weather and
climate modeling capabilities in the same code platform. The quality
of weather predictions is constantly checked against millions of
observations during forecast verification. For climate models
pre--industrial control runs are performed and the model is verified
against historical observations. The quality of the model is therefore
judged on its ability to both produce a good forecast (weather), and
to match Earth's recent climate history (climate). Improvements which
make the underlying model components more representative of the
natural system do not always satisfy both these requirements due to,
for instance, compensatory errors.

The requirement for accurate climate predictions is becoming
increasingly important for Earth as our climate is changing.
Additionally, GCMs are also now used for climate modeling of systems
other than Earth's future climate. For these cases there is no data
assimilation and few independent validating observations. For studies
of Earth's palaeoclimate, observational constraints become more
uncertain with increasing temporal distance from the present
\citep[see for example][]{lenton_2008}. GCMs have also been used to
model the climates of other Solar--system planets \citep[see for
example models of Jupiter, Saturn, Mars and
Venus:][respectively]{yamazaki_2004,muller_wodarg_2006,hollingsworth_2011,lebonnois_2011}
where observations exist but are often much harder to interpret and
dramatically less numerous than for our own planet. Finally, in the
most extreme case, recent detections and observations of exoplanets,
or planets outside our own Solar--system, have prompted many groups to
begin exploring the possible climate regimes of very distant worlds
with GCMs originally designed for the study of Earth's climate
\citep[see for
example][]{cho_2008,showman_2009,zalucha_2012}. Accordingly, for such
cases the primary means of assessing model quality is via a focus on
the nature and statistics of the longer term simulated model flow
\citep[see Section 2 in][]{held_2005}.

This combination of the increasing importance of long term predictions
for our own climate, and the extension into new modeling regimes,
means that simple testing of climate modeling applications of GCMs is
becoming increasingly important. In these cases the exact predictions
at a given time are not the best analysis of the quality of the model
(unlike weather prediction). The more important aspect of climate
models is whether they self--consistently capture the dominant aspects
of a climate system under varying conditions, approaching those of the
target system (or planetary atmosphere to be
studied). \citet{held_2005} has already explained the increasing need
for a hierarchy of tests performed on components, or modules, of GCMs
as the complexity of models we can feasibly run increases with
increasing computing power. This hierarchy includes analytical tests,
such as normal mode analysis and the reproduction of analytic flows,
as well as more prescriptive tests targeting specific atmospheric
phenomena, and extends to statistical analysis of model differences
for detailed climate models. Bridging these regimes are tests such as
the Held--Suarez test \citep{held_1994}, which is a simplified and
idealised experiment isolating the dynamical core (the section which
models the evolution of the resolved dynamical flow) of a GCM. This
test, and others like it, allow the exploration of model differences
or similarities, whilst exploring realistic three dimensional flows
run over long periods of elapsed model time. They incorporate a set of
simple parameterisations allowing comparison free of the details of,
for instance, complicated radiative transfer or boundary layer
codes. Such tests increase our confidence in the predictions of GCMs,
which is paramount if they are to be used to explore systems where
observational constraints are sparse. Furthermore, using idealised
tests one can begin to alter aspects of the model to approach the
regime we are ultimately interested in. 

Tests like the Held--Suarez tests are not, in themselves, completely
satisfactory tests of the accuracy of a dynamical core. Firstly, no
analytical or reference solution is available to verify the model
results. Secondly, the sensitivity of the test is low. The diagnostic
plots used to determine a satisfactory result are constructed using
temporal and zonal averages and usually compared `by eye' resulting in
a coarse measure of agreement. Therefore satisfying the
\cite{held_1994} test does not guarantee the details of the
atmospheric solution between two models will closely match. Therefore,
idealised tests such as the Held--Suarez test are complementary, but
not a replacement for more simplified or prescriptive tests, such as
tests of intermediate complexity targeting specific physical phenomena
\citep[see for example][]{reed_2011}, or the reproduction of
analytical flows. Several tests have already been successfully
performed using the UM. Most recently, \citet{wood_2013} performed a
subset of tests detailed in the Dynamical Core Model Intercomparison
Project (DCMIP, see
\url{http://earthsystemcog.org/projects/dcmip-2012/}) and the
deep--atmosphere baroclinic instability test of
\citet{ullrich_2013}. However, these tests evaluate the modeling of
specific atmospheric responses, such as gravity waves induced by
orography, whereas tests such as \citet{held_1994} evaluate the
modeled state of the entire atmosphere over long integration times.

We have recently begun a project to model a subset of the most
observationally constrained exoplanets using the UM. The subset is
termed hot Jupiters as it consists of gas giant planets (of order the
mass of Jupiter) which orbit close to their parent star (closer than
Mercury is to our Sun). Torques from tidal forces between the star and
planet force the planet orbit and rotation into a synchronous state
i.e. one year equals one day. This results in a permanent `day' and
`night' side \citep[for a review see][]{baraffe_2010}. Their relative
brightness and proximity to their host star make observations of some
aspects of their atmospheres possible. Most existing GCMs applied to
hot Jupiters solve simplified equations of motion, most commonly the
so--called primitive equations
\citep[e.g.][]{showman_2009,heng_2011}. 

The derivation of the primitive equations incorporates several
simplifications including the assumption of vertical hydrostatic
equilibrium and the adoption of the `shallow--atmosphere'
approximation. Adopting the nomenclature of \citet{white_2005} the
`shallow--atmosphere' approximation is actually a term combining three
assumptions, that of a constant (with height) gravity, the
`shallow--fluid' and the `traditional' approximation. The effect of
these assumptions on the equations of motion is stated explicitly in
Table \ref{assume}. The `shallow--fluid' approximation is the
assumption that the atmosphere is a thin layer, when compared to the
radius of the planet, and can be justified with a small ratio of the
modeled atmospheric extent to the planetary radius, termed the aspect
ratio. However, the `traditional' approximation, taken with the
`shallow--fluid' approximation, involves the neglect of several metric
and rotation terms and, critically, is not strongly justified by a
physical argument but adopted to allow energy, angular momentum and
potential vorticity conservation in the final equation set
\citep{white_1995}.

It is probable that several important aspects of hot Jupiter systems,
for instance the day--night side heat redistribution and the radius of
the hot Jupiter itself \citep{showman_2002,baraffe_2010} depend on the
detailed dynamics of the atmosphere over many pressure
scale-heights. Consequently `shallow--atmosphere', hydrostatic models
may be too simplified to correctly interpret the observations of hot
Jupiter atmospheres. For example, \citet{tokano_2013} shows that GCMs
adopting the primitive equations do not correctly represent the
dynamics of Titan's (and Venus's) atmosphere, which has a similar
aspect ratio to hot Jupiters ($\sim$ 0.1). Although
\citet{tokano_2013} focuses on the assumption of hydrostatic
equilibrium, the term they indicate is dominant, $(u^2+v^2)/r$, is
neglected as part of the `traditional'
approximation. \citet{kaspi_2009} present models of Jupiter using an
adapted form of the MITgcm, including the effects of a deep
atmosphere. However, the models of \citet{kaspi_2009} are based on the
anelastic approximation which assumes the flow is incompressible and
filters out sound waves (as well as breaking down for flows with Mach
numbers of close to one).

The Met Office UM solves the deep, non--hydrostatic equations of
motion for the rotating atmosphere, and as part of its continuing
development the UM is currently transitioning to a new dynamical core,
from New Dynamics \citep[ND,][]{davies_2005} to ENDGame
\citep{wood_2013}. The ENDGame dynamical core provides several
improvements on the ND core. For our purposes the most important of
these improvements are: better handling of flow across the poles of
the latitude-longitude coordinate system; an iterated semi-implicit
scheme, providing reduced temporal truncation error; better scaling on
multiple processor computer architecture; and an overall improvement
of model stability and robustness \citep{wood_2013}. Additionally, the
code now includes a set of `switchable' physical assumptions
\citep[for instance it can run both with and without the
`shallow--atmosphere' approximation, as defined by][ and explained in
Table \ref{assume}]{white_2005}. Additionally, a novel mass conserving
transport scheme has been developed (SLICE), although for our purposes
a standard semi--Lagrangian scheme is used and mass is conserved via a
correction factor.

The ability of the UM to solve the non--hydrostatic deep--atmosphere
equations means it is uniquely suited to the study of hot
Jupiters. Additionally, the capability of the ENDGame dynamical core
to incorporate different simplifications to the dynamics, provides an
exceptional tool with which to explore hot Jupiter systems, and
determine the importance of the approximations made by previous works
modeling such atmospheres. The governing equations of the UM are those
best suited (of available GCMs) to modeling hot Jupiters. However, the
flow regimes expected in hot Jupiter atmospheres are particularly
under constrained, and very different from Earth. Furthermore, the
ENDGame dynamical core is not yet operational i.e. used for weather
prediction\footnote{ENDGame will be used for operational forecasts in
  early 2014.}. Therefore, given the exotic nature of the flow and the
use of a developmental code, we require extensive testing. Detailed
analytical analysis of the equation set used for the ND and ENDGame
dynamical cores has been performed and published \citep[see for
example][]{thuburn_2002a,thuburn_2002b}, alongside prescriptive tests
of atmospheric phenomena \citep{wood_2013}. However, little published
testing exists in the regime of idealised three--dimensional flows
integrated over long periods, as described previously and in
\citet{held_1994} and \citet{held_2005}. Moreover, existing testing
has not been performed on flow regimes with aspects in common with hot
Jupiters.

Therefore, we have performed a suite of test--cases using both the ND
and ENDGame dynamical cores of the UM ranging from an Earth--type
system to a full hot Jupiter system. In this work we present the
results for the Earth--type tests namely, the Held-Suarez test
\citep{held_1994}, the Earth-like test case of \citet{menou_2009} and
the Tidally Locked Earth of \citet{merlis_2010}. These tests progress
an Earth--like system, from a simple system, essentially driven by an
equator--to--pole temperature difference, to the inclusion of a
stratosphere and culminate with the modeling of a longitudinal
temperature contrast, which is expected for hot Jupiters. Further
development and alterations to the code are required for the modeling
of hot Jupiter atmospheres and, therefore, these results will be
presented in a subsequent publication.

The rest of this paper is structured as follows. Section \ref{code}
details the key formulations within the ND and ENDGame cores. Then in
Section \ref{models} we present the results of the test cases and
compare the results across the dynamical cores (ND to ENDGame) and
after adoption of the various simplifications to the dynamical
equations supported by the ENDGame formulation. We also compare with
results from literature using independent GCMs. Finally, in Section
\ref{conclusions} we discuss our results and conclude that the
dynamical cores of the UM are both self--consistent and consistent
with literature results obtained using other GCMs. As expected
invoking the `shallow--atmosphere' approximation does not
significantly alter the results for the flow regimes in our
Earth--like cases. We find, however, that the eddy kinetic energy over
the polar region, for the tidally locked Earth test case, increases
moving from the ND to ENDGame models. We also find a more symmetric
circulation pattern for the ENDGame models. These differences in the
ENDGame and ND flow are most likely caused by improvements in the
discretisation and numerical scheme used in the ENDGame model.

\section{Details of dynamical cores}
\label{code}

The dynamical cores of the UM, both the ND and ENDGame versions are
based on the Non-Hydrostatic Deep formulation (NHD) as described in
\citet{staniforth_2003,staniforth_2008,white_2005,wood_2013}. The
cores both use a latitude--longitude grid with a terrain following
height--based vertical coordinate\footnote{Although for this work we
  include no orography.}. The cores also have the same underlying
horizontal \citep[i.e. an Arakawa--C grid,][]{arakawa_1977}, and
vertical \citep[Charney--Phillips grid,][]{charney_1953} grid
structure, and both are semi--implicit and semi--Lagrangian.

\subsection{Improvements from ND to ENDGame}

Although the equation set and grid staggering are the same in ENDGame
and ND, the development of the ENDGame dynamical core includes a large
number of changes. In this paper we focus only on the details
pertinent to running a set of temperature forced test cases using the
dynamical core. The main changes from ND to ENDGame, with respect to
this aim, are explained in this section \citep[a more detailed
description of the ENDGame core can be found in][]{wood_2013}.

\subsubsection{Changes to the formulation}
\label{delta_formulation}

The ND dynamical core has been used operationally for several years
and results of simulations run using this core have been presented and
discussed in the literature \citep[for example
see][]{walters_2011}. The full equation set solved is the NHD
incorporating three momentum equations for the zonal, meridional and
vertical winds, $u$, $v$ and $w$, the continuity and thermodynamic
equation, and (in the absence of heating) the
equation--of--state. These are:

\begin{multline}
F^{u}=\frac{Du}{Dt}-\frac{uv\tan\phi}{r}+\frac{uw}{r}-2\Omega v\sin\phi+2\Omega w\cos\phi\\
+\frac{C_{p}\theta}{r\cos\phi}\frac{\partial \Pi}{\partial \lambda},\\
F^{v}=\frac{Dv}{Dt}+\frac{u^2\tan\phi}{r}+\frac{vw}{r}+2\Omega u\sin\phi+\frac{C_{p}\theta}{r}\frac{\partial \Pi}{\partial \phi},\\
\delta F^{w}=\delta\frac{Dw}{Dt}-\frac{u^2+v^2}{r}-2\Omega u\cos\phi+g(r)+C_{p}\theta\frac{\partial \Pi}{\partial r},\\
0=\frac{D \rho}{Dt}+\rho \left[ \frac{1}{r\cos\phi}\frac{\partial u}{\partial \lambda}+\frac{1}{r\cos\phi}\frac{\partial (v\cos\phi)}{\partial \phi}+\frac{1}{r^2}\frac{\partial (r^2w)}{\partial r} \right],\\
\frac{D\theta}{Dt}=0,\\
\Pi^{\frac{1-\kappa}{\kappa}}=\frac{R\rho\theta}{p_0},\\
\label{full}
\end{multline}
where, $\lambda$, $\phi$, $r$ and $t$ are the longitude, latitude
(measured from equator to pole), radial distance from the centre of
the planet and time, respectively. $\Omega$, $g(r)$, $R$, $C_{p}$ and
$\kappa$ are the rotation rate, gravitational acceleration, gas
constant, the heat capacity at constant pressure, and the ratio
$R/C_{p}$, respectively. $F^{u,v,w}$ represent sink or source terms
for the momenta, $p_0$ is the reference pressure, conventionally
chosen to be $10^5$ Pa, and $\delta$ is a `switch' ($\delta=0$ or $1$)
to enable a quasi-hydrostatic equation set \citep[not studied here,
see][for explanation]{wood_2013}. $\rho$, $\theta$ and $\Pi$ are the
density, potential temperature and Exner function (or Exner
pressure). $\theta$ is given by,
\begin{equation}
\theta = T\left(\frac{p_0}{p}\right) ^{R/C_p},\\
\end{equation}
where $T$ is temperature, and $p$ is pressure. $\Pi$ is given by,
\begin{equation}
\Pi=\left(\frac{p}{p_0}\right)^{R/C_{p}} = \frac{T}{\theta}.\\
\end{equation}
Finally, the material derivative ($\frac{D}{Dt}$) is given by,
\begin{equation}
\frac{D}{Dt}\equiv \frac{\partial}{\partial t}+\frac{u}{r\cos\phi}\frac{\partial}{\partial \lambda}+\frac{v}{r}\frac{\partial}{\partial \phi}+w\frac{\partial}{\partial r}.\\
\end{equation}

Despite solving a set of dynamical equations close to the
fully-compressible Euler equations (transformed to a rotating
reference frame), i.e. involving very few approximations, some
simplifications still remain including:

\begin{itemize}
\item{Spherical Geopotential (spherical symmetry): $\Phi
    (\lambda,\phi,r)=\Phi (r)$, where $\Phi$ is the geopotential
    (i.e. the gravitational potential plus the centrifugal
    contribution). Here the geopotential is constant at a given height
    \citep[i.e. the latitude and, much smaller, longitude dependencies
    are dropped, the effect of this assumption is small for the Earth,
    for a full discussion on geopotentials see][]{white_2008}.}
\item{Constant \textit{apparent} Gravity: $g(r)=g_{\rm surf}$, where
    $g_{\rm surf}$ is the gravitational constant at the Earth's
    surface and is adopted throughout the atmosphere (and ocean). As
    this value is that measured on the Earth's surface (at the
    equator) the magnitude of the centrifugal component is
    incorporated. This neglects the contribution of the atmosphere
    itself to the gravitational potential (self--gravity).}
\end{itemize}

In the ENDGame dynamical core the geopotentials are still approximated
as spheres but \textbf{the acceleration due to gravity may vary with
  height}.  It is unclear what effect either of these assumptions has
on the reliability of weather or climate
predictions. \citet{white_2005} classify four consistent
(i.e. conservative of energy, axial angular momentum and vorticity)
equations sets for global atmosphere models. Each equation set
involves a different combination of approximations, as detailed in
\citet{white_2005}. Table \ref{assume} summarises the main
approximations, their effect on the equations of motion and their
validity.

\begin{table*}[t]
  \caption{Table showing approximations made to the equations of motion (or
    associated geometry), the actual effect on the terms of Equation
    (\ref{full}) and the validity criteria. Here $R_{\rm p}$ is the radius
    of the planet, $z$ is the distance from the surface of the planet,
    i.e. $r=z+R_{\rm p}$, $M_{\rm p}$ is the mass of the planet, in this
    case Earth, and $N$ is the buoyancy (or Brunt-V\"ais\"al\"a)
    frequency. (1) This validity criterion is from \citet{phillips_1968},
  however, the validity of the `traditional' approximation is debatable
  and may break down for planetary scale flows
  \citep[see][for a discussion]{white_1995}. \label{assume}}
\vskip4mm
\centering
\begin{tabular}{lccc}
\hline
\multicolumn{2}{l}{Assumption}&Mathematical effect&Validity\\
\hline
\multicolumn{2}{l}{Spherical geopotentials}&$\Phi(\lambda,\phi,r)=\Phi (r)$&$\Omega^2r\ll g$\\
\multirow{3}{*}{`Shallow--atmosphere'}\ldelim\{{3}{0mm}[]&Constant gravity&$g(r)=g_{\rm surf}=\frac{GM_{\rm p}}{R_{\rm p}^2}$&$z\ll R_{\rm p}$\\
&`Shallow--fluid'&$r\rightarrow R_{\rm p}$ \& $\frac{\partial}{\partial r}\rightarrow\frac{\partial}{\partial z}$&$z\ll R_{\rm p}$\\
&`Traditional'&$\frac{uw}{r}$, $\frac{vw}{r}$, $\frac{u^2+v^2}{r}$, $2\Omega u\cos\phi$, $2\Omega w\cos\phi\rightarrow 0$&$N^2\gg \Omega^2$$^{(1)}$\\
\hline
\end{tabular}
\end{table*}

If one approximates the atmosphere as a `shallow--fluid' then in order
to retain a consistent equation set one must also adopt the
`traditional' approximation \citep{white_2005}. \citet{white_2005},
therefore, define the `shallow--atmosphere' approximation as the
combination of the `shallow--fluid' and traditional' approximations
\citep[the `traditional' approximation is not invoked based on
physical arguments and in fact may be invalid for planetary scale
flows, see discussion in][]{white_1995}, and also include the
assumption of constant gravity, a nomenclature we adopt (see Table
\ref{assume}). This results in a consistent equation set termed the
non--hydrostatic shallow--atmosphere equations (NHS). Although the ND
dynamical core is based on the NHD equations the constant gravity
approximation is still made, essentially meaning the core is based on
a pseudo--NHD system. When moving to a shallow, NHS type system the
omission of gravity variation is not as immediately inconsistent as
adopting a `shallow--fluid' without the `traditional'
approximation. \citet{white_2012} explain, in the NHS framework,
approximating geopotentials to be spherical leads to a spurious
divergence of this potential (which should be zero), which is
increased if gravity is allowed to vary with height. A more detailed
comparison of the NHS and NHD atmosphere equations and their
conservative properties can be found in
\citet{staniforth_2003,white_2005}.

One unique and scientifically useful capability of the ENDGame core is
the ability to `switch' the underlying equation set solved, without
changing the numerical scheme. ENDGame is capable of solving, within
the same numerical framework, either the NHS or NHD equations and
further invoking constant or varying gravity (with height). Almost all
of the GCMs applied to the study of exoplanets have solved the
Hydrostatic Primitive Equations \citep[HPEs][]{white_2005}, involving
the assumption of vertical hydrostatic equilibrium and a
`shallow--atmosphere'. For the test cases studied in this work the
assumptions listed in Table \ref{assume} are generally valid, or at
least have a small effect on the results. When modeling hot Jupiters
however, one might expect such approximations to break down, for
example, the ratio of the modeled atmospheric extent to planetary
radius is much larger (i.e. aspect ratio in this work $\sim 10^{-3}$,
but for hot Jupiters $\sim 0.1$). Therefore, the ability of ENDGame to
relax or invoke the canonically made approximations, and thereby
cleanly test their impact, will prove vital.

\subsection{Changes to the numerical scheme}
\label{delta_num_scheme}

The ND and ENDGame dynamical cores are both semi--implicit and based
on a Crank-Nicolson scheme, where the the temporal weighting between
the $i\,th$ and the $i+1\,th$ state is set by the coefficient
$\alpha$. This leads to a non--linear set of equations which must be
solved. The key change to the numerical scheme from ND to ENDGame has
been the method of overcoming the non--linearity of the problem, for
each atmospheric timestep. A nested iteration structure is now
used. The \textit{outer} iteration performs the semi--Lagrangian
advection (including calculation of the departure points) and the
\textit{inner} iteration solves the Helmholtz problem to obtain the
pressure increments. The Coriolis and nonlinear terms are updated and
the pressure increments from the \textit{inner} iteration are back
substituted into the \textit{outer} loop to obtain updated values for
each prognostic variable. There has also been a change in the spatial
discretisation, such that the meridional wind is stored at the
poles. Consequently pressure is not stored at the poles, thus removing
the polar problem from the semi-implicit solver
\citep{wood_2013}\footnote{\citet{thuburn_2004} also show that mass,
  angular momentum and energy are much more readily conserved using
  grid staggering such that presure is not stored at the poles.}. The
values of meridional wind stored at a pole serve as boundary values
for that field in an infinitessimal approach to the pole. Such
boundary values are required for the determination of semi-Lagrangian
departure points close to the pole, and for interpolation of the
meridional wind field to those points.

\begin{figure}[t]
\vspace*{2mm}
\begin{center}
\includegraphics[width=8.5cm,angle=0.0,origin=c]{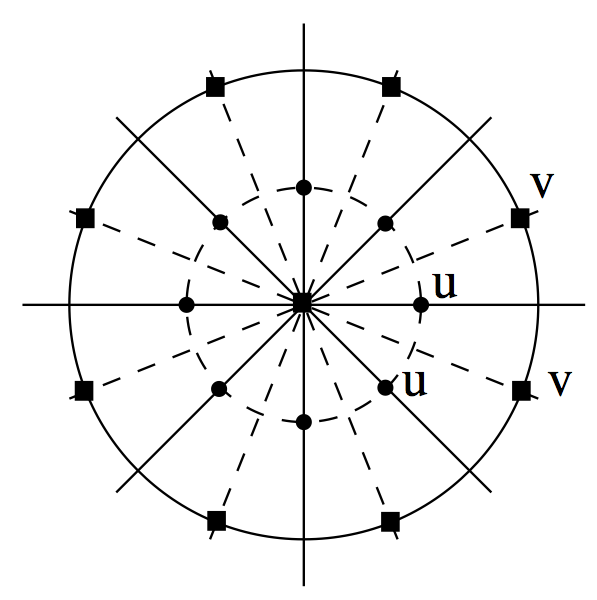}
\end{center}
\caption{Schematic depicting arrangement of winds around a pole in the
  latitude-longitude grid. Zonal wind components, $u$, are used to
  determine a horizontal wind vector at the pole, using a
  least-squares best fit to an assumed solid-body
  rotation.\label{pole}}
\end{figure}

Figure \ref{pole} shows the arrangement of zonal and meridional wind
components around a pole. Circles show the location of the zonal wind
($u$) and squares the location of the meridional wind ($v$). The polar
values of $v$ are obtained by assuming that the wind across the pole
is that of a solid-body rotation; the magnitude and direction of this
polar wind being determined by a least-squares best fit to the zonal
wind on the grid-row closest to the pole. The changes to the spatial
and temporal discretisation included in the ENDGame dynamical core
have led to greater stability at the pole, and have removed the need,
in most cases, for polar filters. For cases where $v$ becomes
significant (as demonstrated in Section \ref{tle}) a `sponge layer'
\citep {klemp_2008,melvin_2010} has been implemented which allows
damping of vertical velocity (usually from gravity or acoustic waves),
which can be used as part of the upper boundary condition and extend
down to the surface at each pole.

\section{Test Cases}
\label{models}

As part of our project to model exoplanets we have installed the
externally released UM VN7.9, using the ND dynamical core and VN8.2,
adapted to use the developmental ENDGame dynamical core. We have, in
order to check the veracity of our version of the code and test
regimes approaching our target systems of hot Jupiters, then run each
version through a set of test cases. These test cases isolate the
dynamical core and solve for the atmosphere only, in the absence of
orography. The test cases presented in this work are the original
(simple) Held-Suarez test \citep[HS,][]{held_1994}, a simple
Earth-Like test case including a stratosphere \citep[EL,][]{menou_2009}
and a hypothetical tidally locked Earth, allowing the opportunity to
explore the model performance with a longitudinal temperature contrast
\citep[TLE,][]{merlis_2010,heng_2011}.

For these tests radiative transfer is parameterised using simple
temperature forcing to a prescribed temperature profile or `Newtonian
cooling', and the heating rate is therefore set by the Newtonian
heating rate, $Q_{\rm Newton}$. Practically, however, the codes uses
potential temperature as a prognostic, thermodynamic variable and
therefore the heating rate is prescribed by
\begin{equation}
  Q=Q_{\rm Newton}=-\Pi\left(\frac{\theta-\theta_{\rm eq}}{\tau_{\rm rad}}\right),
  \label{newt_cool}
\end{equation}
where $\tau_{\rm rad}$ the characteristic radiative or relaxation
timescale and can be set as constant or as a function of position
(latitude) and pressure or height. $\theta_{\rm eq}$ is the
equilibrium potential temperature and is derived from the equilibrium
temperature profile ($T_{\rm eq}$) using
\begin{equation}
\theta_{\rm eq}^{i}=\frac{T_{\rm eq}}{\Pi^{i}},
\end{equation}\textbf
where superscript $i$ denotes the current timestep. Practically, the
potential temperature is adjusted explicitly within the
semi--Lagrangian scheme using
\begin{equation}
\theta^{i+1}=\theta_{D}^{i}-\frac{\Delta t}{\tau_{\rm rad}}\left(\theta^{i}-\theta_{\rm eq}^{i}\right)_{D},
\end{equation}
where the superscript $i+1$ denotes the next timestep and $\Delta t$
is the length of the timestep. The subscript $D$ denotes a quantity at
the departure point of the fluid element \citep[see explanation in
Section \ref{delta_num_scheme} and][for a full discussion]{wood_2013}
\footnote{From the equations in this section one can recover, $Q_{\rm
    Newton}=\frac{T_{\rm eq}-T}{\tau_{\rm rad}}$ and
  $T^{i+1}=T^{i}-\frac{\Delta t}{\tau_{\rm rad}}(T^{i}-T_{\rm eq})$ as
  shown, for example in \citet{heng_2011}.}. Boundary layer friction
is also represented using a simple `Rayleigh friction' scheme, where
the horizontal winds are damped close to the surface (again
explicitly),
\begin{equation}
u^{i+1}=u^{i}-\frac{\Delta t}{\tau_{\rm fric}}u^{i},\\
\end{equation}
(and similarly for $v$) where $\tau_{\rm fric}$ is the characteristic
friction timescale, and as with $\tau_{\rm rad}$ can be a constant or
a function of position and pressure or height.  Therefore, each test
case prescribes three `profiles': an equilibrium temperature,
relaxation or radiative timescale and horizontal frictional timescale
profile.

Finally, each simulation has also been run including a very simple dry
static adjustment of $\theta$ to remove any convective instability. As
the condition for convective instability is $\frac{d\theta}{dz}<0$,
each column is examined for negative vertical potential temperature
gradients after each timestep. If a column is found to be convectively
unstable $\theta(z)$ is re-arranged, i.e. the temperature in the
column is just rearranged to ensure stability. Practically, this
routine only operates over the pole where the atmosphere can become
unstable to convection. The original Held--Suarez test does not
include a dry static adjustment scheme, and the atmosphere is close to
being neutrally stable over the poles, meaning our results will differ
slightly. However, the effect of including a convective adjustment
scheme has been explored for several Earth--like test cases by
\citet{heng_2011b}, and been shown to be negligible.

\subsection{Models run}
\label{models_run}

We have run each test case using ND and ENDGame. We have also run each
test case using ENDGame but varying the set of simplifications or
assumptions to the dynamical equations. Table \ref{model_names} shows
the names we use to refer to different model setups, the dynamical
core used, the underlying equation set and the associated
approximations (the approximations are as discussed in Section
\ref{delta_formulation} and presented in Table \ref{assume}).

\begin{table*}[t]
  \caption{Table showing the name used in this work with the dynamical core, the name for the equation set \citep[as described in][]{white_2005} and the main included assumptions. For a full description of the underlying equations see \citet{white_2005} \label{model_names}}
  \vskip4mm
  \centering
\begin{tabular}{lcccc}
\hline
Short--Name&EG$_{\rm sh}$&EG$_{\rm gc}$&EG&ND\\
\hline
Dynamical core&ENDGame&ENDGame&ENDGame&New Dynamics\\
\citet{white_2005} equation set&NHS&NHD&NHD&NHD\\
Spherical geopotentials&Yes&Yes&Yes&Yes\\
Constant gravity&Yes&Yes&No&Yes\\
`Shallow--atmosphere'&Yes&No&No&No\\
\hline
\end{tabular}
\end{table*}

The model EG$_{\rm gc}$ setup was chosen explicitly to match the ND
equations, and thereby allow us to potentially isolate differences in
solution caused by changes in the numerical scheme between the
dynamical cores. These runs are compared and discussed for each test
case in turn, alongside comparison to the original test, in this
section. These practical tests complement the analysis of normal modes
in \citet{thuburn_2002a,thuburn_2002b}, and standardised flow tests
\citep[e.g.][]{ullrich_2013,wood_2013}. The general parameters for the
model runs are listed in Table \ref{gen_par}.

\begin{table}[t]
  \caption{Table showing the general parameters adopted for the calculations. G72N45 is notation for 144 longitude points and 90 latitude points and $N_{\rm z}$ is the number of vertical levels. T$_{\rm init}$ is the temperature adopted for our initial hydrostatically stable isothermal atmosphere (as explained in Section \ref{init_cond}) and $\Delta T_{\rm sample}$ is the temporal distance between model outputs.\label{gen_par}}
  \vskip4mm
  \centering
\begin{tabular}{lc}
  \tophline
  Parameter&Value\\
  \middlehline
  Horizontal Resolution&G72N45\\
  N$_{\rm z}$&32\\
  Timestep (s)&1200\\
  T$_{\rm init}$ (K)&264\\
  $\Delta T_{\rm sample}$ (days)&10\\
  Temporal weighting, $\alpha$&0.7 (ND), 0.55 (EG)\\
  \bottomhline
\end{tabular}
\end{table}

\subsection{Vertical coordinate \& methods of model comparison}
\label{comp}

The literature sources which we compare our results with all used GCMs
which adopt pressure or $\sigma$ as their vertical coordinate ($\sigma
= \frac{p}{p_{\rm surf}}$, where $p_{\rm surf}$ is the surface
pressure), whereas the UM is height-based \citep[the MCore is another
example of a dynamcial core adopting a height--based coordinate,
see][for a description]{ullrich_2012}. This creates some barriers to a
clean comparison between our models and the literature
examples. Firstly, the boundary conditions (and therefore model
domain) can only be approximately matched. Secondly, our vertical
resolutions, and more specifically, level placements will be
different. Finally, to explicitly compare the results we must
transform our results to $\sigma$ space.

Our upper boundary, being constant in height, will experience
fluctuations in pressure\footnote{In most pressure--based models the
  inner boundary is still a constant height surface.}. Practically, the
initial pressure of the inner boundary (or surface) is set and a
domain large enough so as to reach the lowest required pressure is
selected. Therefore, if the horizontal or temporal pressure gradients
are significant our model domain will not match that of a pressure
based model, where the upper boundary is a constant pressure
surface. While this is not the case for the tests in this work, for
our work on hot Jupiters changes in the pressure on the top boundary
can lead to a significant change in the physical size of the domain
(Mayne et al, submitted). The distribution of levels within our domain
can then be selected to sample the associated $\sigma$ space evenly to
match the literature models. Practically, for each test case we run a
model with a (moderate resolution) uniform grid over a domain
extending to pressures lower than sampled in the original, literature,
$\sigma$ model. Zonal and temporal averages are then used to create a
set of level heights (and an upper boundary position) to emulate even
$\sigma$ sampling. We have also, when compared to the literature
models we examine, increased our number of vertical levels to ensure
sufficient resolution. The resulting level heights for each test case
are presented in Table \ref{vert_levels} in dimensionless height
coordinates, alongside the approximate $\sigma$ value of each level.

Comparison of our models with literature results then requires
additional conversion. Although our level and boundary placements have
been selected to better sample the required $\sigma$ space we still
use geometric height as our vertical coordinate. Therefore, for each
completed test case, the pressure (and therefore $\sigma$) values are
found and the prognostic variable is interpolated (at every output
timestep) into $\sigma$ space.

To determine a satisfactory match of the mean, large--scale, long term
structure of our modeled atmospheres with literature results, we
compare the prognostic fields of velocity and temperature. These
fields are averaged (using a mean) in the diagnostic plots of the
original publications in both time and space. Additional care must be
taken when performing spatial averaging and comparing models across
different vertical coordinates \citep[as discussed in the Appendix
of][]{hardiman_2010}. Where we are comparing directly to a literature
figure or result we perform the spatial averaging in $\sigma$
space. The required prognostic field is (as discussed above)
interpolated from the height grid onto a $\sigma$ grid, and then the
average performed along constant $\sigma$ surfaces, to allow the most
consistent comparison with literature, $\sigma$--based models. To
further enhance the comparison of our results with those in the
literature, where possible the line contours (solid lines for positive
values and dotted lines for negative) presented in the plots of our
model results have been chosen to match the original publications. We
have then, to aid a qualitative interpretation of our models,
complemented the line contours with additional (more numerous) colour
contours. For plots showing wind or circulation patterns the coloured
contours are separated at zero (where blue represents negative flow,
and red positive\footnote{The splitting means that the red and blue
  colour scales need not be symmetric about zero.}), again to aid
visual presentation of the flow. Each of the original publications
introducing the tests we have performed include the comparison of
additional quantities \citep[for example the eddy temperature and wind
variance in][]{held_1994}. In this work, however, for brevity (as we
are performing several tests) we compare only the prognostic variable
fields, i.e. wind and temperature, complemented by comparison of the
Eddy Kinetic Energy (EKE) defined as
\begin{equation}
  {\rm EKE}=\frac{\left( u^{\prime 2}+v^{\prime 2}\right)}{2},\\
\end{equation}
where the prime denotes a perturbation such that
$u^{\prime}=u-\overline{u}^{\lambda_{\rm z},t}$, where
$\overline{u}^{\lambda_{\rm z},t}$ is the variable averaged (mean) in
longitude ($\lambda$) and time ($t$). One critical difference with
this quantity (compared to the others we plot) however, is that the
spatial (zonal) average is performed in height coordinates (hence the
subscript $z$). Therefore, plots of EKE will be presented in height
not $\sigma$ space. This is done as we compare the zonal and temporal
mean of the EKE, i.e. $\overline{{\rm EKE}}^{\lambda_{\rm
    z},t}$. Given that the perturbation itself is constructed from a
spatial and temporal mean, we are performing several averaging
processes and it is simpler and more intuitive to keep the variable in
the natural coordinate system of the model. Moreover, in the case of
EKE, we are actually comparing only our own models with each other,
not with a literature $\sigma$-based model. The EKE then allows us to
explore differences in the eddy structures of the models,
complementary to the plots depicting the relatively insensitive means
of the wind and temperature fields. Additional details regarding the
comparison between our work and that of \citet{heng_2011} can be found
in Appendix \ref{heng_comp}.

\subsubsection{Initial conditions}
\label{init_cond}

As stated in \citet{held_1994}, for their HS test an initial spin--up
time of 200 days is used to effectively allow the system to reach a
statistically steady--state and erase the initial conditions. This is
why \emph{temporal average} (whenever it is stated as being performed)
means the average of the field from 200 to 1200 days. Our adopted
initial conditions were a simple, hydrostatically balanced, isothermal
atmosphere (temperature presented in Table \ref{gen_par}) with zero
$u$,$v$ and $w$ velocities.

\subsection{Held--Suarez}
\label{hs}

The HS test prescribes an equilibrium temperature profile of
\begin{equation}
T_{\rm eq}={\rm max} \{ T_{\rm stra},T_{\rm HS} \},\\
\end{equation}
where, 
\begin{multline}
T_{\rm HS}=\\
\left[ T_{\rm surf}-\Delta T_{\rm EP}\sin^2\phi - \Delta T_{z}\ln\left( \frac{p}{p_0}\right)\cos^2\phi\right] \left( \frac{p}{p_0}\right) ^{\kappa},
\end{multline}
and, $T_{\rm stra}=200$ K, $T_{\rm surf}=315$ K, $\Delta T_{\rm
  EP}=60$ K, $\Delta T_{z}=10$ K and $p_0=1\times 10^5$
Pa\footnote{All units used are SI units.}.  The radiative timescale is
modeled as,
\begin{multline}
\frac{1}{\tau_{\rm rad}}=\\
\frac{1}{\tau_{\rm rad,d}}+\begin{cases}
0\mbox{,}\,&\sigma \leq \sigma_{\rm b}\mbox{,}\\
\left( \frac{1}{\tau_{\rm rad,u}}-\frac{1}{\tau_{\rm rad,d}} \right) \left( \frac{\sigma - \sigma_{\rm b}}{1-\sigma_{\rm b}}\right)\cos^4 \phi \mbox{,}\,&\sigma > \sigma_{\rm b}\mbox{,}
\end{cases}
\end{multline}
where, $\tau_{\rm rad,d}=40$ days, $\tau_{\rm rad,u}=4$ days and
$\sigma_{\rm b}=0.7$ (the top of the surface friction boundary layer).

The boundary layer horizontal wind damping enforces a damping on a
timescale, $\tau_{\rm fric}$ given by:
\begin{align}
\frac{1}{\tau_{\rm fric}}=\begin{cases}
0\mbox{,}\,&\sigma \leq \sigma_{\rm b}\mbox{,}\\
\left( \frac{1}{\tau_{\rm fric,f}}\right) \left(\frac{\sigma - \sigma_{\rm b}}{1-\sigma_{\rm b}}\right) \mbox{,}\,&\sigma > \sigma_{\rm b}\mbox{,}
\end{cases}
\end{align}
where, $\tau_{\rm fric,f}=1$ day.

\begin{figure}[t]
\vspace*{2mm}
\begin{center}
\includegraphics[width=7.8cm,angle=0.0,origin=c]{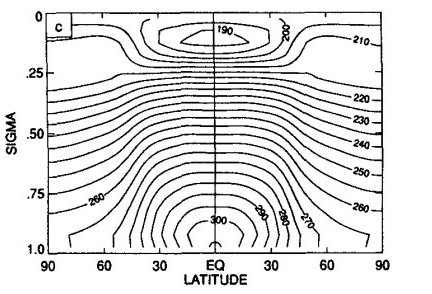}
\hspace*{-1.0cm}\includegraphics[width=6.5cm,angle=90.0]{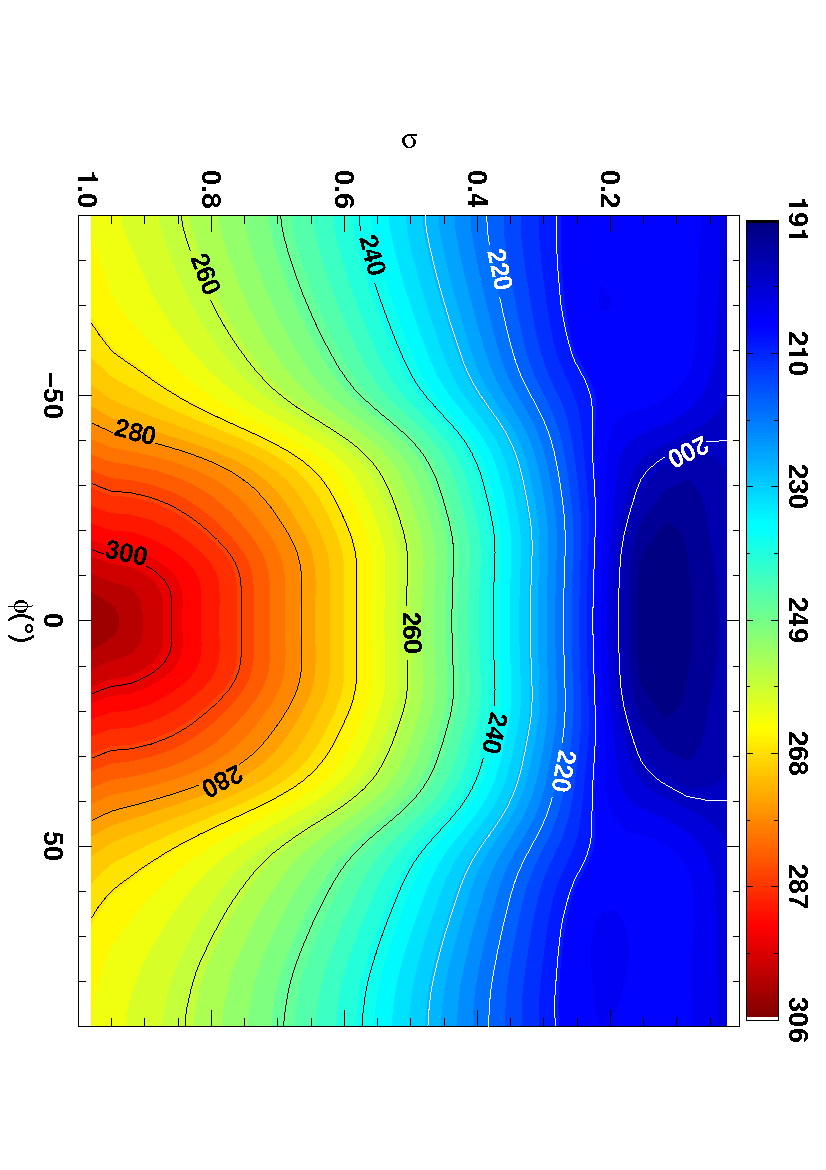}
\hspace*{-1.0cm}\includegraphics[width=6.5cm,angle=90.0]{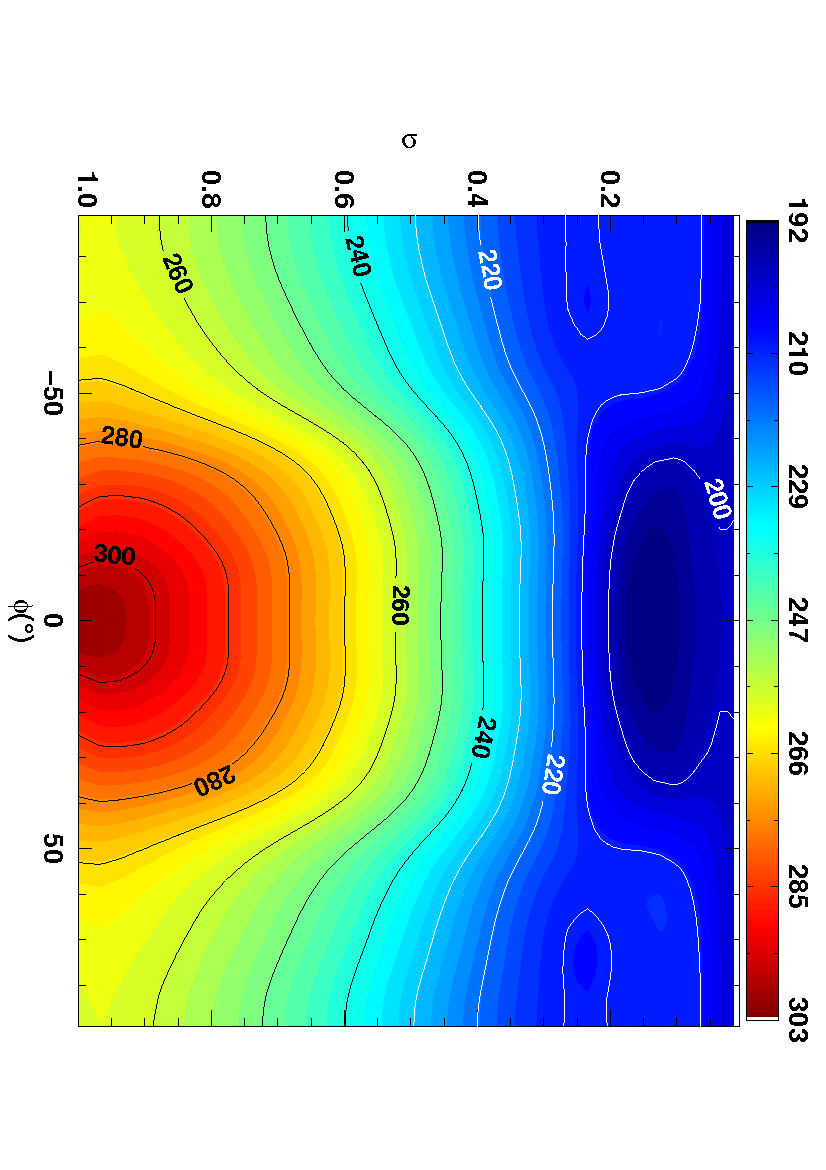}
\end{center}
\caption{Figure showing, for the Held--Suarez test \citep{held_1994},
  the zonally and temporally averaged (i.e. mean from 200 to 1200
  days, see Section \ref{init_cond}) temperature (K) as a function of
  latitude and $\sigma$. \textit{Top panel:} original finite
  difference model Figure 1 from \citet{held_1994}, (c) American
  Meteorological Society. Used with permission. \textit{Middle panel:}
  ND version. \textit{Bottom Panel:} EG version (see Table
  \ref{model_names} for explanation of model types).\label{HS_T}}
\end{figure}

\begin{figure}[t]
\vspace*{2mm}
\begin{center}
\includegraphics[width=7.7cm,angle=0.0,origin=c]{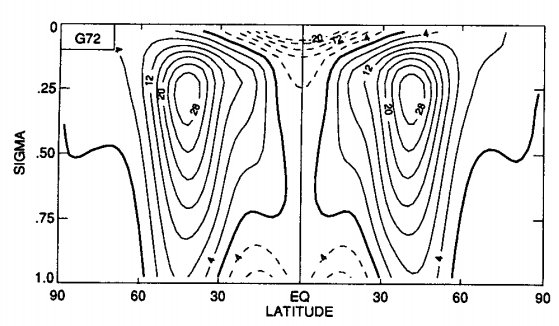}
\hspace*{-1.0cm}\includegraphics[width=6.5cm,angle=90.0]{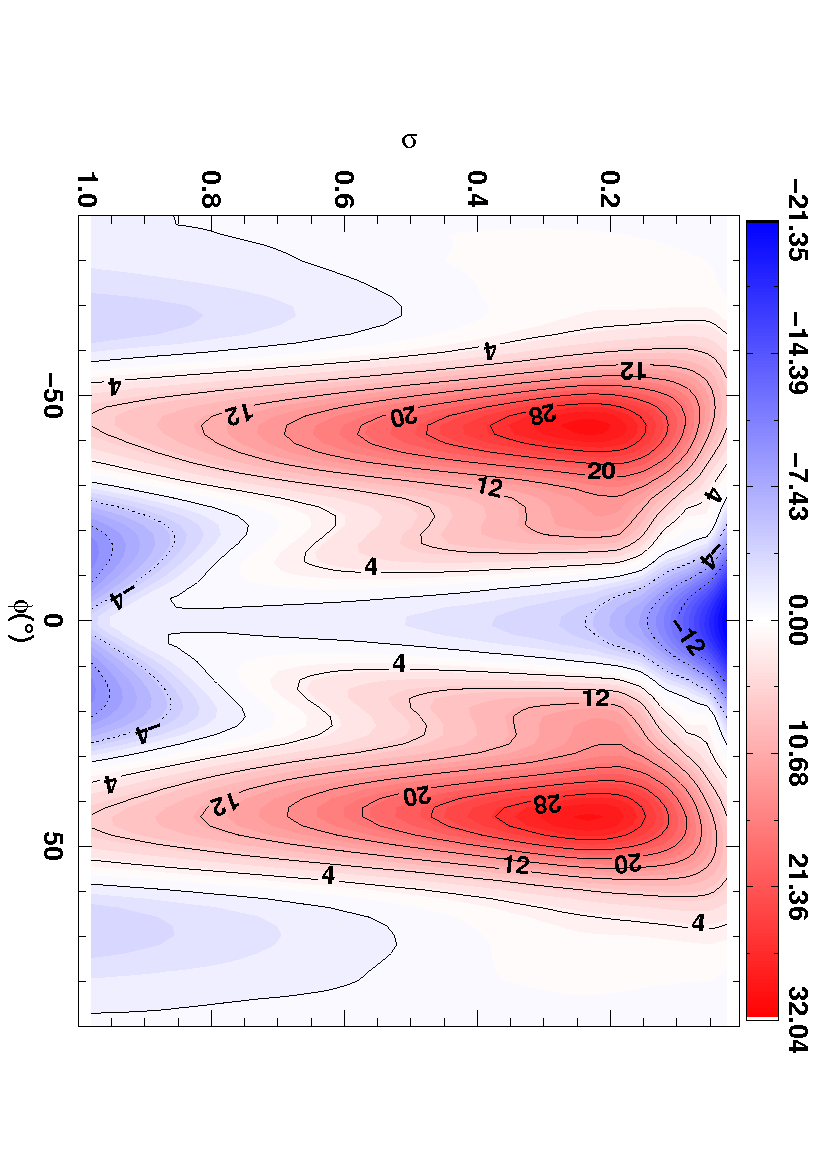}
\hspace*{-1.0cm}\includegraphics[width=6.5cm,angle=90.0]{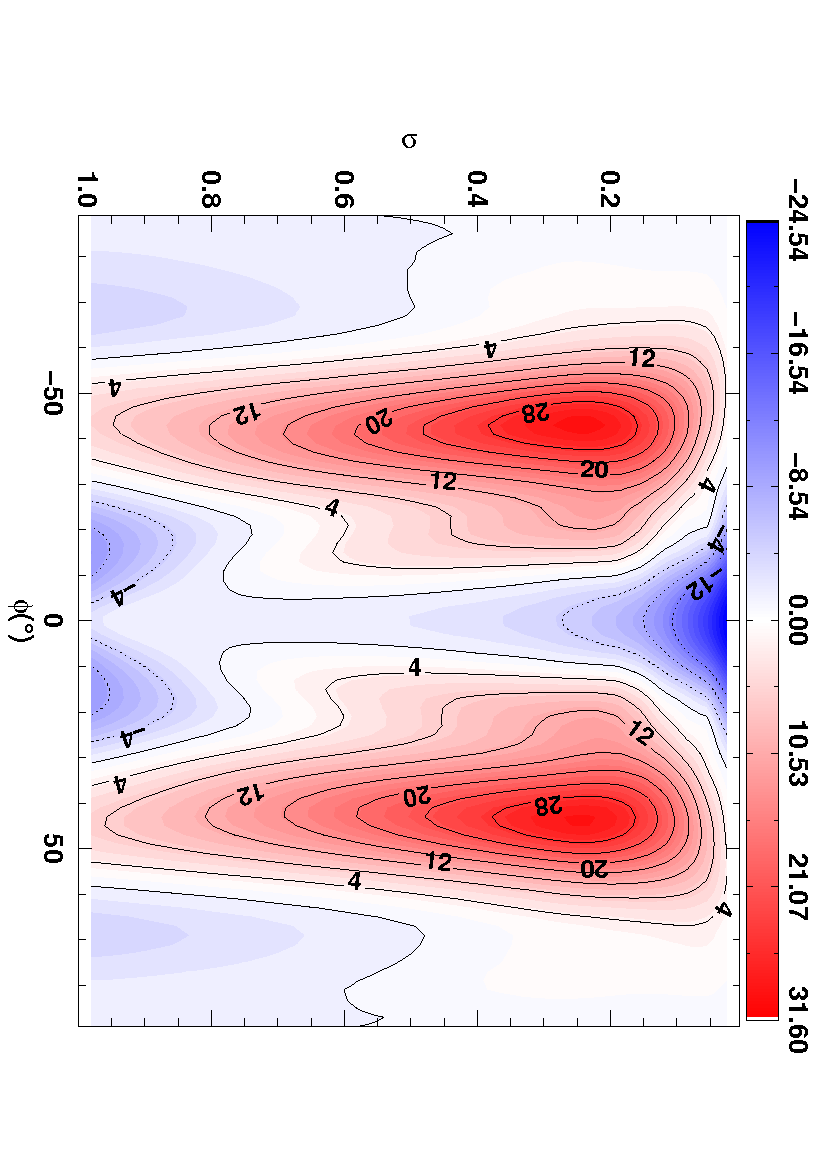}
\end{center}
\caption{Same as Figure \ref{HS_T} but for zonal wind
  (ms$^{-1}$).\label{HS_U}}
\end{figure}

Figures \ref{HS_T} and \ref{HS_U} show the zonally (along constant
$\sigma$ surfaces) and temporally averaged zonal wind and temperature
($\overline{u}^{\lambda_{\sigma},t}$ and
$\overline{T}^{\lambda_{\sigma},t}$), respectively, from the original
\citep{held_1994} publication, and from our ND and ENDGame setups.

Qualitatively, both the ND (\textit{middle panel}) and the EG
(\textit{bottom panel}) temperature and zonal wind fields (when
averaged zonally and temporally) match the original \citet{held_1994}
(\textit{top panel}) results of the finite difference model. However,
the 210 K contour (Figure \ref{HS_T}), and the wind contours extending
over the poles, and over the equator (Figure \ref{HS_U}) show a
slightly better match with \citet{held_1994} when moving from the ND
to the ENDGame models (however these flows represent very small
velocities $\lesssim 1$ ms$^{-1}$). The ND model shows a slightly
different vertical temperature profile for the lowest levels, when
compared to the EG model. This is caused by differences in the
temperature modeled in the lowest grid cell. The ENDGame model records
the temperature, in the atmosphere array, down to the surface, whereas
ND does not. Therefore, for display purposes the potential temperature
across the bottom cell has been estimated to be constant in the ND
model, resulting in a slight increase of temperature (as $T=\Pi\theta$
and the lowest $\sigma\sim0.97$, and by definition $\sigma_{\rm
  surf}\equiv 1$, see Table \ref{vert_levels}).

\begin{figure}[t]
\vspace*{2mm}
\begin{center}
\hspace*{-1.0cm}\includegraphics[width=6.5cm,angle=90.0]{./Figs/HS_EG_Deep_Uvel.png}
\hspace*{-1.0cm}\includegraphics[width=6.5cm,angle=90.0]{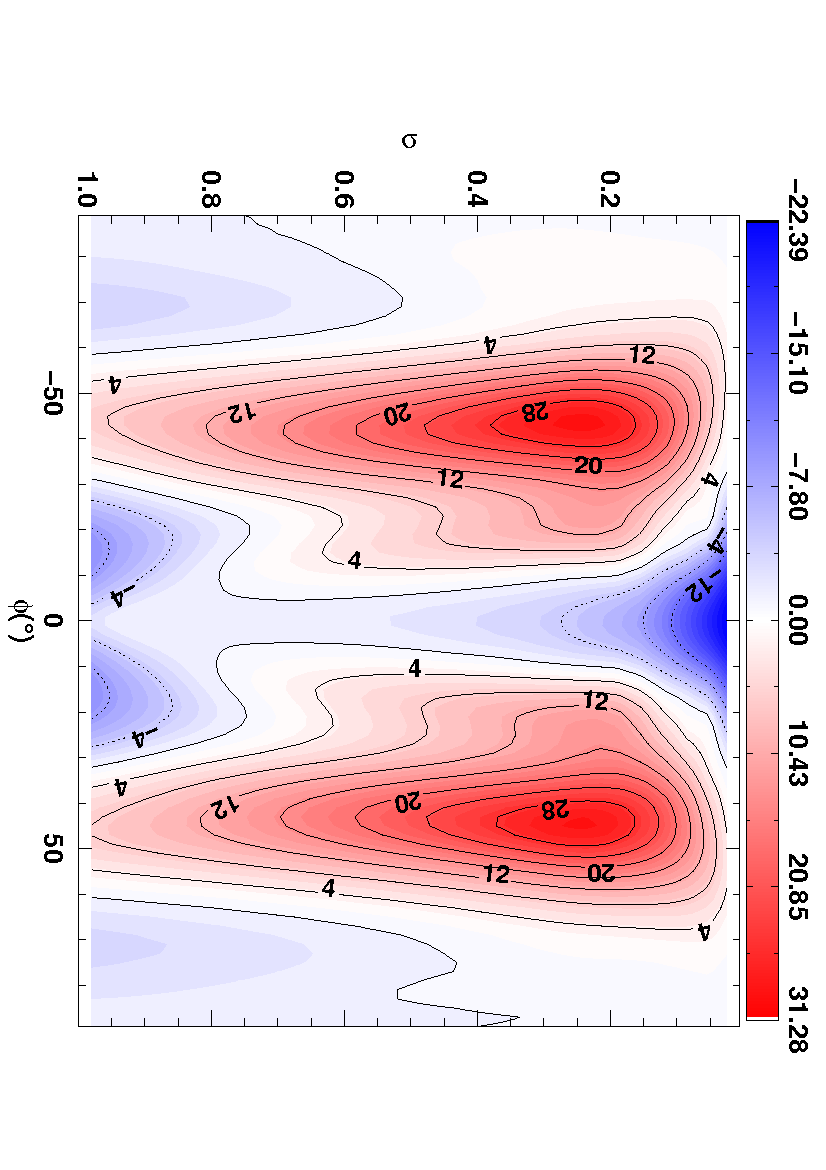}
\hspace*{-1.0cm}\includegraphics[width=6.5cm,angle=90.0]{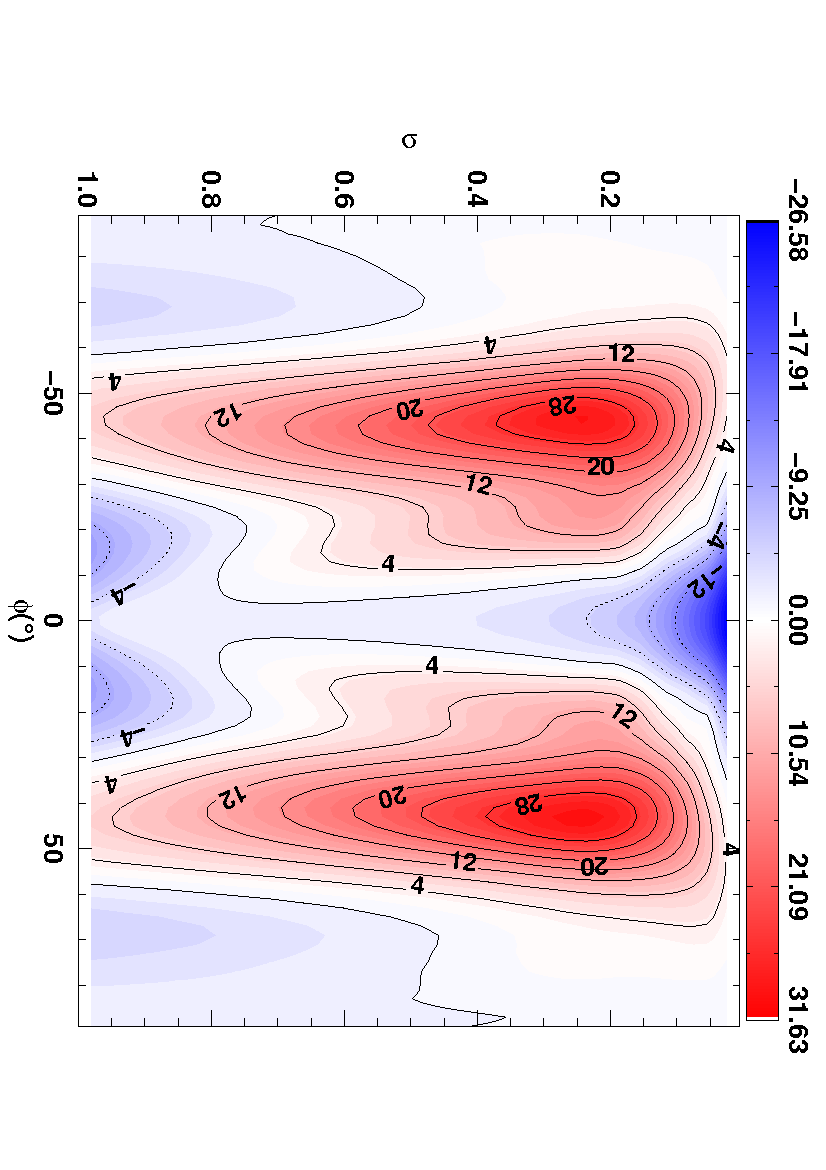}
\end{center}
\caption{Figure, for the Held--Suarez test \citep{held_1994}, showing
  the zonally and temporally averaged zonal wind (ms$^{-1}$) as a
  function of latitude and $\sigma$. \textit{Top panel:} EG model
  (also shown in Figure \ref{HS_U} but reproduced here to aid
  comparison). \textit{Middle panel:} EG$_{\rm gc}$
  model. \textit{Bottom panel:} EG$_{\rm sh}$ model (see Table
  \ref{model_names} for explanation of model types). \label{HS_EG}}
\end{figure}

Figure \ref{HS_EG} shows zonally and temporally averaged zonal wind
plots for all of the ENDGame models (namely, EG, EG$_{\rm gc}$ and
EG$_{\rm sh}$, where EG has been presented already in Figure
\ref{HS_U} but is reproduced in Figure \ref{HS_EG} to aid visual
comparison). The similarity of the \textit{panels} of Figure
\ref{HS_EG} shows that, as expected for such a domain and flow regime
(i.e. the lack of large, in vertical extent, circulation cells),
making the `shallow--atmosphere' approximation (or approximating
gravity as a constant only) does not significantly affect the
resulting long--term large--scale flow. There is tentative evidence,
if one scrutinises the flow over the pole, for the subsequent
simplification of the model moving it towards the \citet{held_1994}
result, however, the velocities in these regions are small ($<1$
ms$^{-1}$). These results also match the spectral and grid--based
models of \citet{heng_2011} \citep[see Figures 1 \& 2
of][]{heng_2011}. Another important point to note is that in
\citet{held_1994} the model was run using 20 vertical levels.  We have
adopted 32 vertical levels, and the agreement between our results and
those of \citet{held_1994} is a promising indication that we have used
sufficient resolution.

\begin{figure}[t]
\vspace*{2mm}
\begin{center}
\hspace*{-1.0cm}\includegraphics[width=6.5cm,angle=90.0]{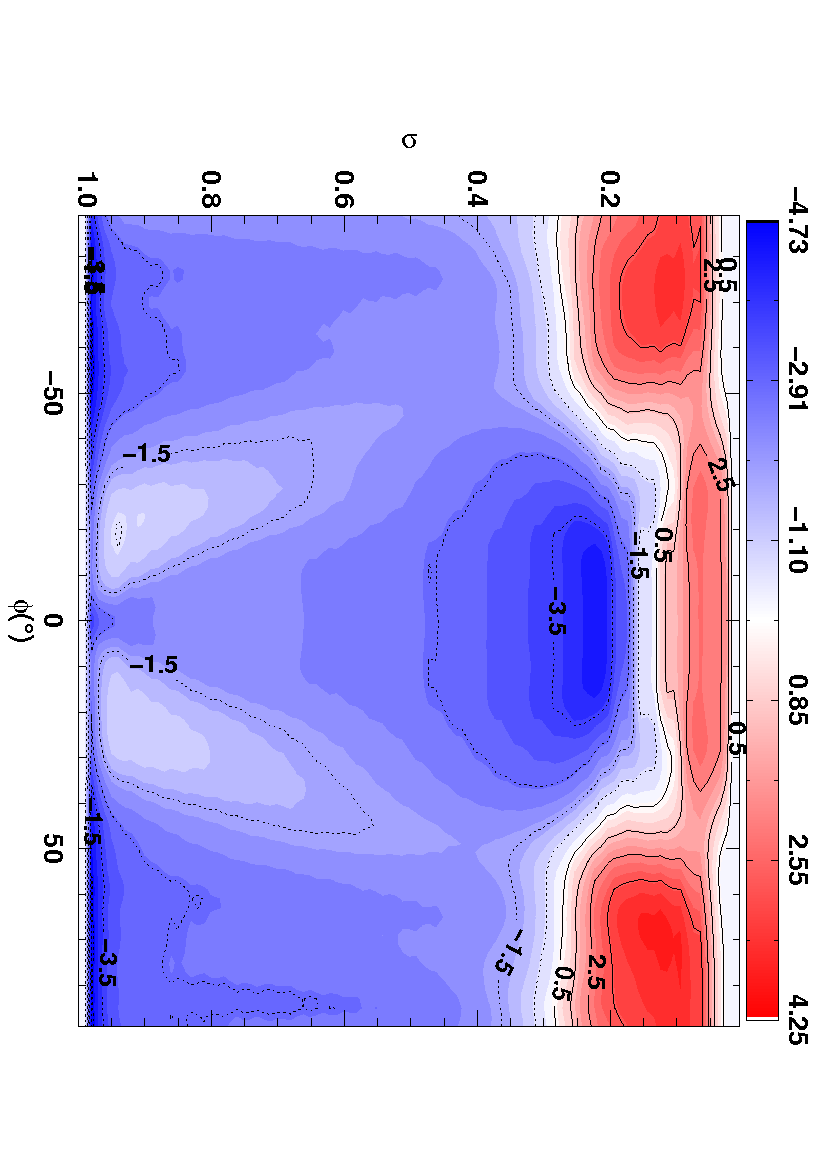}
\hspace*{-1.0cm}\includegraphics[width=6.5cm,angle=90.0]{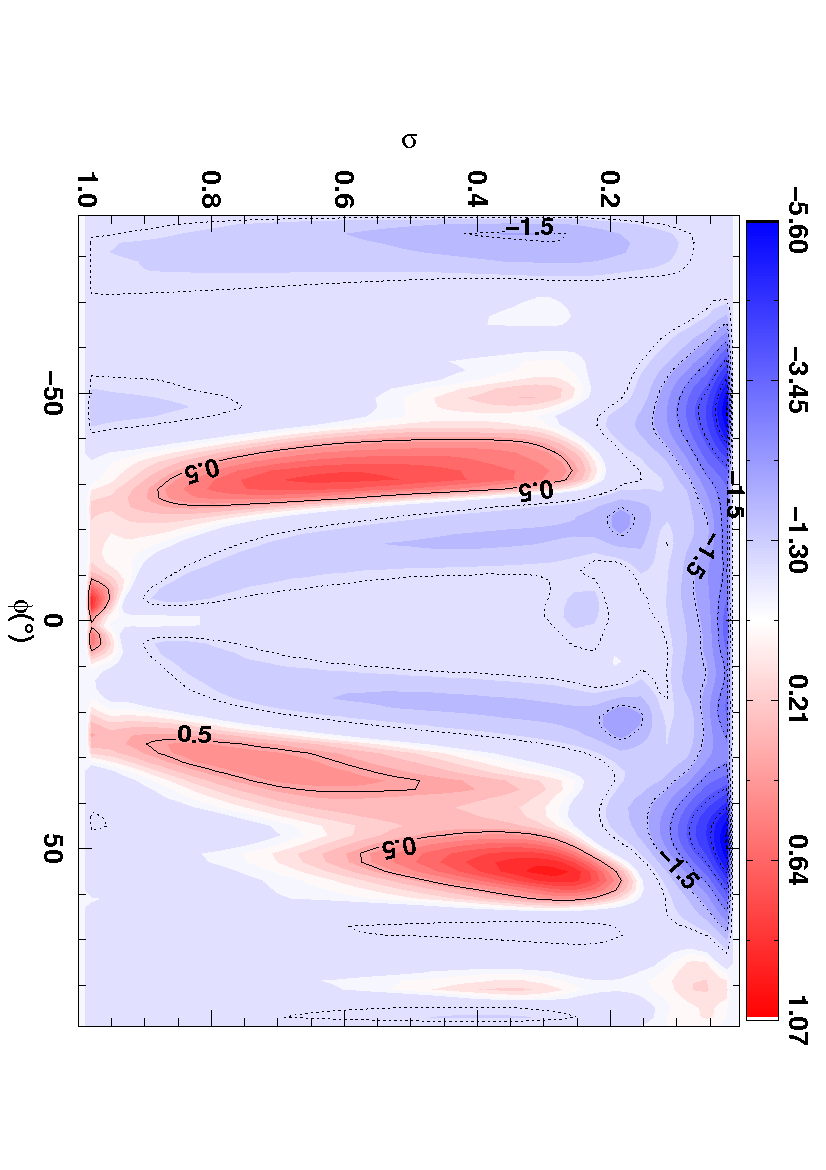}
\end{center}
\caption{Figure, for the Held--Suarez test \citep{held_1994}, showing
  the differences EG$-$ND of the zonally and temporally averaged zonal
  temperature (K), \textit{top panel}, and wind (ms$^{-1}$),
  \textit{bottom panel} (see Table \ref{model_names} for explanation
  of model types). \label{diff_HS}}
\end{figure}

Figure \ref{diff_HS} shows, explicitly, the differences between the
temperature and wind structures between the EG and ND models,
i.e. EG$-$ND from Figures \ref{HS_T} and \ref{HS_U} as the
\textit{top} and \textit{bottom panels}, respectively. Similar plots
have been constructed for EG$-$EG$_{\rm gc}$ and EG$-$EG$_{\rm sh}$
but the differences are negligible ($\Delta T\lesssim 1$ K and $\Delta
u\lesssim 2.5$ ms$^{-1}$).

Figure \ref{diff_HS} shows that the ND model has a cooler upper
atmosphere than the EG model (\textit{top panel}), and a warmer
lower atmosphere, although the differences are only $\sim 3$ K. The
prograde jets in the EG model are faster than those in the ND model,
and the retrograde flow in the upper atmosphere is enhanced
(\textit{bottom panel} of Figure \ref{diff_HS}), however, the
changes are small $\sim 1$ ms$^{-1}$.

Figures \ref{HS_T}, \ref{HS_U}, \ref{HS_EG} and \ref{diff_HS} show
that the overall large--scale, long--term flow for the HS test case
are relatively consistent both across all of our models, and with
literature results (only modest departures are evident in the wind and
temperature structures of the atmosphere). The diagnostics used
i.e. zonal and temporally averaged prognostic variables are, however,
relatively insensitive. Therefore, as discussed in Section \ref{comp}
we now explore the EKE found in each model to illustrate differences
in the eddy component of the flow.

\begin{figure*}[t]
\vspace*{2mm}
\begin{center}
\hspace*{-1.0cm}\includegraphics[width=6.5cm,angle=90.0]{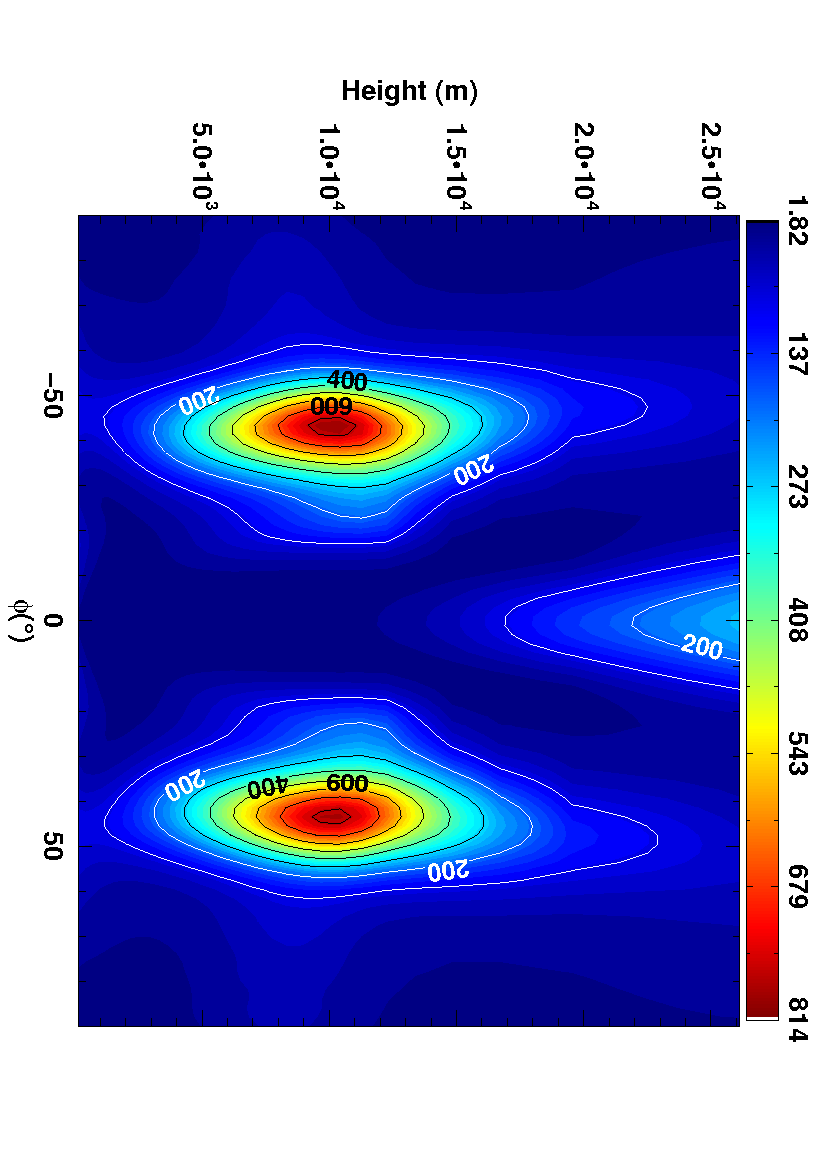}
\hspace*{-1.0cm}\includegraphics[width=6.5cm,angle=90.0]{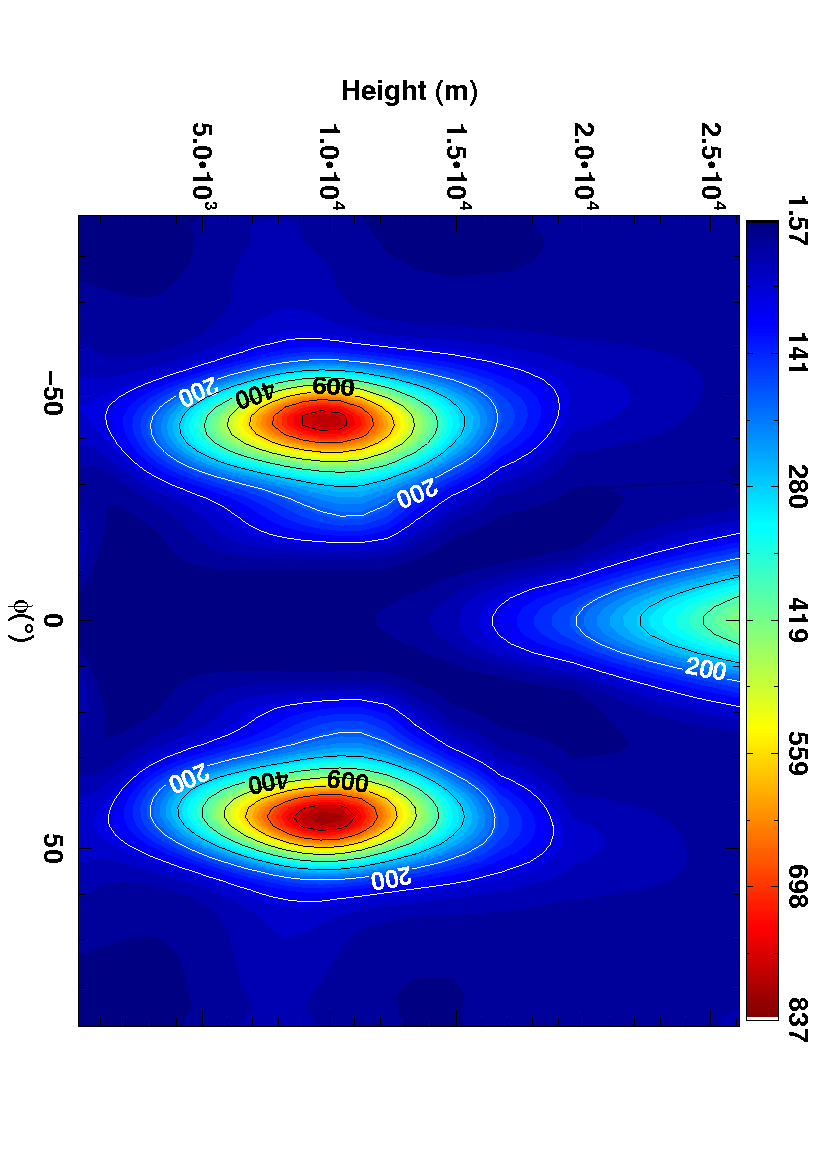}
\hspace*{-1.0cm}\includegraphics[width=6.5cm,angle=90.0]{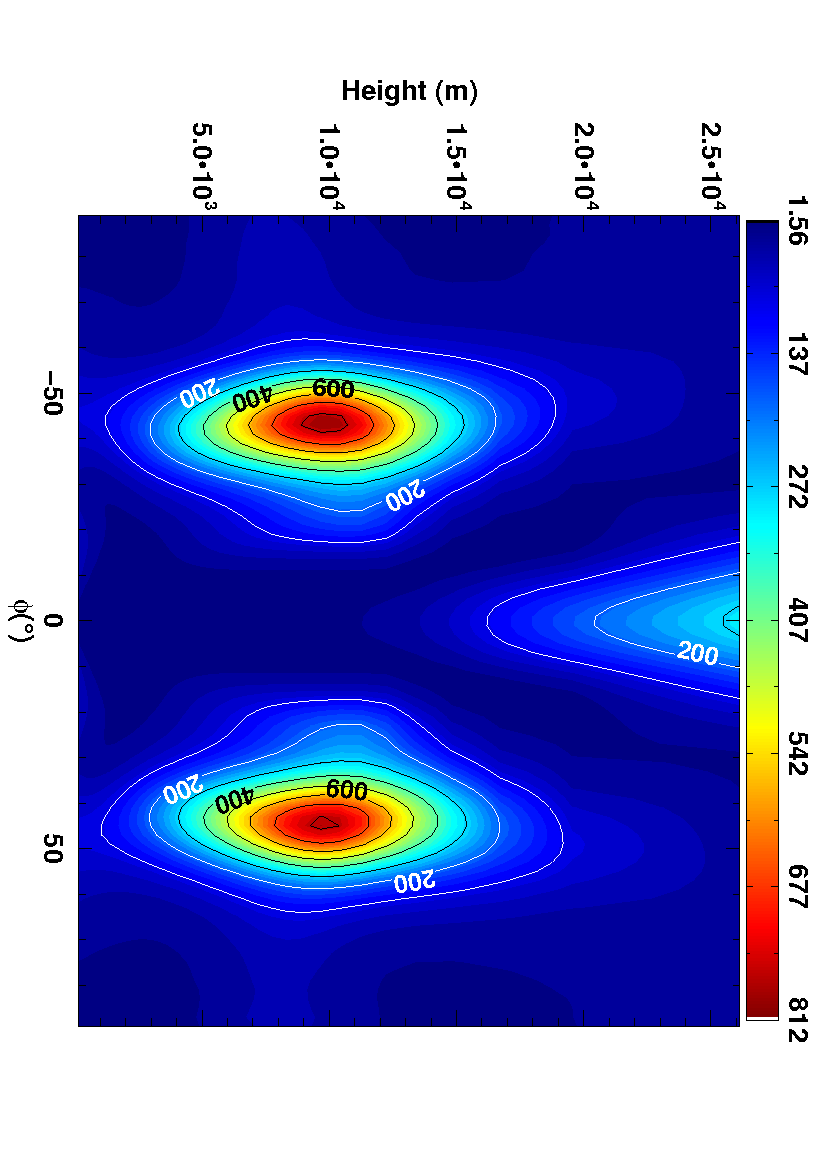}
\hspace*{-1.0cm}\includegraphics[width=6.5cm,angle=90.0]{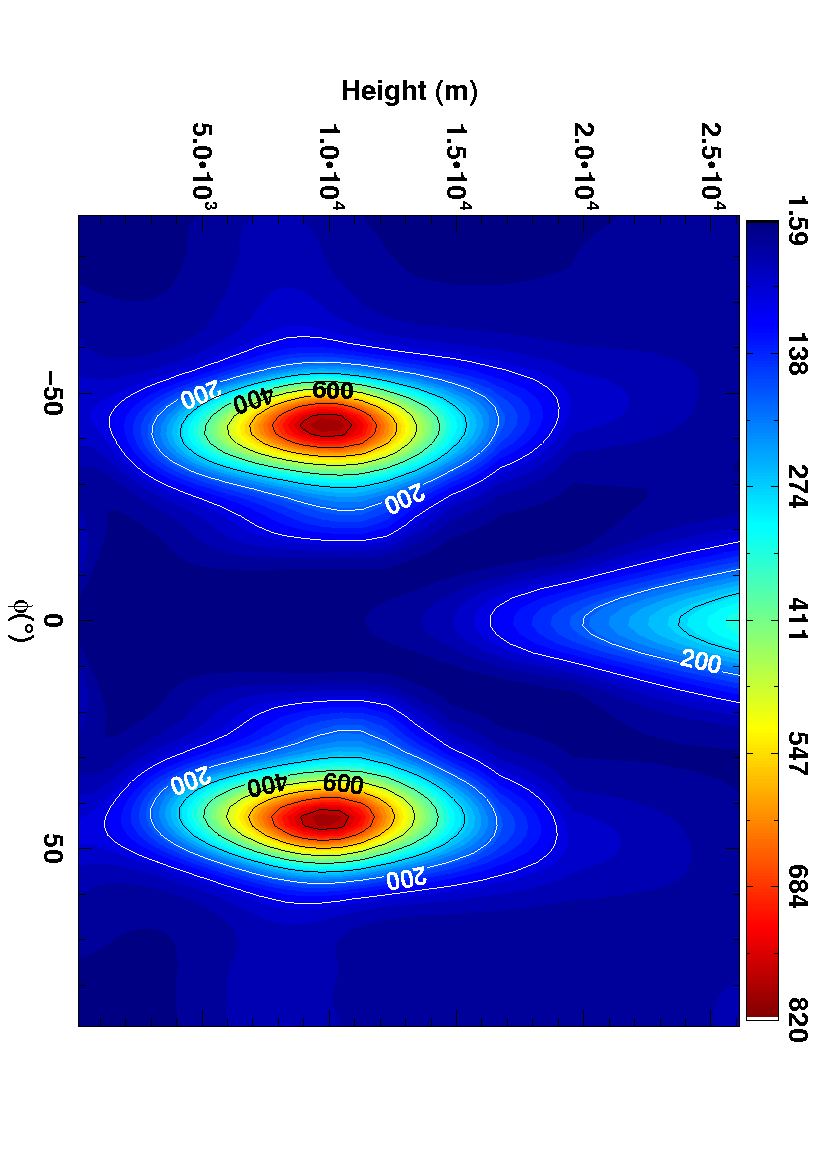}
\end{center}
\caption{Figure, for the Held--Suarez test \citep{held_1994}, showing
  the zonally (in geometric height) and temporally averaged Eddy
  Kinetic Energy (EKE, see Section \ref{models_run}) as a function of
  latitude and height. \textit{Top left panel:} ND, \textit{top right
    panel:} EG$_{\rm sh}$, \textit{bottom left panel:} EG$_{\rm gc}$ and
  \textit{bottom right panel:} EG models (see Table \ref{model_names}
  for explanation of model types). Note the contours (solid lines) are
  the same in all plots. \label{HS_EKE}}
\end{figure*}

\begin{figure}[t]
\vspace*{2mm}
\begin{center}
\hspace*{-1.0cm}\includegraphics[width=6.5cm,angle=90.0]{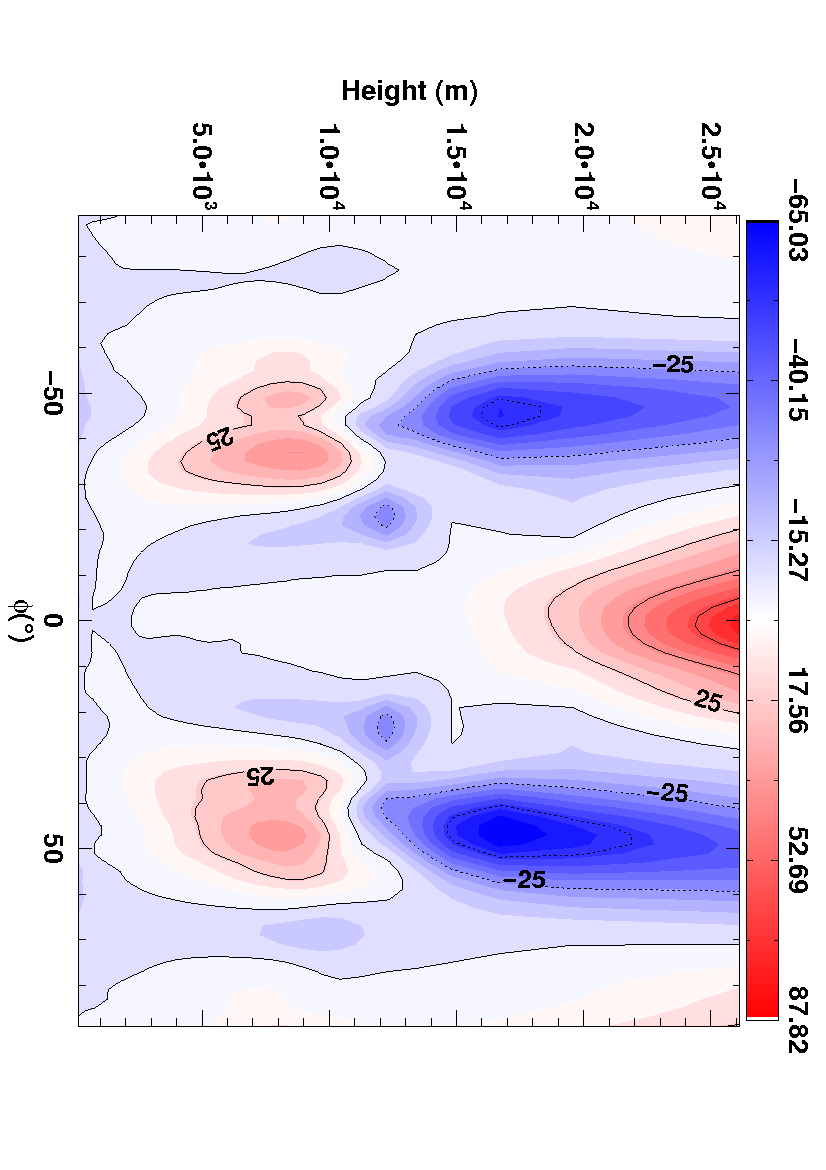}
\hspace*{-1.0cm}\includegraphics[width=6.5cm,angle=90.0]{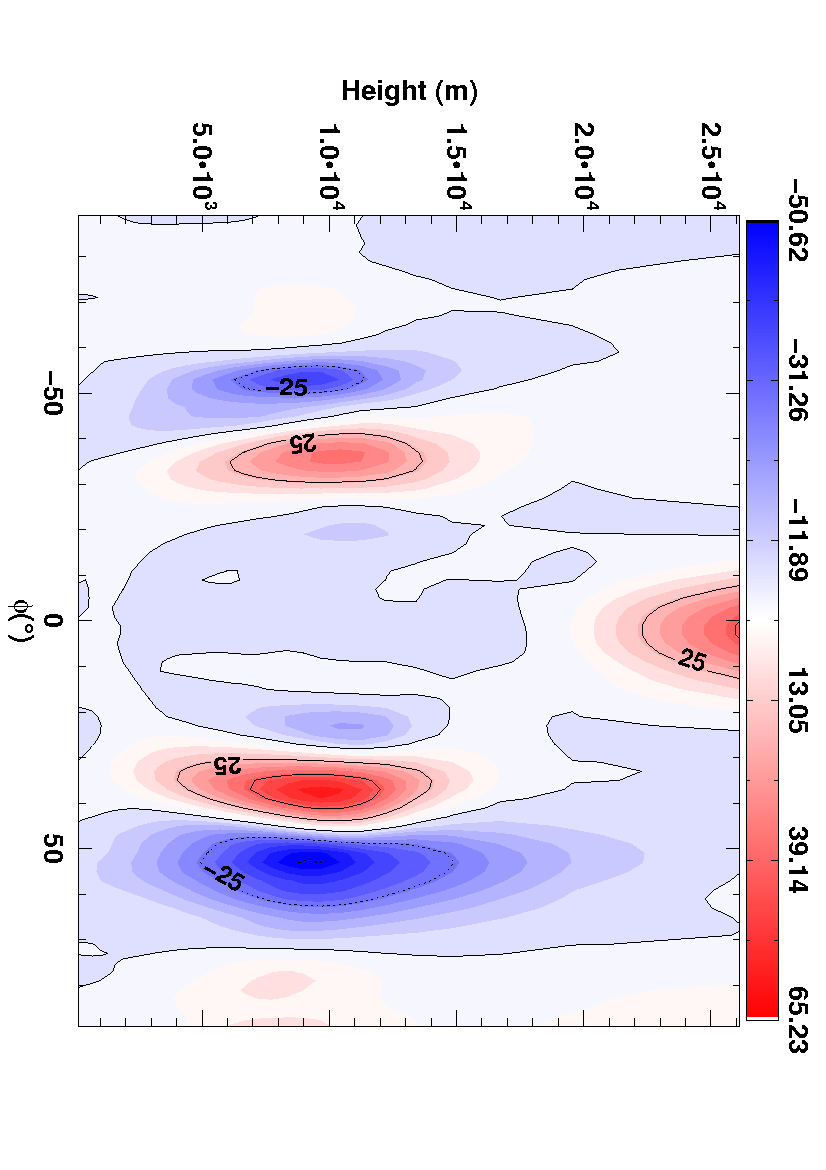}
\hspace*{-1.0cm}\includegraphics[width=6.5cm,angle=90.0]{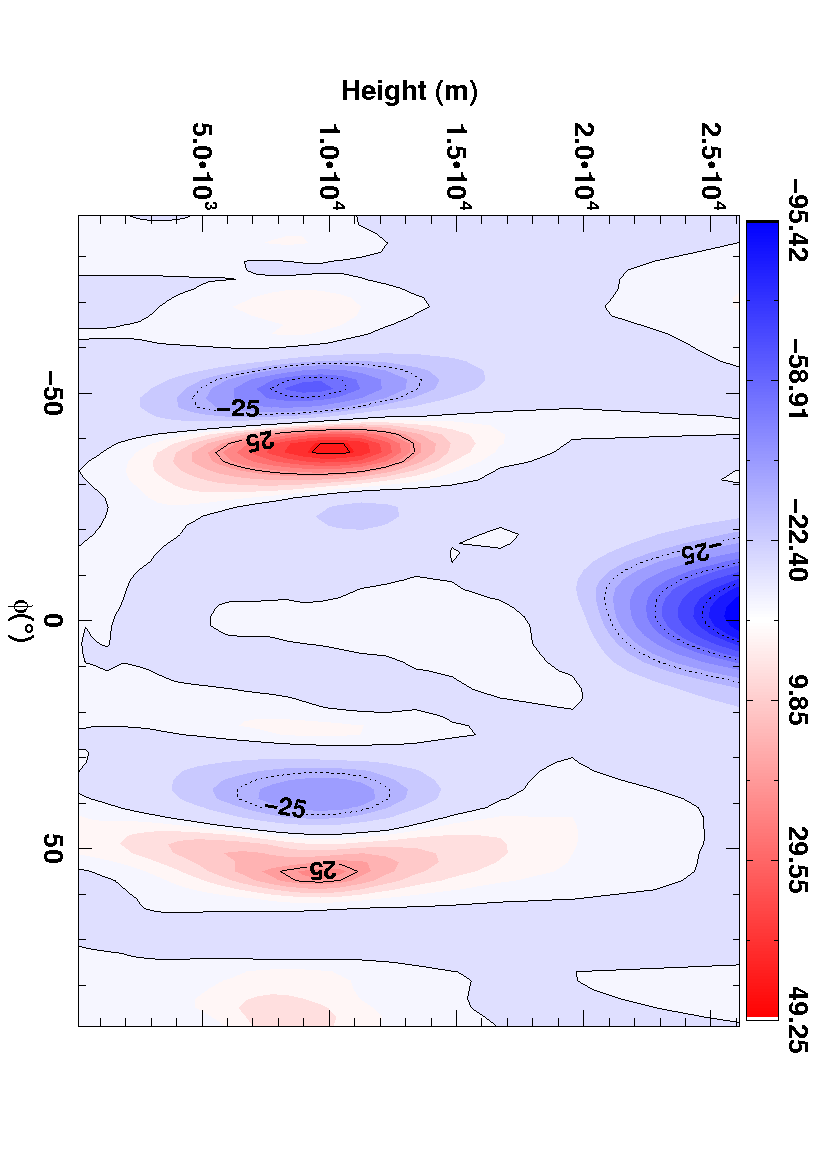}
\end{center}
\caption{Figure, for the Held--Suarez test \citep{held_1994}, showing
  the differences EG$-$ND (\textit{top}), EG$-$EG$_{\rm gc}$
  (\textit{middle}) and EG$-$EG$_{\rm sh}$ (\textit{bottom}), of the
  zonally and temporally averaged EKE. The line contours are the same
  for \textit{all panels} (see Table \ref{model_names} for explanation
  of model types). \label{diff_HS_EKE}}
\end{figure}

Figure \ref{HS_EKE} shows the EKE as defined in Section
\ref{models_run}, zonally (along geometric height surfaces) and
temporally averaged ($\overline{{\rm EKE}}^{\lambda_{\rm z}t}$) as a
function of height (m) and latitude ($^{\circ}$), for the ND and all
ENDGame models. Figure \ref{HS_EKE} shows excellent agreement of the
EKE for all of the models. However, a greater peak level of EKE
associated with the EG$_{\rm sh}$ model, and the least with the
EG$_{\rm gc}$ model. Overall, the structures of the plots are very
similar for all models. However, the results of the ND model shows,
with respect to the ENDGame plots, an increase in the EKE at
$\phi\sim50^{\circ}$ towards the upper boundary (i.e. coincident with
the peak wind speed of the prograde jets). To illustrate the
difference explicitly we show in Figure \ref{diff_HS_EKE}, as with the
temperature and zonal wind fields, the differences of the
$\overline{{\rm EKE}}^{\lambda_{\rm z}t}$ for each
model. Specifically, Figure \ref{diff_HS_EKE} shows difference in
$\overline{{\rm EKE}}^{\lambda_{\rm z}t}$ in the sense EG$-$ND,
EG$-$EG$_{\rm gc}$ and EG$-$EG$_{\rm sh}$, as the \textit{top},
\textit{middle} and \textit{bottom} rows respectively. In Figure
\ref{diff_HS_EKE} the line contours have been chosen to be the same
for \textit{all panels}.

Figure \ref{diff_HS_EKE} shows, for the EG model compared to ND
(\textit{top panel}), more kinetic energy associated with the eddy
component of the flow over the equator, and near the surface at a
latitude associated with the peak zonal wind speed ($\phi\sim
50^{\circ}$). The magnitude of the peak relative differences in
$\overline{{\rm EKE}}^{\lambda_{\rm z}t}$ are $\sim$1.65, 0.36 and
0.42 for the differences EG$-$ND, EG$-$EG$_{\rm gc}$ and EG$-$EG$_{\rm
  sh}$, respectively. There is a decrease in EKE found in the EG model
when compared to the ND model higher in the atmosphere. Comparing EG
to EG$_{\rm gc}$ (\textit{middle panel}) again shows more kinetic
energy associated with eddies in the EG model, over the equator, at
high altitudes, however, the differences associated with the
mid--latitude jets now appear over similar altitudes. Finally, the
difference EG$-$EG$_{\rm sh}$ (\textit{bottom panel}) shows a similar
spatial pattern to EG$-$EG$_{\rm gc}$ but the signs are
reversed. Overall, Figure \ref{diff_HS_EKE} shows that detailed, eddy,
component of the flow, can be quite different, although not affecting
the diagnostic plots (for example Figures \ref{HS_T} and \ref{HS_U})
significantly.

\clearpage

\subsection{Earth-Like}
\label{el}

For the Earth-Like test case of \citet{menou_2009}, the temperature
profile includes a parameterised stratosphere,
\begin{equation}
T_{\rm eq}=T_{\rm vert}+\beta_{\rm trop}\Delta T_{\rm EP}\left( \frac{1}{3}-\sin^2\phi\right)\\
\end{equation}
where
\begin{align}
  T_{\rm vert}=
\begin{cases}
    T_{\rm surf}-\Gamma_{\rm trop}(z_{\rm stra}+\frac{z-z_{\rm stra}}{2})\\
    +\left( \left[ \frac{\Gamma_{\rm trop}(z-z_{\rm stra})}{2}\right] ^2+\Delta T_{\rm strat}^2\right) ^{\frac{1}{2}}\mbox{,}\, &z \leq z_{\rm stra}\mbox{,} \nonumber\\
    T_{\rm surf}-\Gamma_{\rm trop}z_{\rm stra}+\Delta T_{\rm strat}\mbox{,}\, &z > z_{\rm stra} \mbox{,} 
\end{cases}\\
\label{T_vert}
\end{align}
and $T_{\rm surf}=288$ K is the surface temperature, $\Gamma_{\rm
  trop}=6.5\times 10^{-3}$ Km$^{-1}$ is the lapse rate, and $\Delta
T_{\rm strat}=2$ K, an offset to smooth the transition from the
troposphere (finite lapse rate) to the isothermal
stratosphere. $z_{\rm stra}$ and $\sigma_{\rm stra}$ are then the
locations in height and $\sigma$ of the tropopause. $\beta_{\rm trop}$
is defined as
\begin{align}
  \beta_{\rm trop}&=\begin{cases}
    \sin\frac{\pi(\sigma-\sigma_{\rm stra})}{2(1-\sigma_{\rm stra})}\mbox{,}\, &z \leq z_{\rm stra} \,\mbox{ or }\, \sigma \geq\sigma_{\rm stra} \mbox{,} \nonumber\\
    0\mbox{,}\, &z > z_{\rm stra} \, \mbox{ or }\, \sigma < \sigma_{\rm stra} \mbox{.} 
\end{cases}\\
\end{align}
The remaining parameters match those of HS, except, here the radiative
timescale is set as a constant, $\tau_{\rm rad}=15$ days, but,
following \citet{heng_2011} the same `Rayleigh friction' scheme as for
HS is implemented \citep[this differs from the choice of][where only
the bottom level winds are damped which creates a resolution dependent
damping profile]{menou_2009}.

\begin{figure*}[t]
\vspace*{2mm}
\begin{center}
\includegraphics[width=8.5cm,angle=0.0,origin=c]{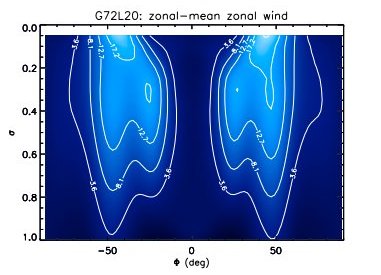}
\includegraphics[width=8.5cm,angle=0.0,origin=c]{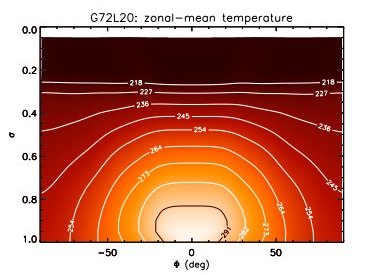}
\hspace*{-1.0cm}\includegraphics[width=6.5cm,angle=90.0]{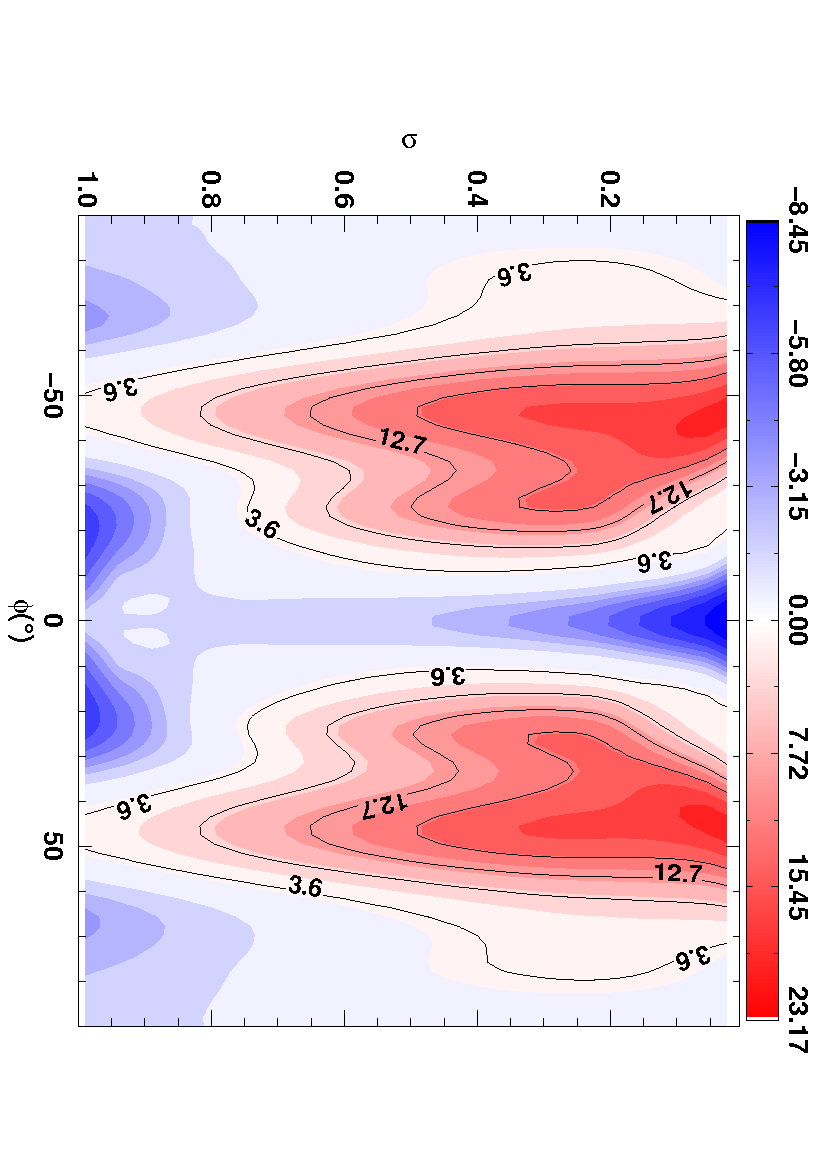}
\hspace*{-1.0cm}\includegraphics[width=6.5cm,angle=90.0]{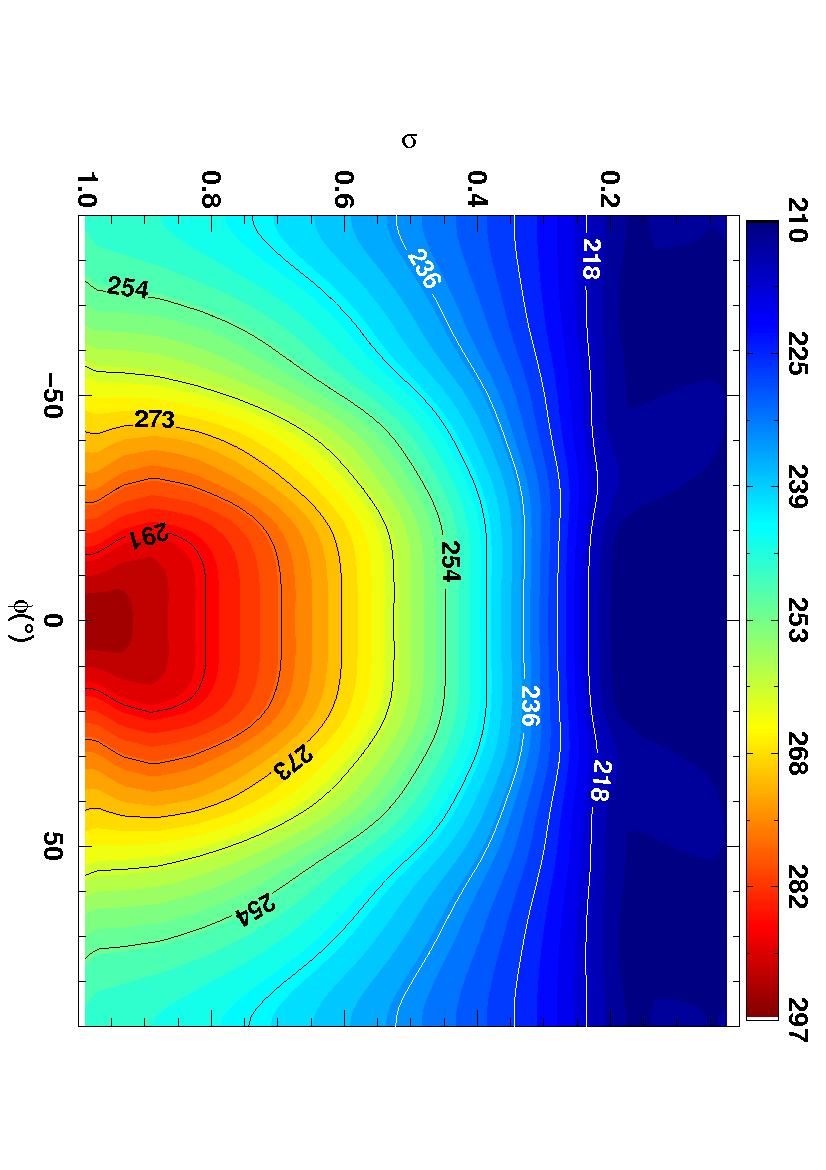}
\hspace*{-1.0cm}\includegraphics[width=6.5cm,angle=90.0]{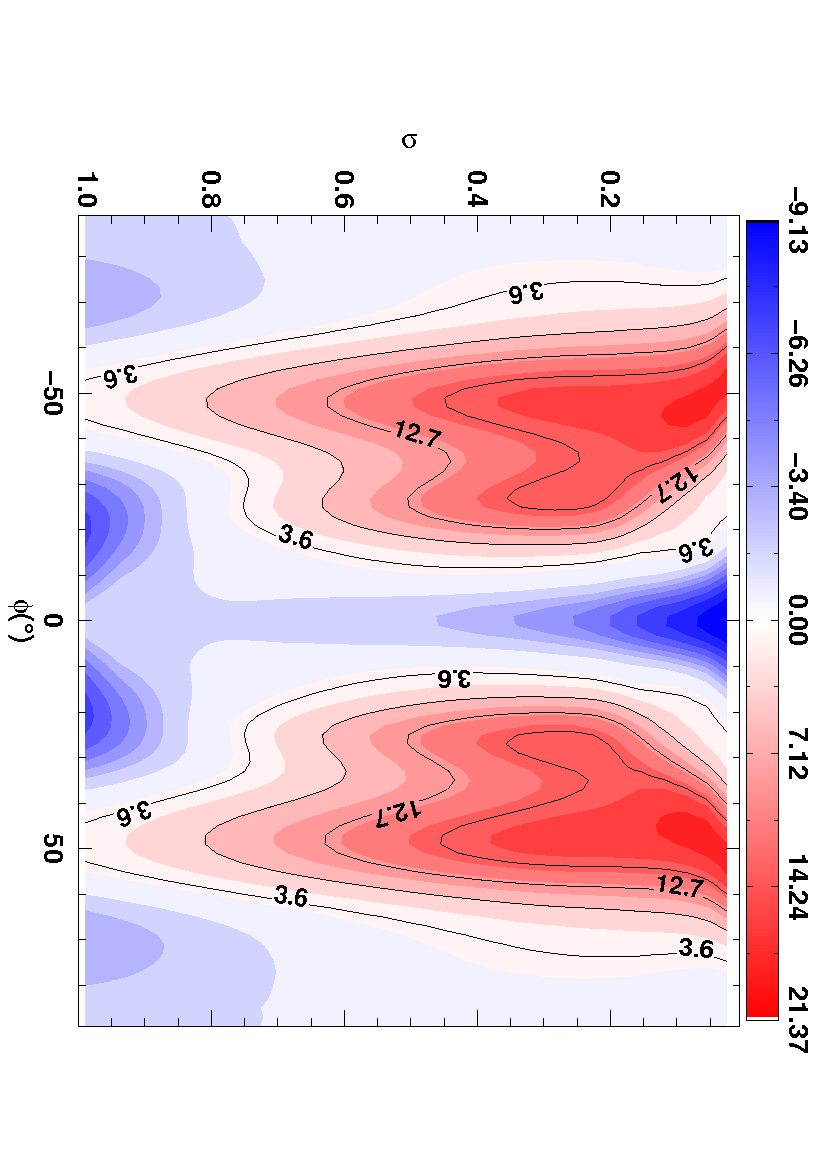}
\hspace*{-1.0cm}\includegraphics[width=6.5cm,angle=90.0]{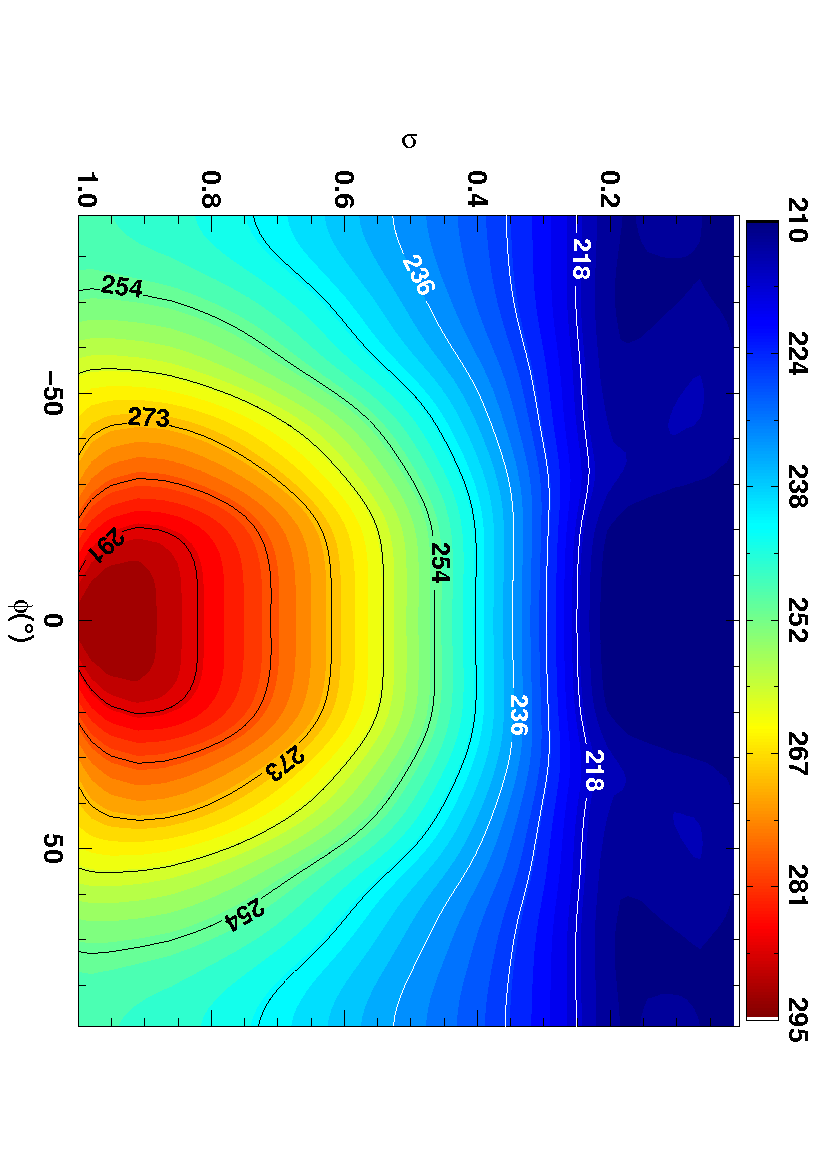}
\end{center}
\caption{Figures showing, for the Earth--Like test \citep{menou_2009},
  the zonally averaged temperature and zonal wind. \textit{Top
    panels:} temporally averaged results from grid--based model of
  \citet{heng_2011} (reproduced by permission of Oxford University
  Press). \textit{Middle} and \textit{bottom panels:} temporally
  averaged results from this work using the ND and EG models,
  respectively (see Table \ref{model_names} for explanation of model
  types). \label{EL_ND}}
\end{figure*}

Figure \ref{EL_ND} shows the zonally averaged (in $\sigma$ space)
zonal wind and temperature fields for our ND and EG models, and the
results from \citet{heng_2011}, both have been temporally averaged
(i.e. $\overline{u}^{\lambda_{\sigma},t}$ and
$\overline{T}^{\lambda_{\sigma},t}$). Our models are in excellent
agreement with the results of \citet{heng_2011} (although we have
slightly stronger high--altitude components of the mid--latitude
jets). Our results also match the `snapshots' of the flow field
presented in \cite{menou_2009}. This agreement again, as found with
the HS test, suggests sufficient vertical resolution \citep[15, 20 and
32 vertical levels used in][and this work,
respectively]{menou_2009,heng_2011}.

\begin{figure}[t]
\vspace*{2mm}
\begin{center}
\hspace*{-1.0cm}\includegraphics[width=6.5cm,angle=90.0]{./Figs/EL_EG_Deep_Uvel.png}
\hspace*{-1.0cm}\includegraphics[width=6.5cm,angle=90.0]{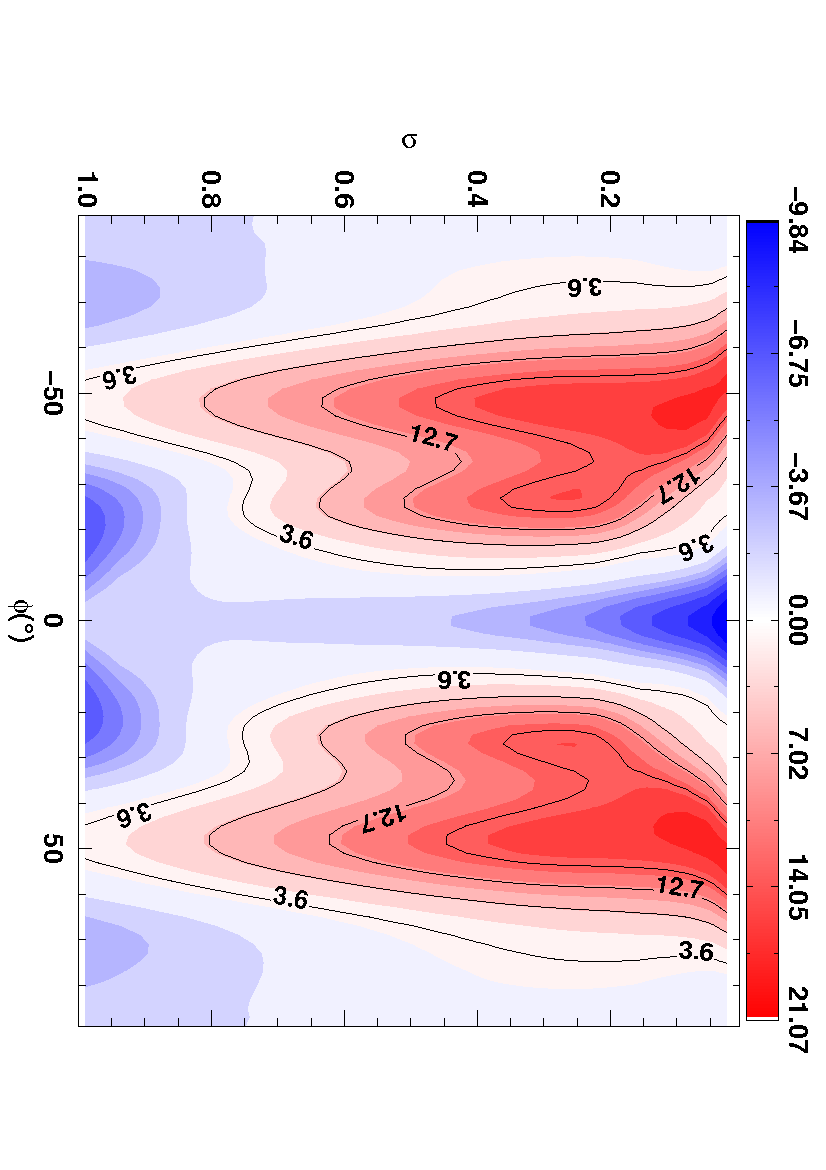}
\hspace*{-1.0cm}\includegraphics[width=6.5cm,angle=90.0]{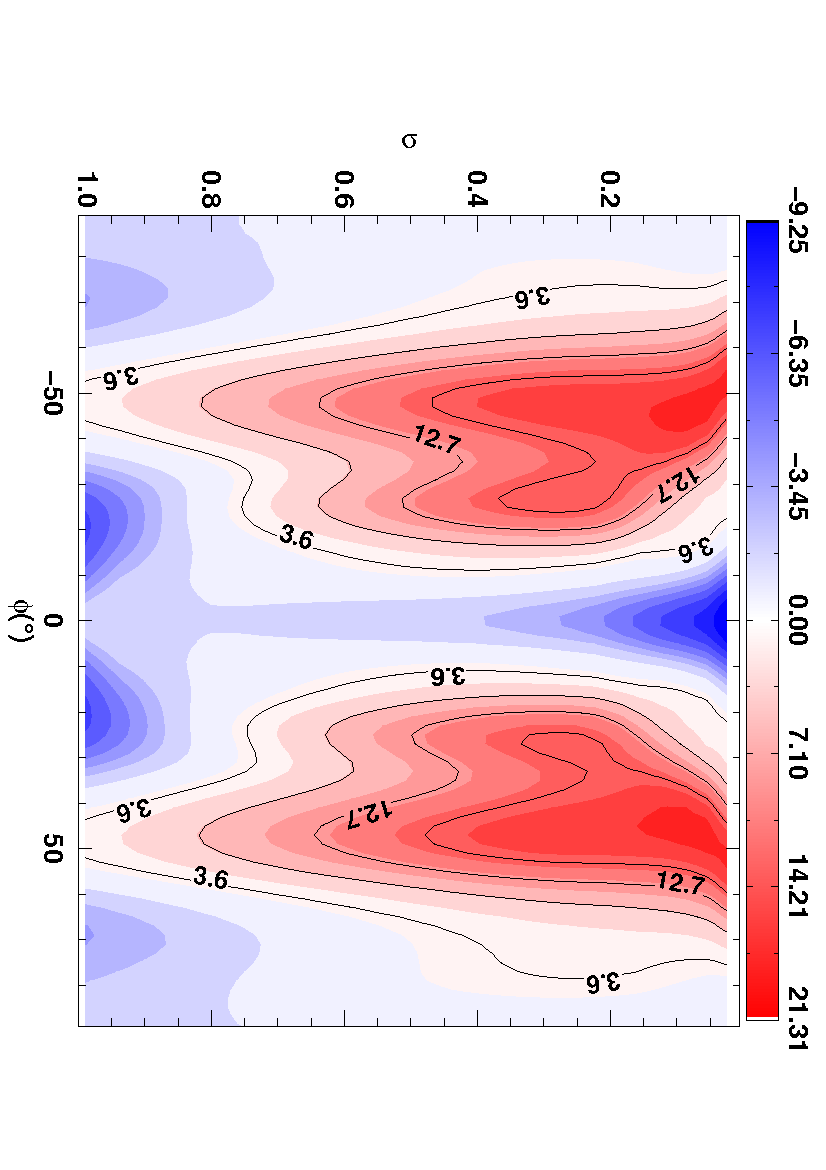}
\end{center}
\caption{Figures showing, for the Earth--Like test \citep{menou_2009},
  the zonally and temporally averaged zonal wind fields for the
  different EG models. \textit{Top panel:}, EG, \textit{middle
    panel:}, EG$_{\rm gc}$ and \textit{bottom panel:}, EG$_{\rm sh}$
  (see Table \ref{model_names} for explanation of model
  types).\label{EL_EG}}
\end{figure}

Further evidence of the extrapolation of the temperature down to the
surface of the ND model, performed as part of the visualisation
process, is apparent in the \textit{right panels} of Figure
\ref{EL_ND}, in the contours close to the surface. The \textit{left
  panels} of Figure \ref{EL_ND} shows a slight improvement in the
agreement of the flow structure at high and low latitudes, between the
results of \citet{heng_2011} and our own model when moving from ND to
EG. Figure \ref{EL_EG} then shows the temporally and zonally averaged
zonal wind for the three versions of the ENDGame models. The
qualitative agreement between all the \textit{panels} in Figure
\ref{EL_EG} again shows that the assumptions are valid, and that the
code is consistently solving for the long--term and large--scale 3D
flow. There are only very slight differences, for example, as we move
towards a more simplified model (i.e. downwards in Figure \ref{EL_EG})
we generally see the edge of 3.6 ms$^{-1}$ contour moving to higher
latitudes, and a slight degradation in the symmetry of the
flow. Additionally, all of the ND and ENDGame models show a greater
hemispherical symmetry in the wind patterns than the finite difference
model presented in \citet{heng_2011}, and, in fact, match the levels
of symmetry present in the results of the spectral code of
\citet{heng_2011} (not shown here).

\begin{figure}[t]
\vspace*{2mm}
\begin{center}
\hspace*{-1.0cm}\includegraphics[width=6.5cm,angle=90.0]{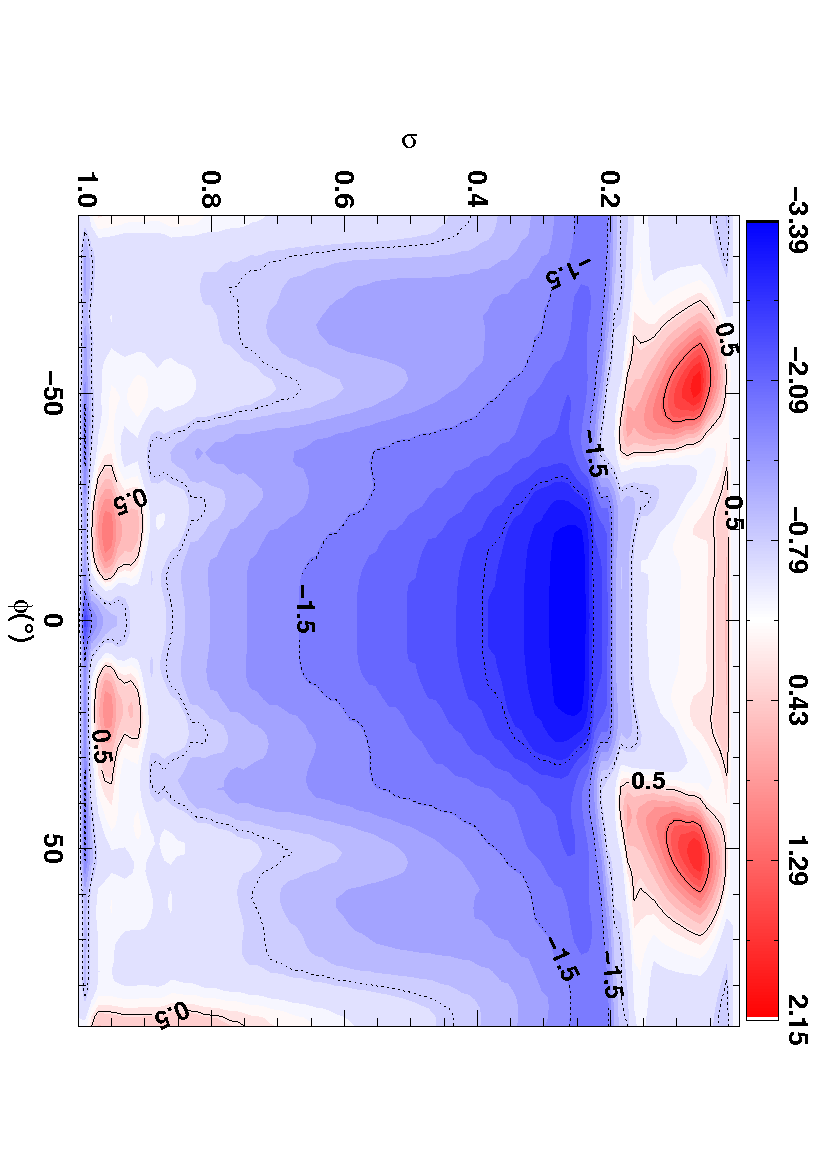}
\hspace*{-1.0cm}\includegraphics[width=6.5cm,angle=90.0]{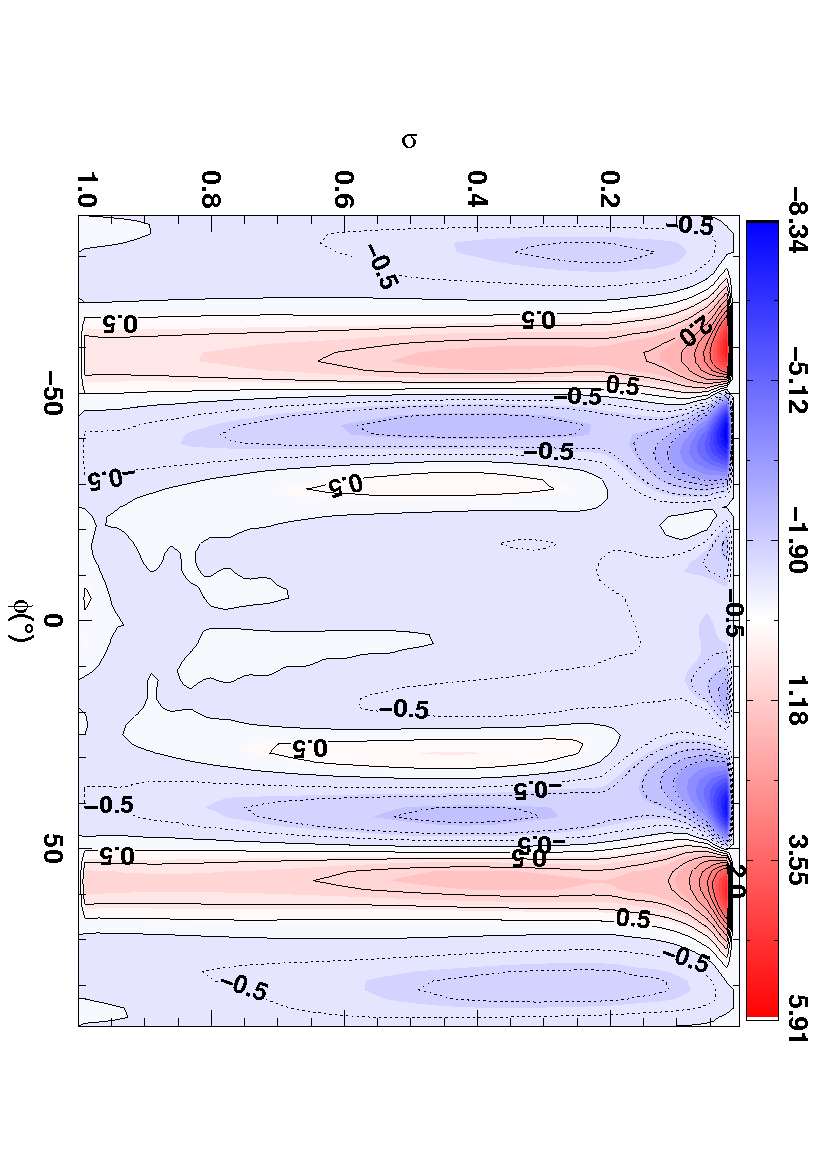}
\end{center}
\caption{Figure, for the Earth--Like test \citep{menou_2009}, showing
  the differences EG$-$ND of the zonally and temporally averaged
  temperature, \textit{top}, and zonal wind (ms$^{-1}$) \textit{bottom
    panel} (see Table \ref{model_names} for explanation of model
  types). \label{diff_EL}}
\end{figure}

Again, as with the HS test case in Section \ref{hs} the different
ENDGame models show negligible differences in the results, so only the
difference EG$-$ND is shown in Figure \ref{diff_EL}. The format of
Figure \ref{diff_EL} matches that of Figure \ref{diff_HS}. Figure
\ref{diff_EL} shows a similar, yet reduced in magnitude, pattern to
that present in Figure \ref{diff_HS}, with a warmer upper atmosphere
showing enhanced flow, and cooler mid--atmosphere, in the EG model
over the ND model. The zonal jets have also shifted closer to the
poles in the EG model. This is caused, largely, by the adverse effects
of the polar filtering used in the ND model (when polar filtering is
applied to the EG model the jets move closer to the location found for
ND).

\begin{figure*}[t]
\vspace*{2mm}
\begin{center}
\hspace*{-1.0cm}\includegraphics[width=6.5cm,angle=90.0]{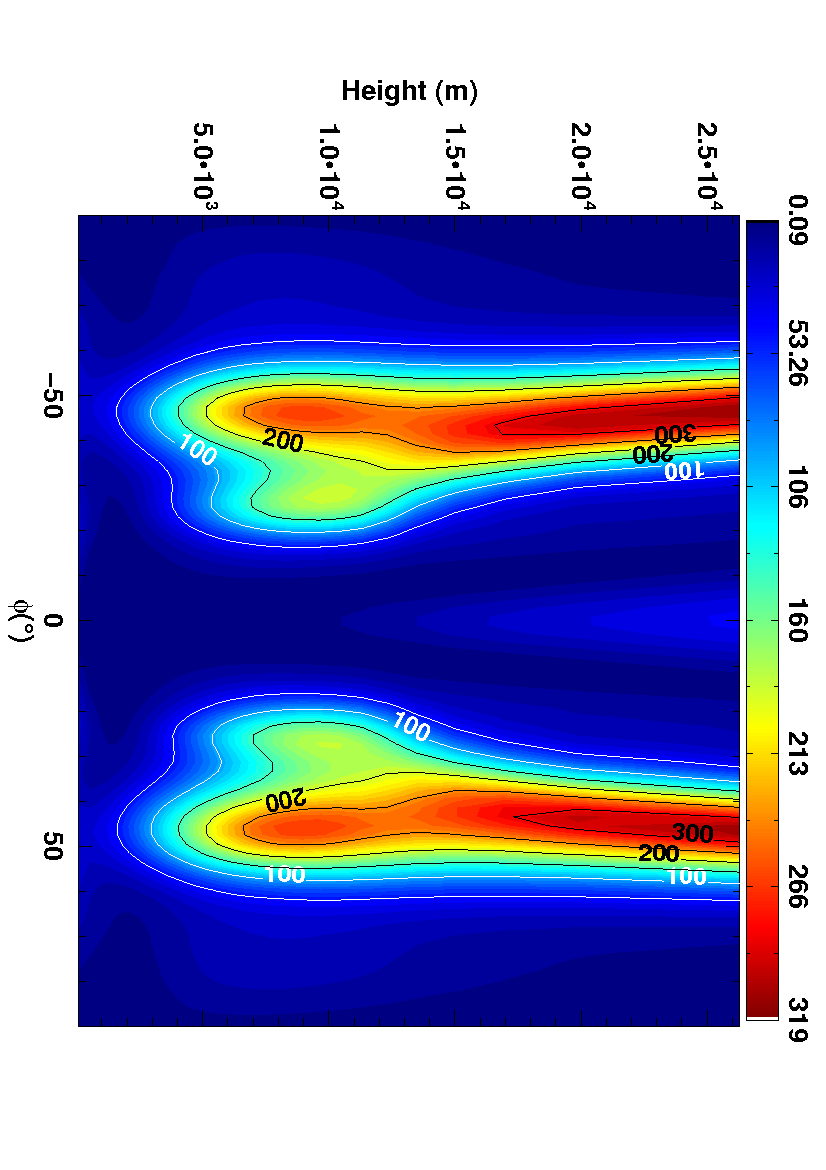}
\hspace*{-1.0cm}\includegraphics[width=6.5cm,angle=90.0]{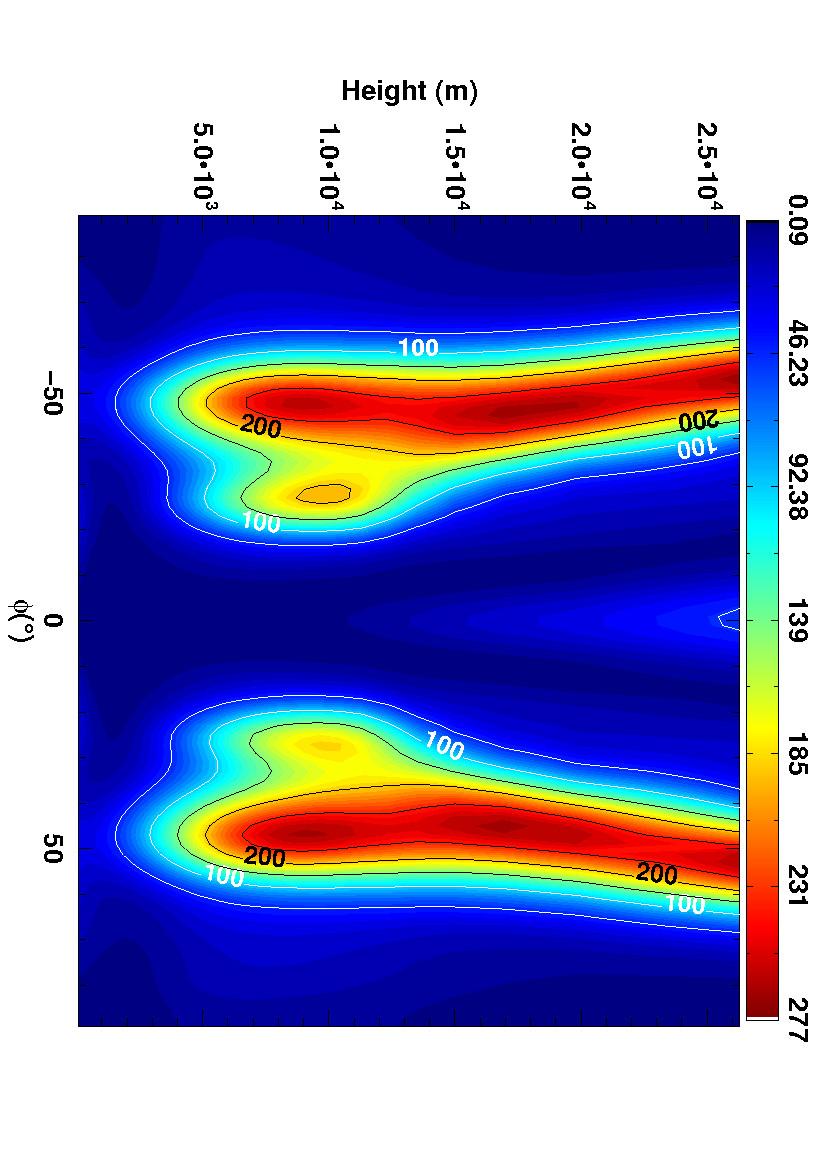}
\hspace*{-1.0cm}\includegraphics[width=6.5cm,angle=90.0]{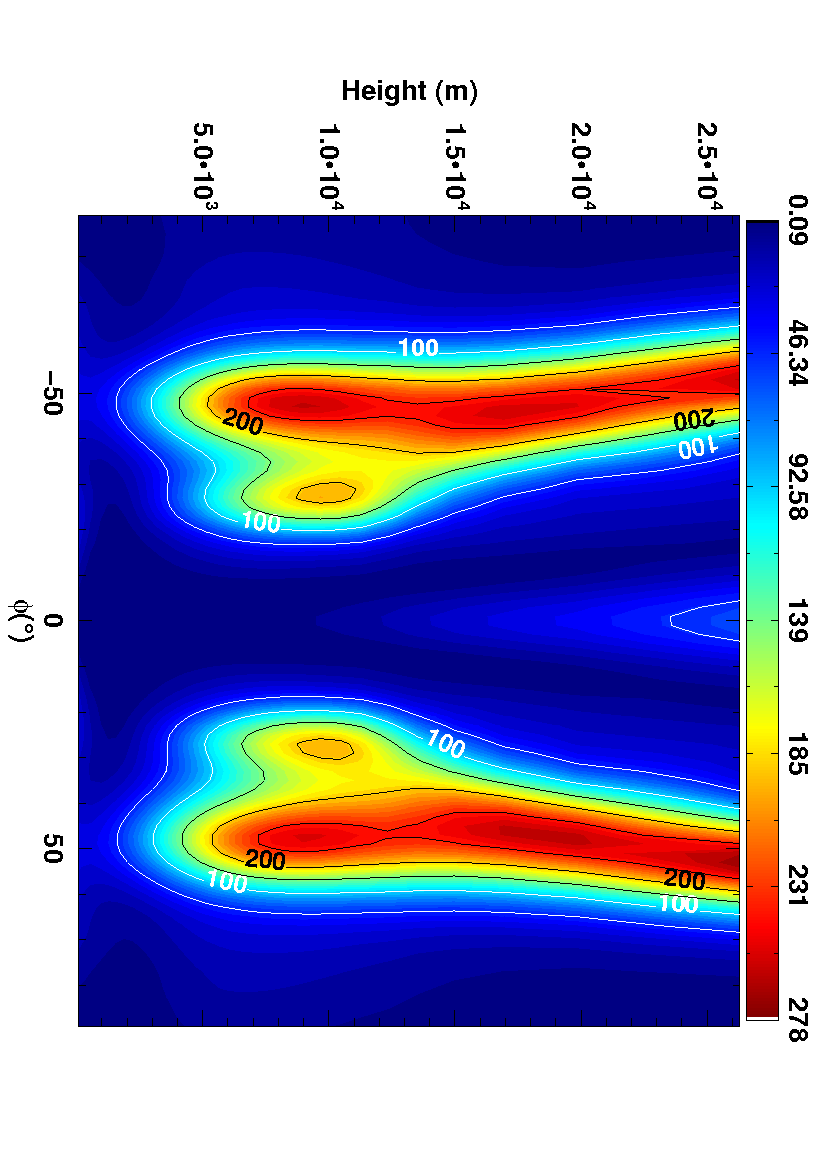}
\hspace*{-1.0cm}\includegraphics[width=6.5cm,angle=90.0]{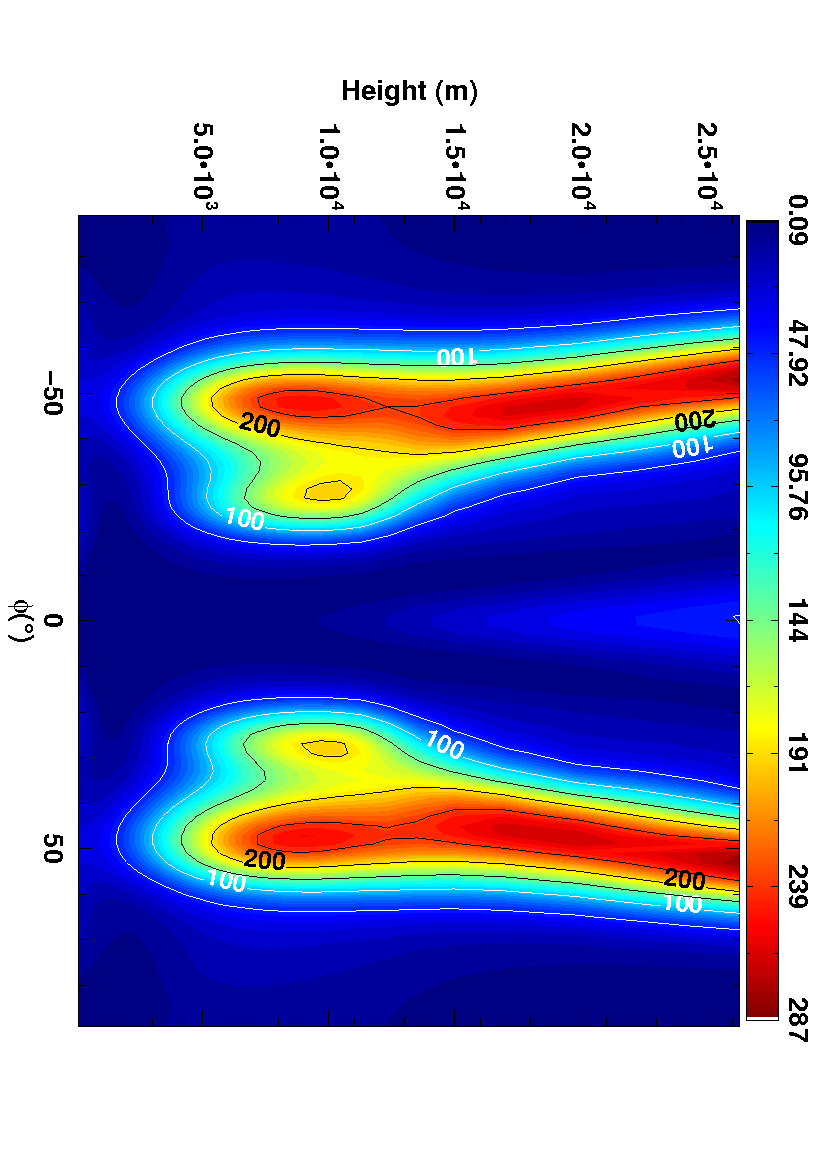}
\end{center}
\caption{Figure, for the Earth--Like test \citep{menou_2009}, showing
  the zonally (in geometric height) and temporally averaged Eddy
  Kinetic Energy (EKE, see Section \ref{models_run}) as a function of
  latitude and height. \textit{Top left panel:} ND, \textit{top right
    panel:} EG$_{\rm sh}$, \textit{bottom left panel:} EG$_{\rm gc}$
  and \textit{bottom right panel:} EG models (see Table
  \ref{model_names} for explanation of model types). Note the contours
  (solid lines) are the same in all plots. \label{EL_EKE}}
\end{figure*}

Again, to explore the eddy component of the flow, Figure \ref{EL_EKE}
shows the EKE, zonally (along geometric height surfaces) and
temporally averaged ($\overline{{\rm EKE}}^{\lambda_{\rm z}t}$), for
the ND and all ENDGame models.  Figure \ref{EL_EKE}, as in Figure
\ref{HS_EKE} shows qualitative agreement with the overall pattern of
$\overline{{\rm EKE}}^{\lambda_{\rm z},t}$, however in this case the
peak value is much larger for the ND model (compared to any ENDGame
model).The magnitude of the peak relative differences in
$\overline{{\rm EKE}}^{\lambda_{\rm z}t}$ are $\sim$2.0, 0.80 and 0.46
for the differences EG$-$ND, EG$-$EG$_{\rm gc}$ and EG$-$EG$_{\rm
  sh}$, respectively, slightly larger than found in the HS case. The
ENDGame models also show more structure along the peak of
$\overline{{\rm EKE}}^{\lambda_{\rm z},t}$ activity and the `lobes'
equator ward of the peak.

\begin{figure}[t]
\vspace*{2mm}
\begin{center}
\hspace*{-1.0cm}\includegraphics[width=6.5cm,angle=90.0]{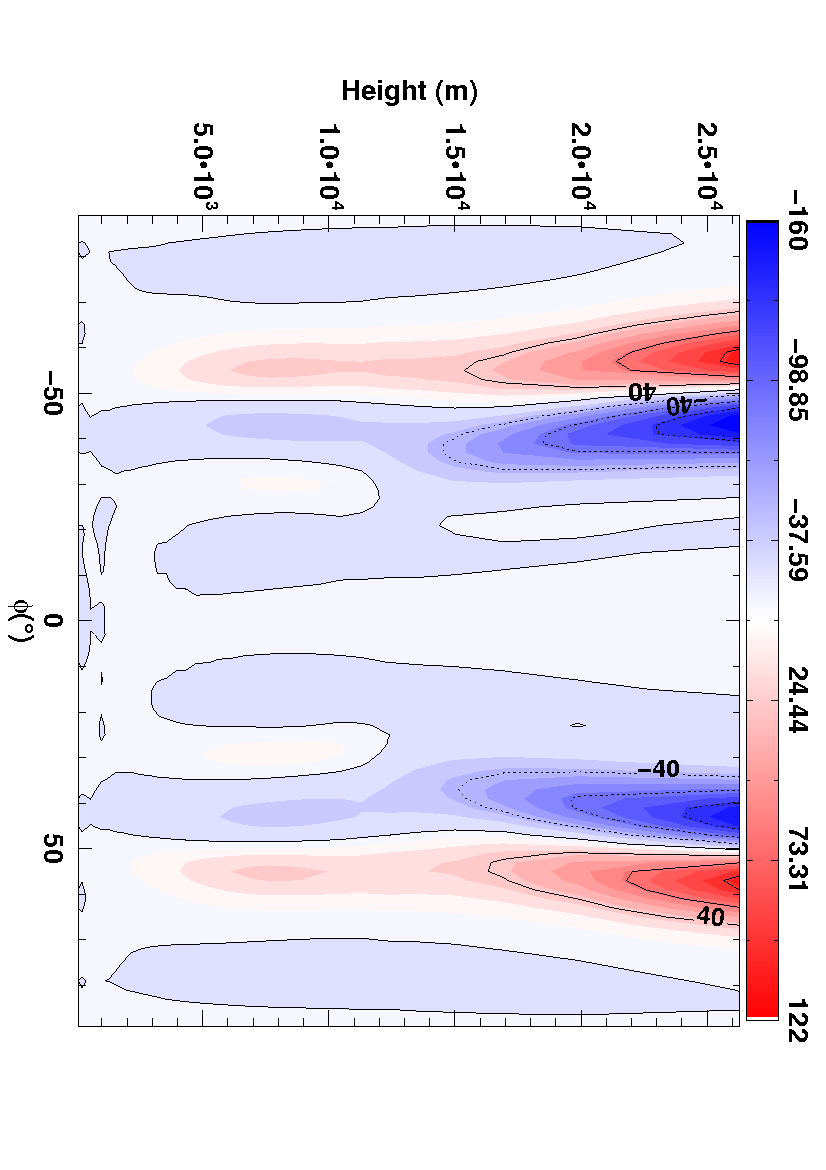}
\end{center}
\caption{Figure, for the Earth--Like test \citep{menou_2009}, showing
  the differences EG$-$ND of the zonally and temporally averaged EKE
  (see Table \ref{model_names} for explanation of model
  types). \label{diff_EL_EKE}}
\end{figure}

To emphasise the slight differences in $\overline{{\rm
    EKE}}^{\lambda_{\rm z},t}$ apparent in Figure \ref{EL_EKE} we
present a difference plot, for EG$-$ND only (as the differences
between the ENDGame models are an order of magnitude smaller), in
Figure \ref{diff_EL_EKE}.

There is a significant reduction in variation in the $\overline{{\rm
    EKE}}^{\lambda_{\rm z},t}$ across all of the EL models, when
compared to the HS test case (see Figures \ref{HS_EKE} and
\ref{EL_EKE}), as the EL test is a simpler flow regime to capture. The
EG$-$ND of $\overline{{\rm EKE}}^{\lambda_{\rm z},t}$, in Figure
\ref{diff_EL_EKE} also shows the peak difference is close to the upper
boundary, coincident in latitude, with the peak of the prograde
jets. As seen in Figure \ref{diff_EL} a shift in the latitudinal
location of the pattern is observed between the EG and ND models. As
before, this is due to the polar filtering applied in the ND model.

\clearpage

\subsection{Tidally Locked Earth}
\label{tle}

For the Tidally Locked Earth (TLE) test of \citet{merlis_2010} we slow
the rotation rate so that a day is now equal to an orbital period
(i.e. a year), $\Omega\to \frac{\Omega}{365}$. This introduces a
longitudinal temperature contrast and allows us to test the model
behaviour in a familiar system (i.e. Earth) but incorporating aspects
found in the hot Jupiter atmospheric regime. We have not included
moisture in the calculation and therefore, have essentially, performed
the simplified version of the test which is described and performed by
\citet{heng_2011}. The equilibrium temperature profile is then a
modified version of the HS profile, enforcing a longitudinal
temperature contrast and `hot spot' at the subsolar point centred at a
longitude of $180^\circ$ (and latitude of zero). It is given by:
\begin{equation}
T_{\rm eq}={\rm max} \{ T_{\rm stra}, T_{\rm TLE} \},\\
\end{equation}
where, 
\begin{multline}
T_{\rm TLE}=\\
\left[ T_{\rm surf}+\Delta T_{\rm EP}\cos(\lambda-180^{\circ})\cos\phi - \Delta \theta_{z}\ln\left( \frac{p}{p_0}\right)\cos^2\phi\right]\\
 \left( \frac{p}{p_0}\right) ^{\kappa}.\\
\end{multline}
The parameters and values in common with the HS model take the same
values.

However, for this model, where significant flow over the pole exists,
we must add a sponge layer into the ENDGame formulation for model
stability (ND incorporates a polar filter). This damps vertical
motions and is explained in \citet{klemp_2008,melvin_2010}. The
damping term $R_w$ (included in the solution for vertical velocity)
is,
\begin{equation}
w^{t +\Delta t}=w^{t \*}+S_w-R_{w}\Delta t w^{t +\Delta t},\\
\end{equation}
where $w^t$ and $w^{t+\Delta t}$ are the vertical velocities at the
current and next timestep, $S_w$ a source term, and $\Delta t$ the
length of the timestep (as before). The spatial extent and value of
the damping coefficient ($R_{w}$) is then determined by the equation
\begin{multline}
R_{w} = \\
\begin{cases}
  {\rm C} \left(\sin^2\left( 0.5\pi(\eta-\eta_{\rm s}) \left( \frac{1.0}{1.0-\eta_{\rm s}}\right) \right) + \sin^{40}(\phi)\right) \mbox{,}\, &\eta \geq \eta_{\rm s}\\
0 \mbox{,}\, & \eta < \eta_{\rm s},\\
\end{cases}
\end{multline}
where, given the absence of orography, $\eta=\frac{z}{H}$
(i.e. non--dimensional height, where $H$ is the height of the upper
boundary), $\eta_{\rm s}$ is the start height for the top level
damping (set to $\eta_{\rm s}=0.75$) and $C$ is a coefficient (set to
0.05).

\begin{figure}[t]
\vspace*{2mm}
\begin{center}
\includegraphics[width=6.5cm,angle=0.0,origin=c]{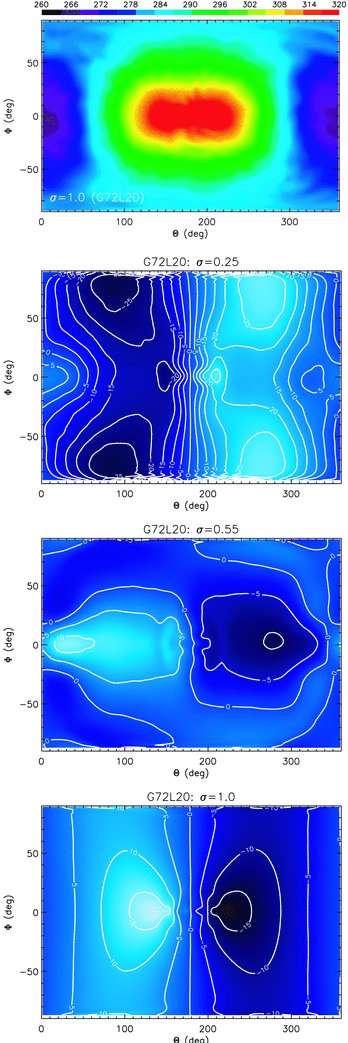}
\end{center}
\caption{Figure reproduced from \citet{heng_2011} of the results from
  the grid--based model of the TLE test case (reproduced by permission
  of Oxford University Press). Showing (from the \textit{top panel} to
  the \textit{bottom panel}) temperature at 1200 days and
  $\sigma=0.975$, then, temporally averaged zonal wind at $\sigma=0.225$,
  $0.525$ and $0.975$. \label{TLE_heng}}
\end{figure}

Figure \ref{TLE_heng} is a reproduction of the grid--based model
results for the TLE test in \citet{heng_2011}. It shows the
temperature at the $\sigma=0.975$ surface at 1200 days (\textit{top
  panel}), the temporally averaged zonal wind ($\overline{u}^{t}$) at
the surfaces $\sigma=0.225$, $0.525$ and $0.975$ (in descending panel
order)\footnote{See discussion in Appendix \ref{heng_comp} for
  explanation of differences in quoted $\sigma$ levels between our
  work and that of \citet{heng_2011}.}.

\begin{figure*}[t]
\vspace*{2mm}
\begin{center}
\hspace*{-1.0cm}\includegraphics[width=5.5cm,angle=90.0]{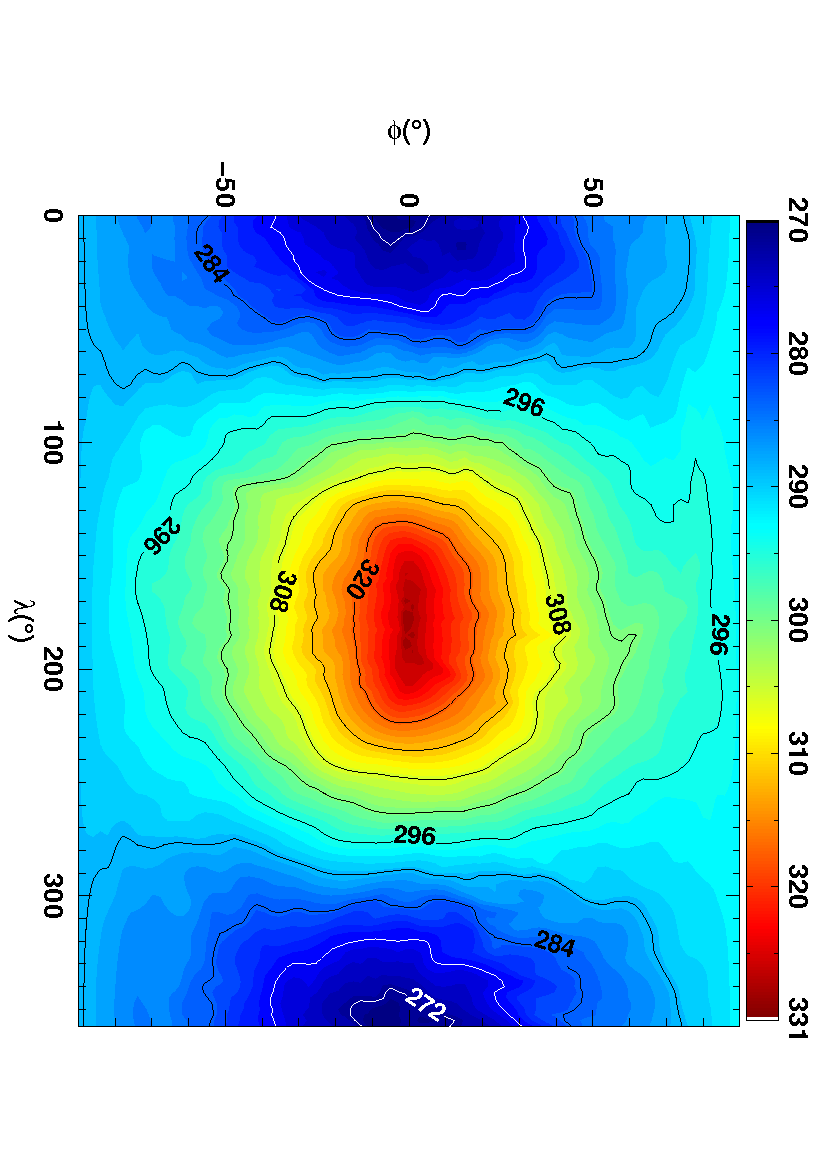}
\hspace*{-1.0cm}\includegraphics[width=5.5cm,angle=90.0]{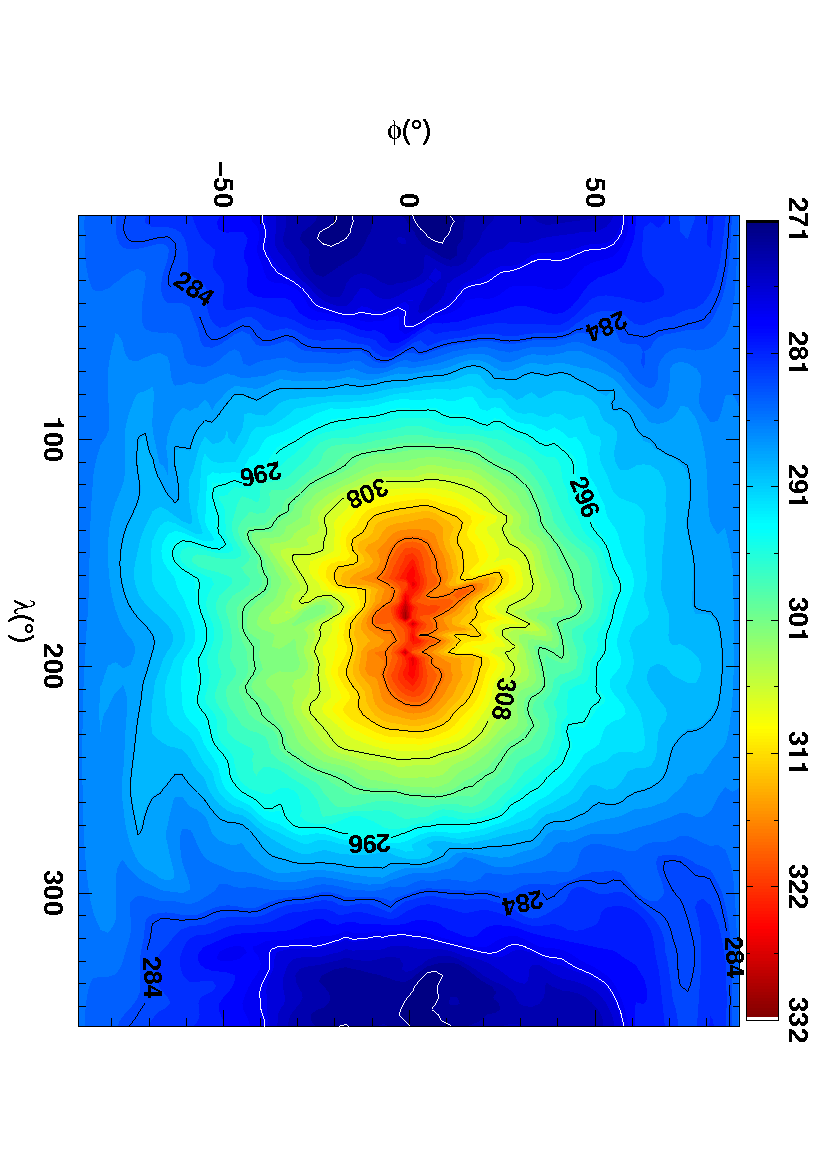}
\hspace*{-1.0cm}\includegraphics[width=5.5cm,angle=90.0]{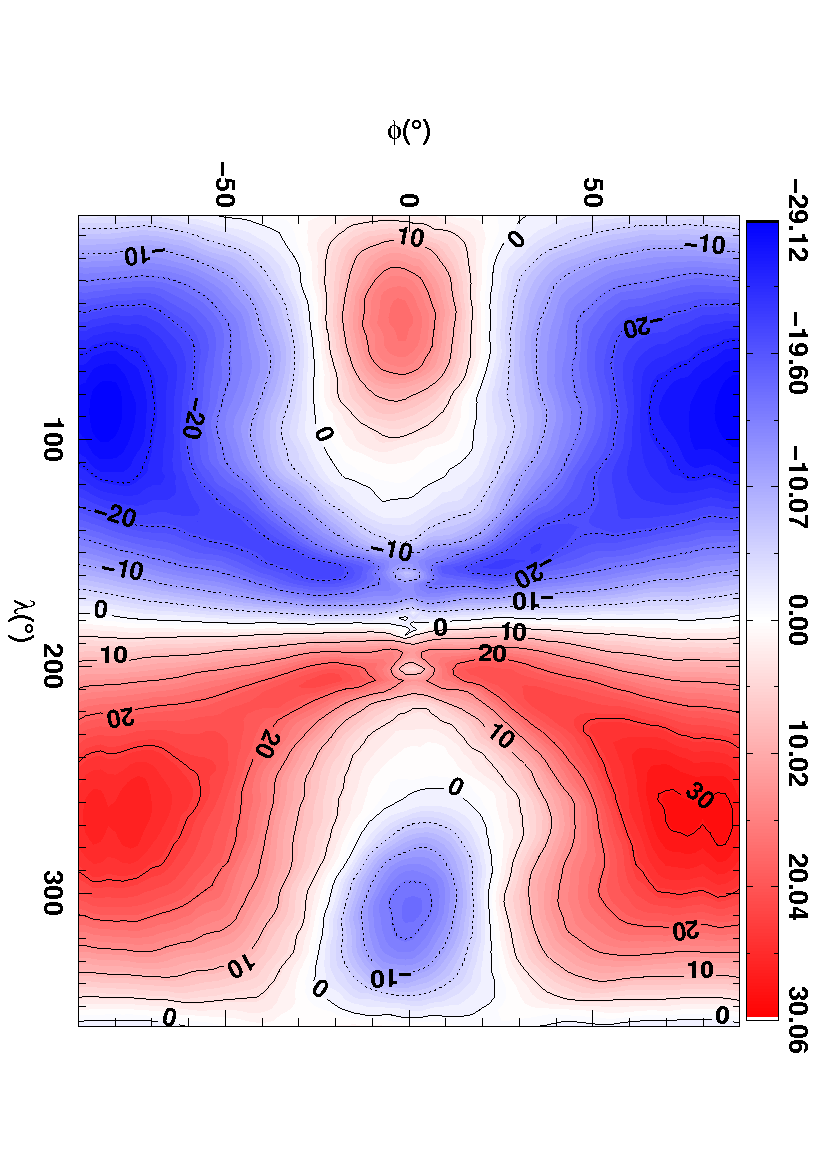}
\hspace*{-1.0cm}\includegraphics[width=5.5cm,angle=90.0]{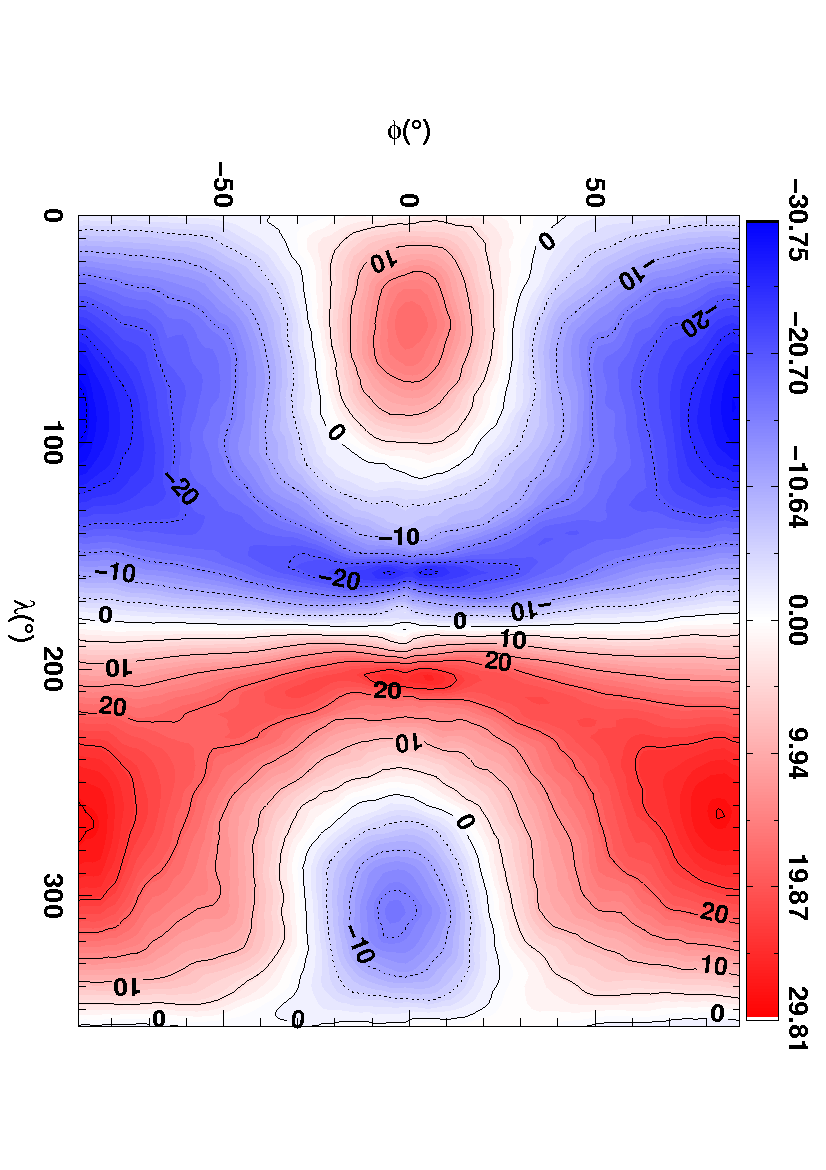}
\hspace*{-1.0cm}\includegraphics[width=5.5cm,angle=90.0]{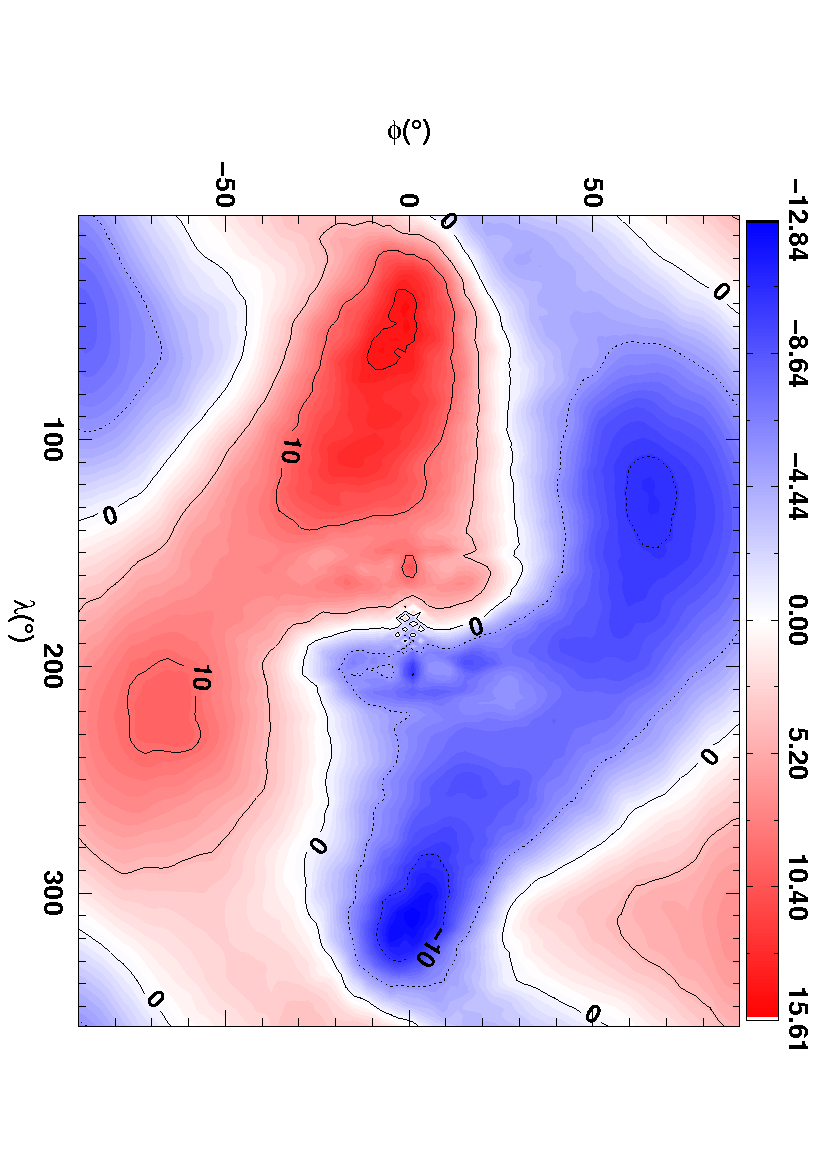}
\hspace*{-1.0cm}\includegraphics[width=5.5cm,angle=90.0]{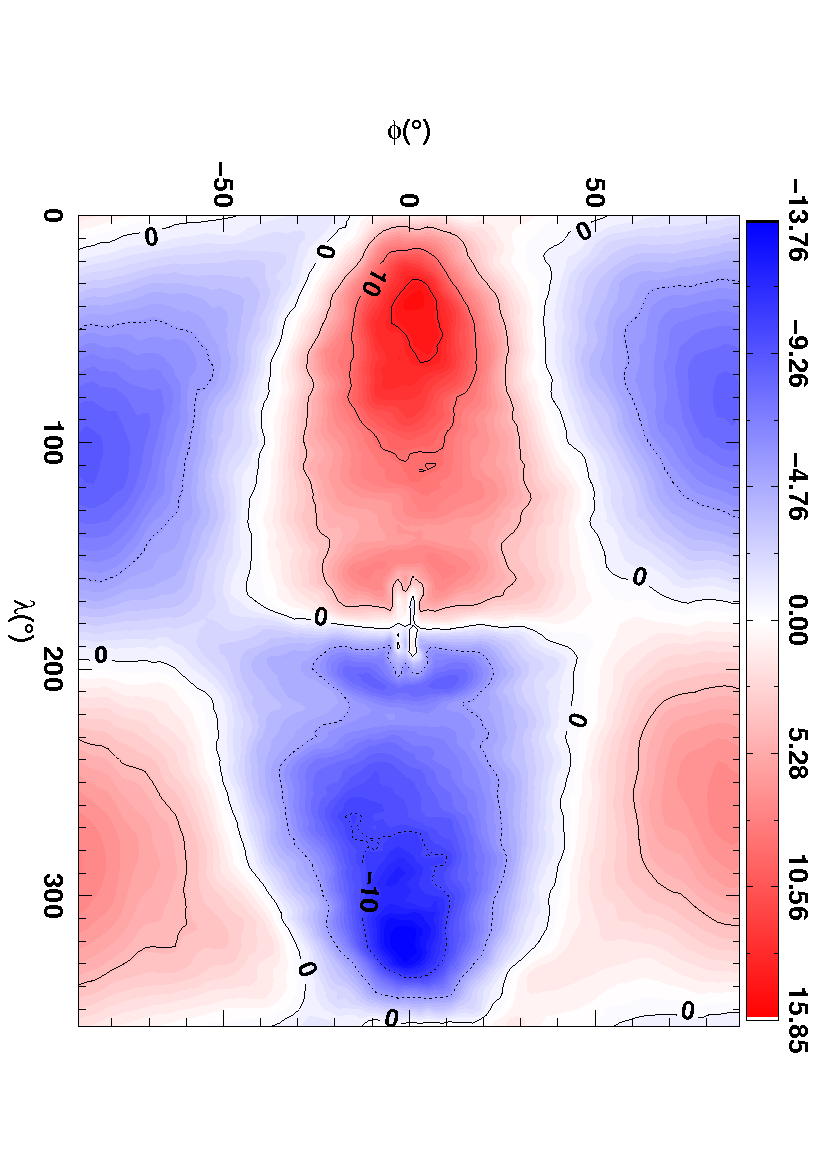}
\hspace*{-1.0cm}\includegraphics[width=5.5cm,angle=90.0]{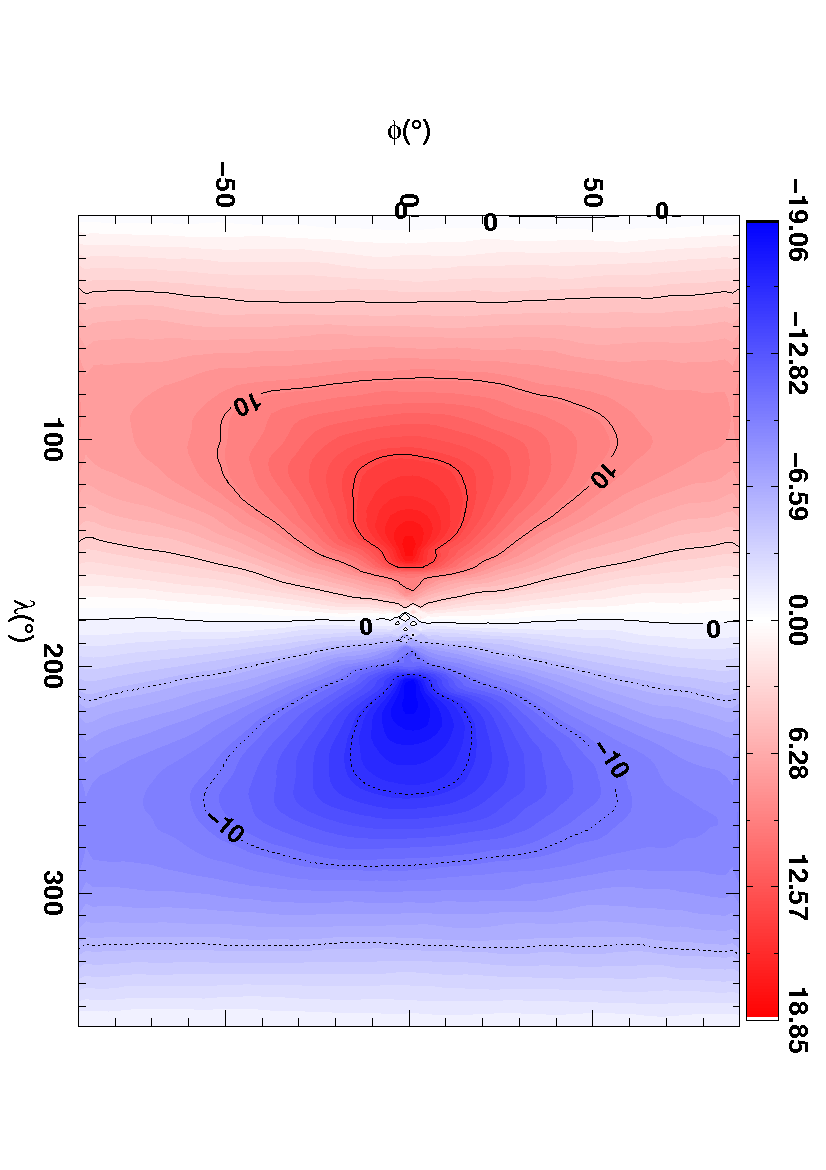}
\hspace*{-1.0cm}\includegraphics[width=5.5cm,angle=90.0]{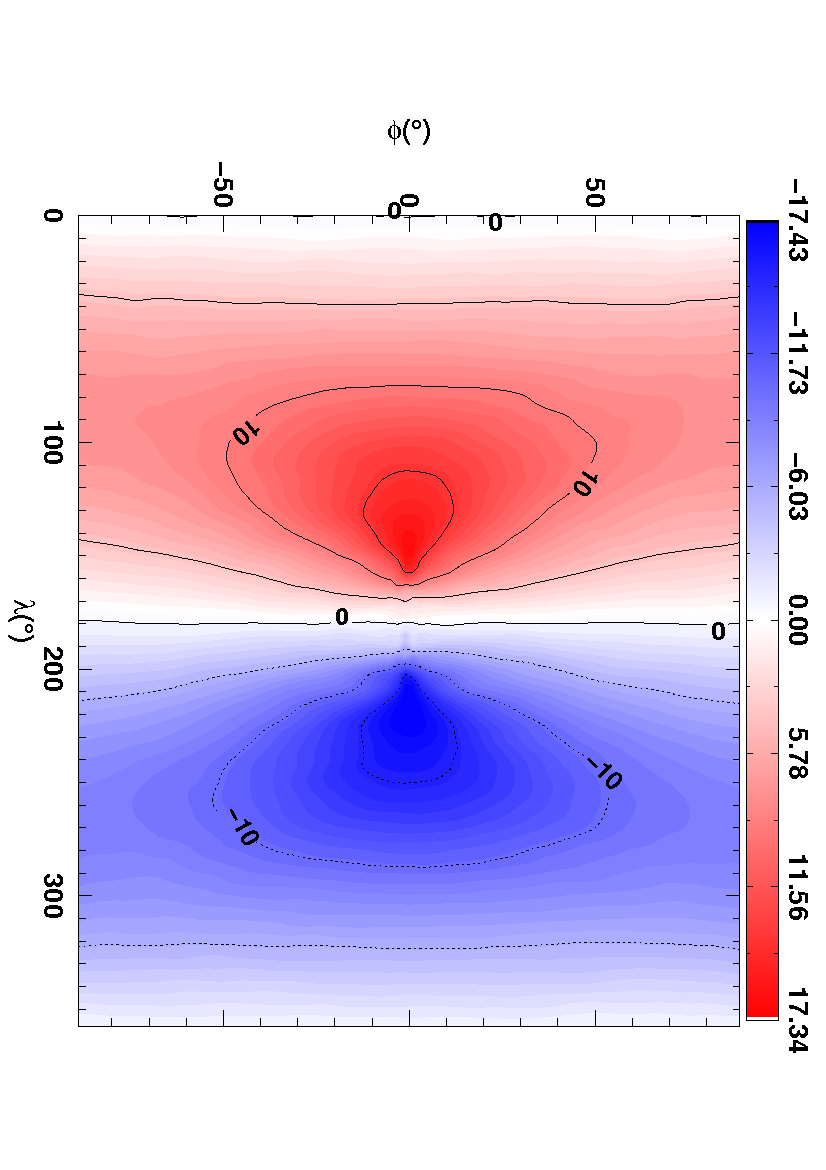}
\end{center}
\caption{Figure showing, for the Tidally Locked Earth test
  \citep{merlis_2010,heng_2011}, (from the \textit{top panels} to the
  \textit{bottom panels}) temperature at 1200 days and $\sigma=0.975$,
  then, temporally averaged zonal wind at $\sigma=0.225$, $0.525$ and
  $0.975$. Results are from the ND (\textit{left panels}) and EG
  (\textit{right panels}) models (see Table \ref{model_names} for
  explanation of model types). \label{TLE_u}}
\end{figure*}

Figure \ref{TLE_u} shows the same type of plots as Figure
\ref{TLE_heng}, but constructed using the ND (\textit{left panels})
and EG (\textit{right panels}) models, where the other ENDGame models
are omitted as the results are negligibly different from the EG model.

\begin{figure}[t]
\vspace*{2mm}
\begin{center}
\includegraphics[width=8.3cm,angle=0.0,origin=c]{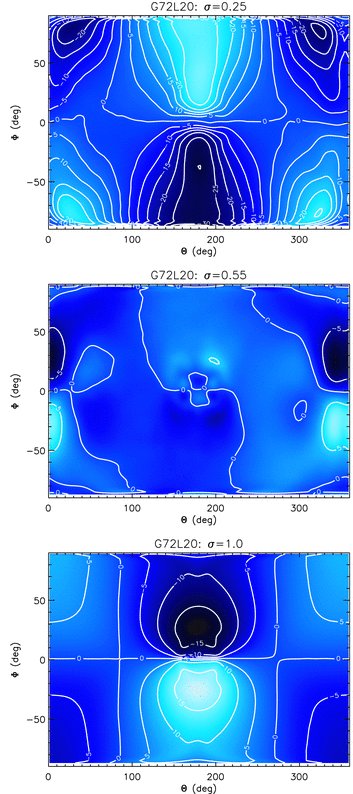}
\end{center}
\caption{Figure reproduced from \citet{heng_2011} of the results from
  the grid--based model of the TLE test case (reproduced by permission
  of Oxford University Press). Showing (from the \textit{top panel} to
  the \textit{bottom panel}) the temporally averaged meridional wind
  at $\sigma=0.225$, $0.525$ and $0.975$. \label{TLE_heng_v}}
\end{figure}

Figure \ref{TLE_heng_v} is a reproduction of the results of the
grid--based model for the TLE test case of \citet{heng_2011}, showing
the temporally averaged meridional velocity ($\overline{v}^{t}$) at
$\sigma=0.225$, $0.525$ and $0.975$ (from \textit{top} to \textit{bottom
  panel}, respectively).

\begin{figure*}[t]
\vspace*{2mm}
\begin{center}
\hspace*{-1.0cm}\includegraphics[width=5.5cm,angle=90.0]{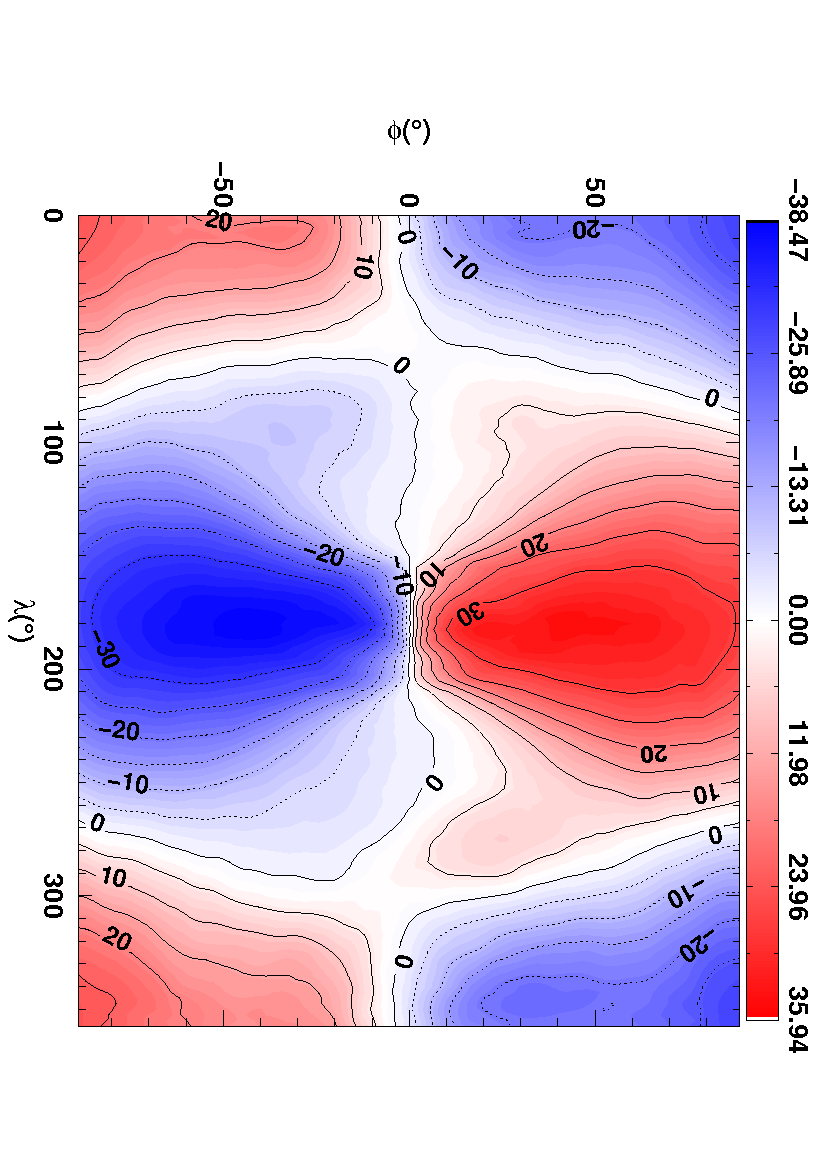}
\hspace*{-1.0cm}\includegraphics[width=5.5cm,angle=90.0]{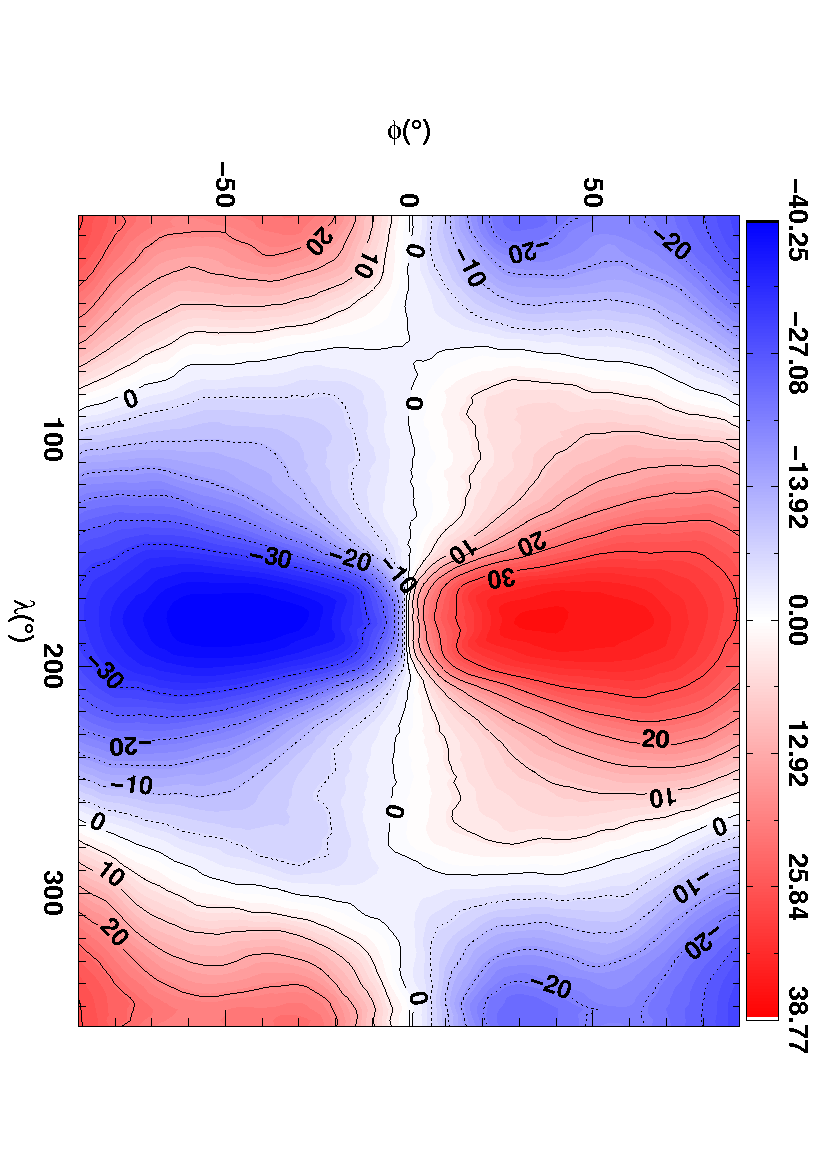}
\hspace*{-1.0cm}\includegraphics[width=5.5cm,angle=90.0]{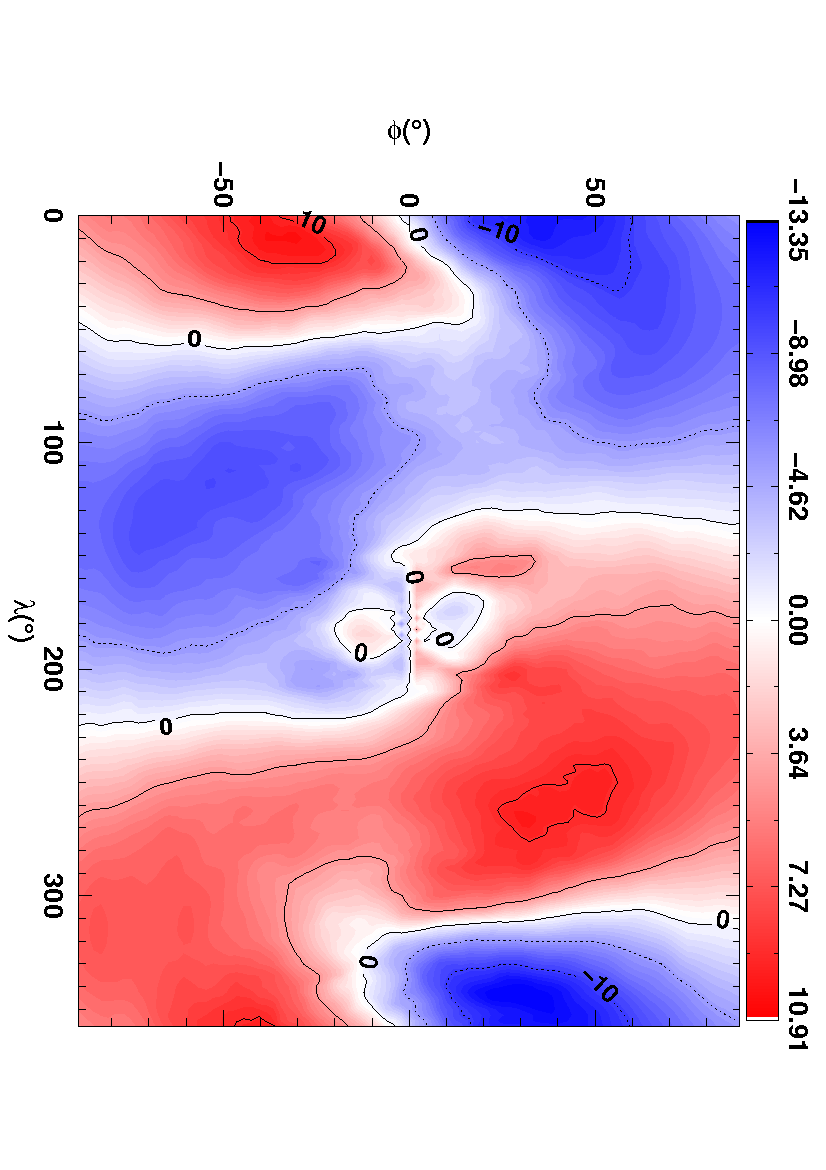}
\hspace*{-1.0cm}\includegraphics[width=5.5cm,angle=90.0]{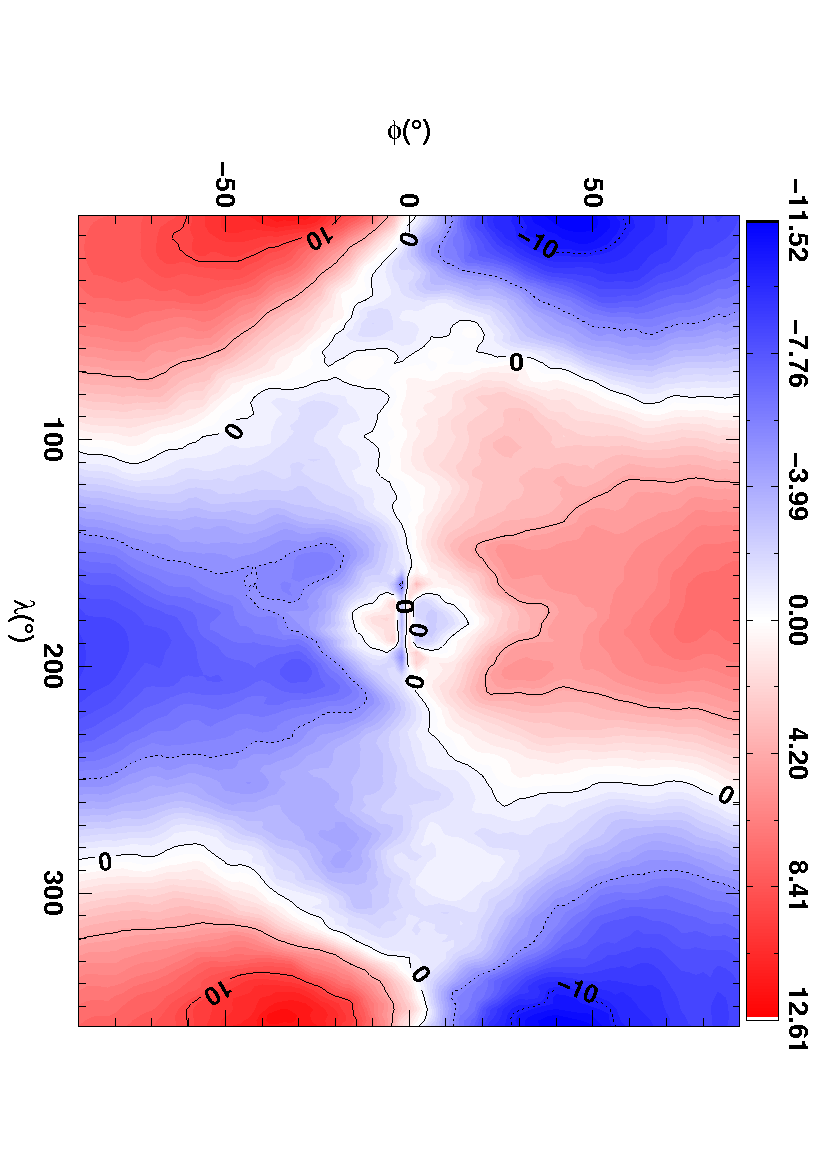}
\hspace*{-1.0cm}\includegraphics[width=5.5cm,angle=90.0]{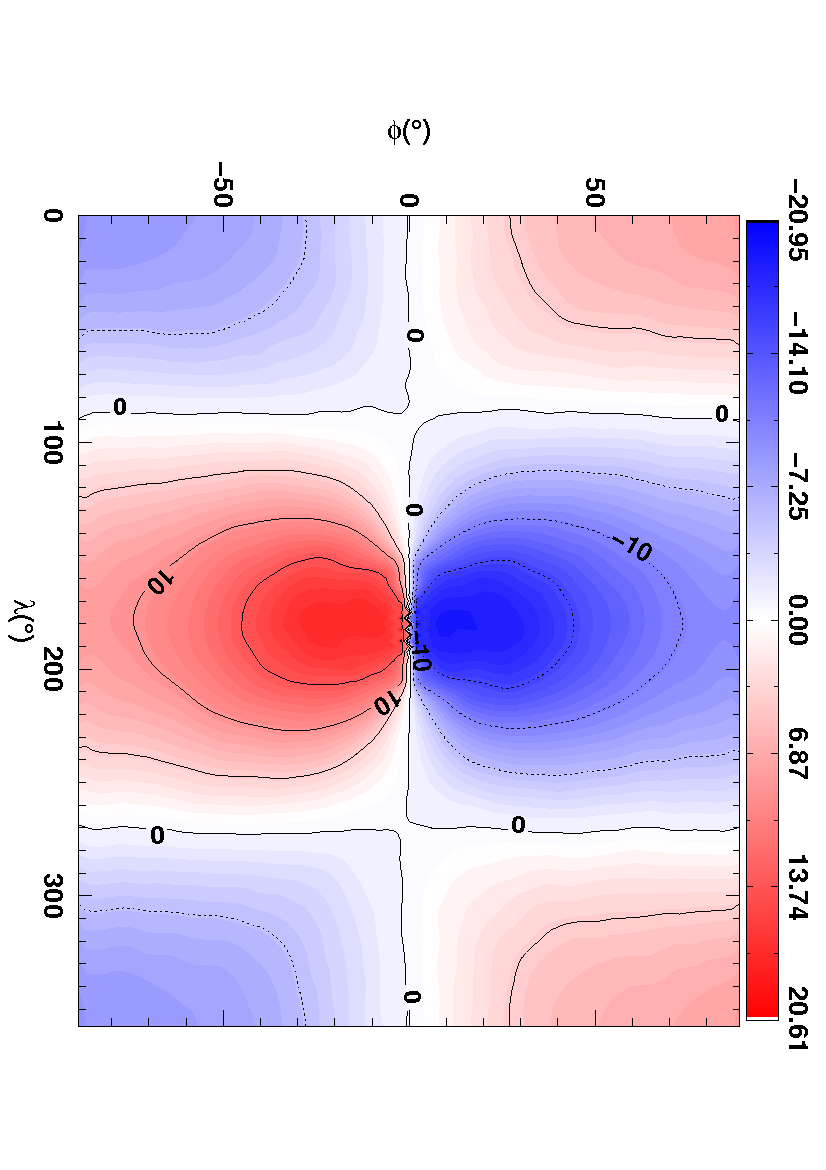}
\hspace*{-1.0cm}\includegraphics[width=5.5cm,angle=90.0]{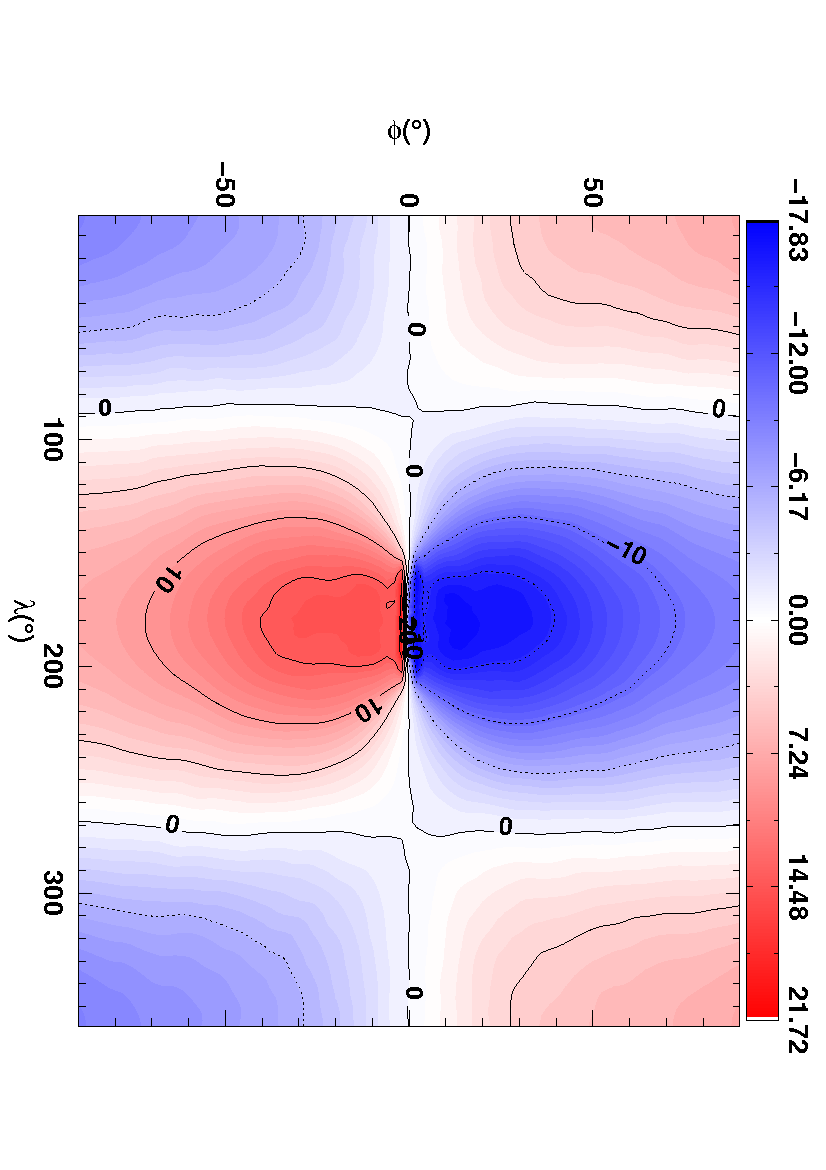}
\end{center}
\caption{Figure showing, for the Tidally Locked Earth test
  \citep{merlis_2010,heng_2011}, (from the \textit{top panel}s to the
  \textit{bottom panels}) the temporally averaged meridional wind at
  $\sigma=0.225$, $0.525$ and $0.975$. Results are from the ND
  (\textit{left panels}) and EG (\textit{right panels}) models (see
  Table \ref{model_names} for explanation of model
  types). \label{TLE_v}}
\end{figure*}

The results for our models are shown in Figure \ref{TLE_v} in the same
vertical format as Figure \ref{TLE_heng_v}. As for Figure \ref{TLE_u}
the figures show the ND (\textit{left panels}) and EG (\textit{right
  panels}) models, where (as with Figure \ref{TLE_u}) the other
ENDGame models are omitted as the results are negligibly different
from the EG model.

Comparison of the results of \citet{heng_2011}, Figures \ref{TLE_heng}
and \ref{TLE_heng_v} with our results, Figures \ref{TLE_u} and
\ref{TLE_v} reveals some disagreement. However, Figures \ref{TLE_heng}
and \ref{TLE_heng_v} show results from the finite difference model,
and our results agree much more closely with those derived from the
spectral code of \citet{heng_2011} (this is discussed in more detail
later in this section). Again, as before our vertical resolution is
higher than that of \citet{heng_2011}, 32 as opposed to 20
levels. Tentative evidence for a smoother modeling of the meridional
flow can also be seen by comparing our results for the $v$ field
(Figure \ref{TLE_v}) at a $\sigma$ of $0.225$ and $0.525$ to that of
\citet{heng_2011} (Figure \ref{TLE_heng_v}). Our figures produce flow
contours less featured than those of \citet{heng_2011} (in fact our
model matches more closely the spectral model results not reproduced
here which we expect to be more accurate for large--scale flows,
compared to the finite-difference model). Additionally, as with the
previous cases, given the model domain one would expect little
difference in results whether the `shallow--atmosphere' approximation
is made or not (given the aspect ratio, height over the length scale,
$H/L\sim \frac{3.2\times 10^4}{2.0\times 10^7}\sim10^{-3}$, where the
length scale is chosen as half the perimeter of the planet due to the
presence of hemispherical circulation cells), and gravity does not
vary much over the atmosphere ($g_{\rm surf}\sim 9.8$ ms$^{-1}$ at the
surface to $g(r_{\rm top})=g_{\rm surf}(R_{\rm p}/r_{\rm
  top})^2\sim9.8\times \left( \frac{6.4\times 10^6}{[3.2\times
    10^4+6.4\times 10^6]}\right)^2\sim 0.990\times 9.8$ ms$^{-1}$, at
the top of the atmosphere ignoring self--gravity and using the inverse
square--law).

\begin{figure}[t]
\vspace*{2mm}
\begin{center}
\hspace*{-1.0cm}\includegraphics[width=6.5cm,angle=90.0]{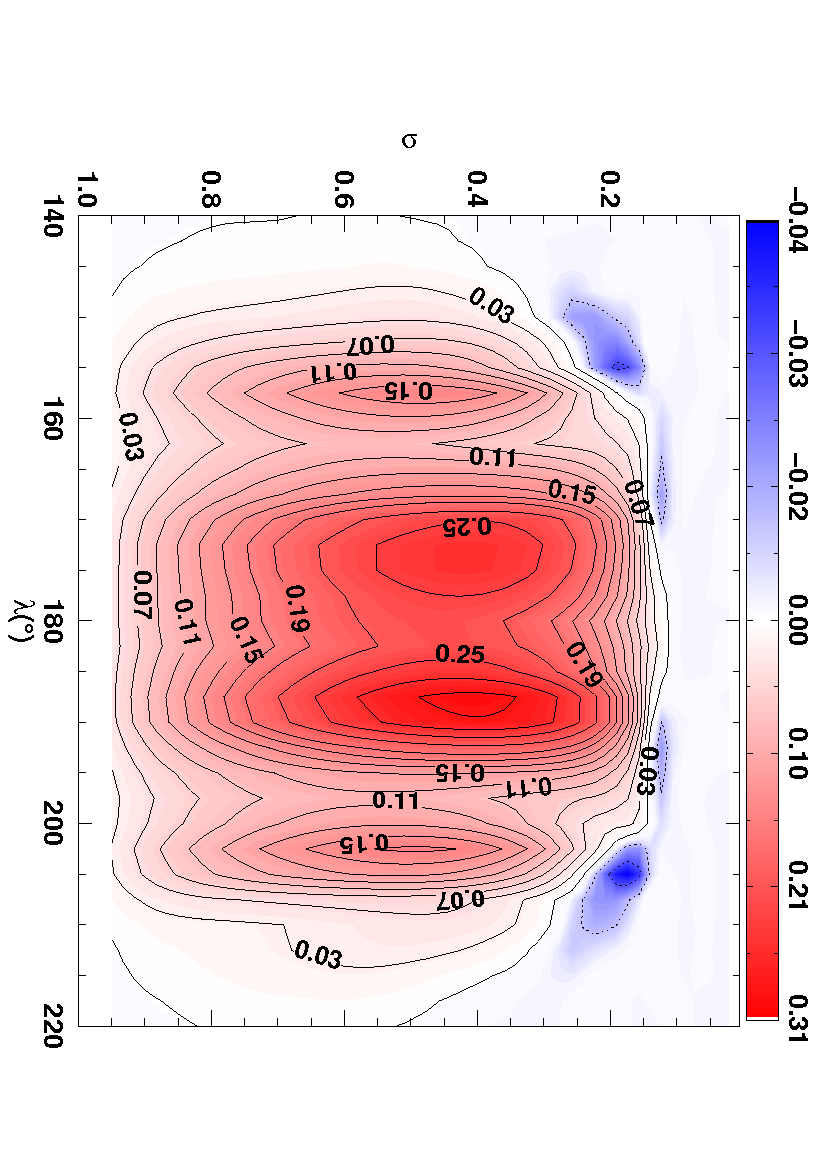}
\hspace*{-1.0cm}\includegraphics[width=6.5cm,angle=90.0]{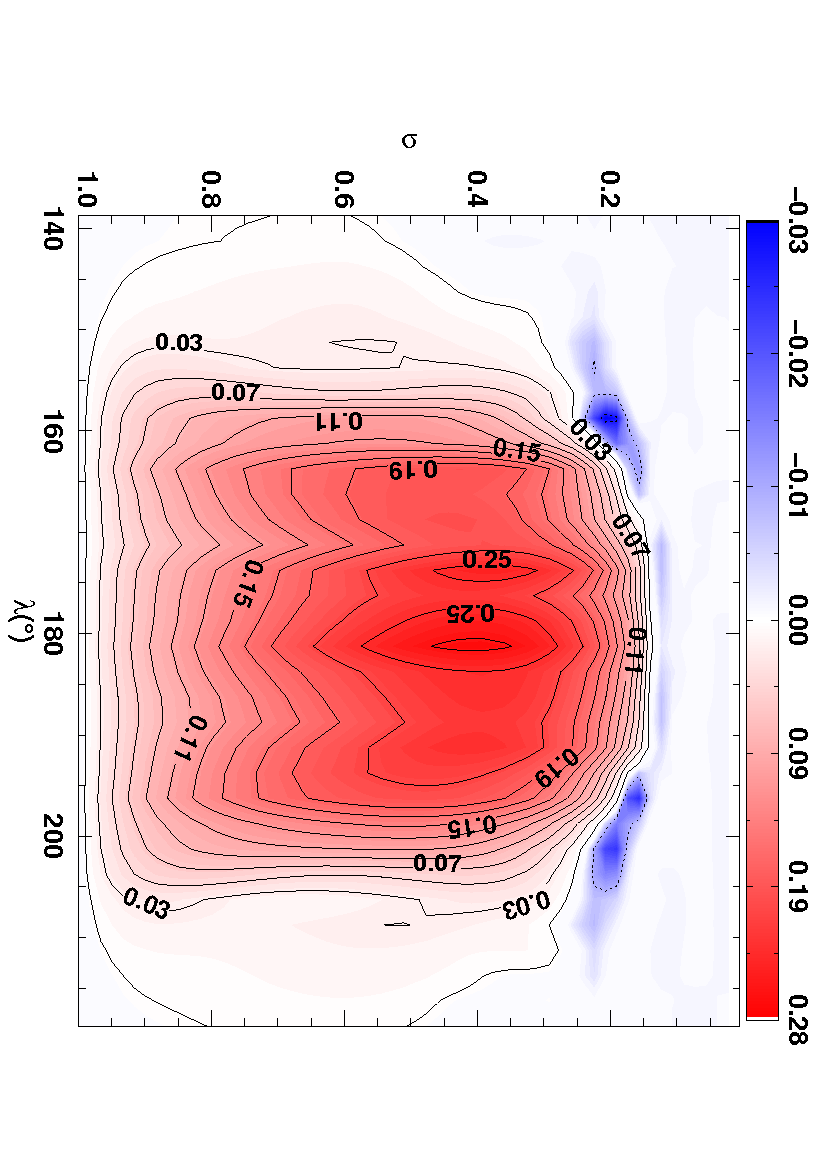}
\hspace*{-1.0cm}\includegraphics[width=6.5cm,angle=90.0]{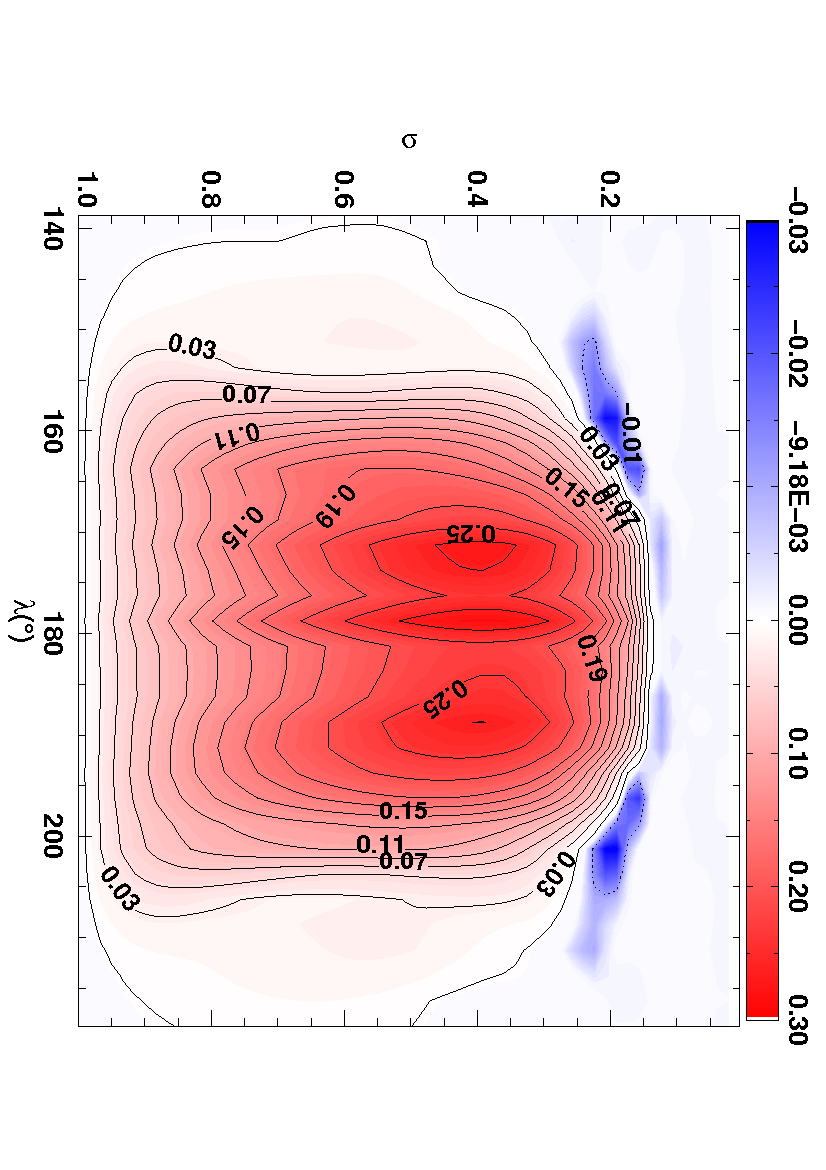}
\end{center}
\caption{Figure showing, for the Tidally Locked Earth test
  \citep{merlis_2010,heng_2011}, the temporally averaged vertical
  velocities (ms$^{-1}$) over the `hot spot' or subsolar point for the
  ND (\textit{top panel}), EG$_{\rm sh}$ (\textit{middle panel}) and
  EG (\textit{bottom panel}) models (see Table \ref{model_names} for
  explanation of model types).\label{hot_spot_W}}
\end{figure}

The horizontal flow, across all of the TLE ENDGame models is
consistent. Further evidence for a consistent solution can be found in
the similarity of the time averaged vertical velocities over the `hot
spot', shown in Figure \ref{hot_spot_W}. Figure \ref{hot_spot_W} shows
the results from the ND, EG$_{\rm sh}$ and EG models as the
\textit{top}, \textit{middle} and \textit{bottom panels},
respectively.

Figure \ref{hot_spot_W} shows a broad updraft over the `hot spot'
rising to $\sigma\sim 0.2$. The maximum difference in vertical
velocity between the EG and EG$_{\rm sh}$ models are $\sim 0.1$
ms$^{-1}$, and these are localised to regions directly above the area
of most intense heating, with negligible differences elsewhere. This,
as is expected suggests that the simplifications of the dynamical
equations are not changing the resulting circulation. The structure of
the updraft is marginally different in the ND compared to either of
the EG models.

\begin{figure*}[t]
\vspace*{2mm}
\begin{center}
\hspace*{-1.0cm}\includegraphics[width=6.5cm,angle=90.0]{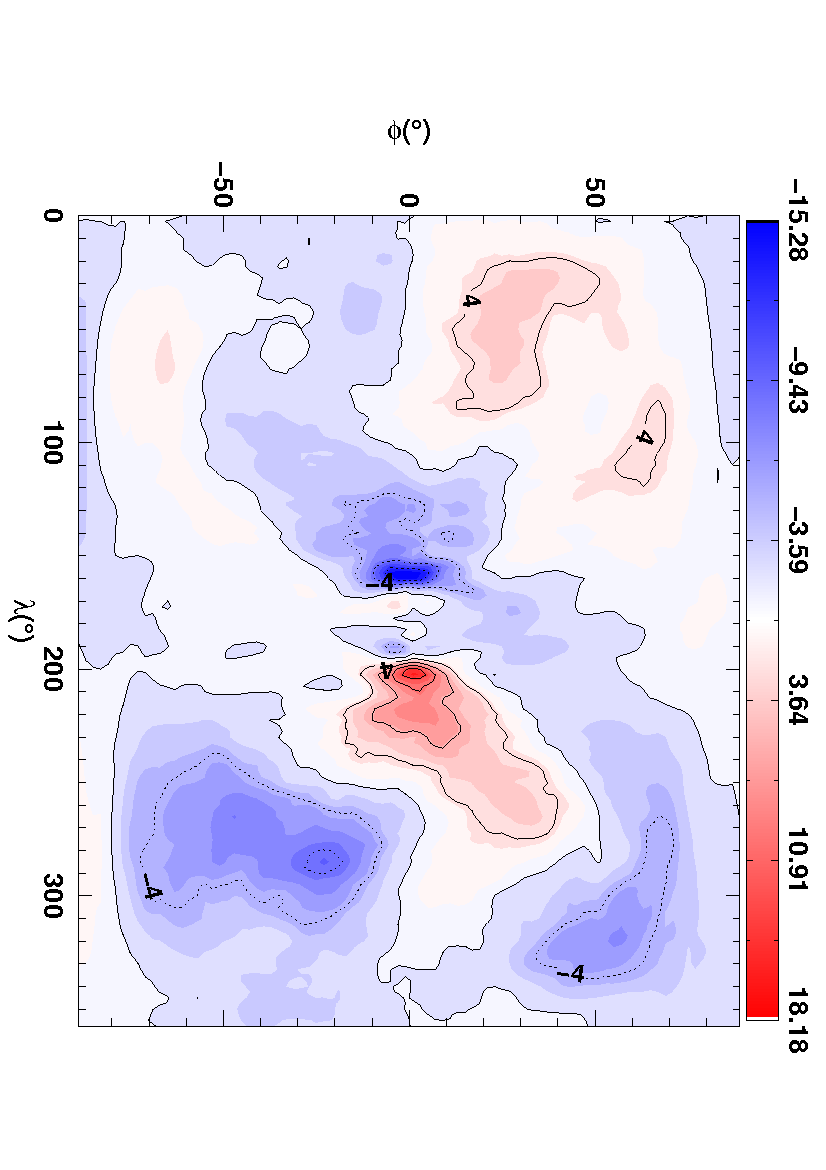}
\hspace*{-1.0cm}\includegraphics[width=6.5cm,angle=90.0]{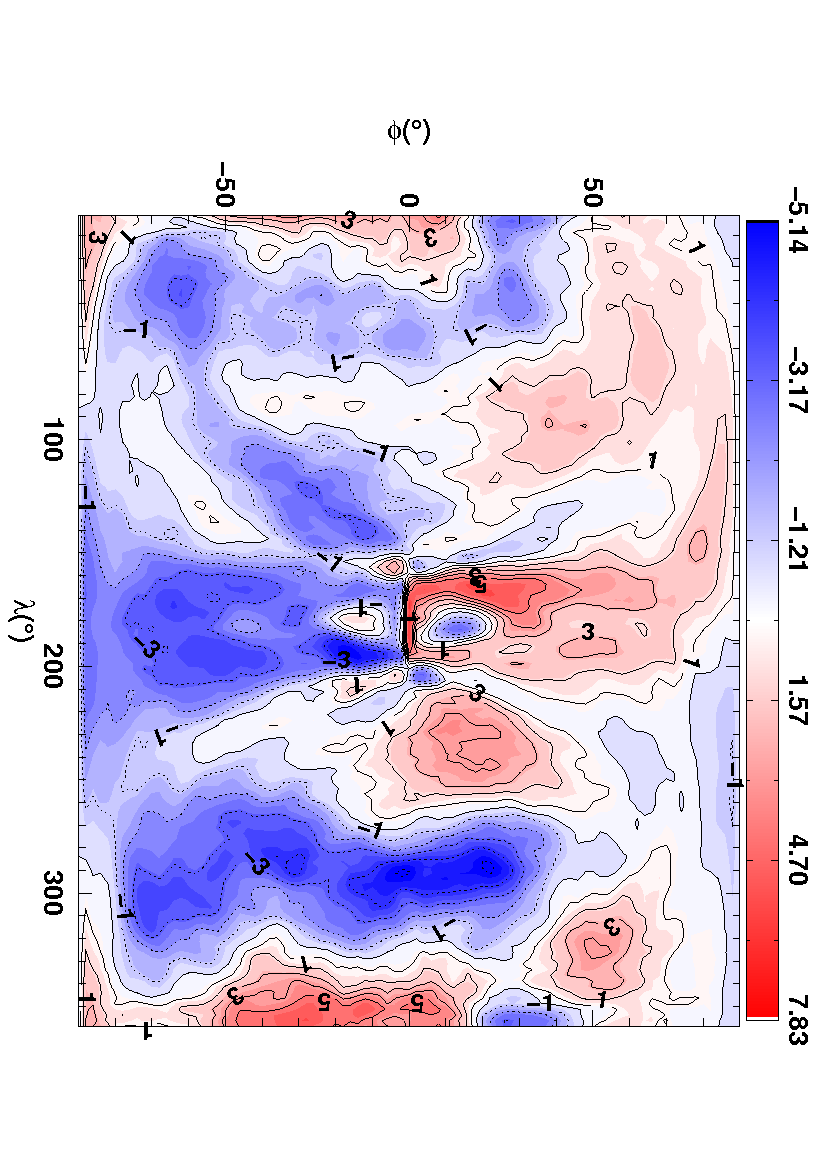}
\hspace*{-1.0cm}\includegraphics[width=6.5cm,angle=90.0]{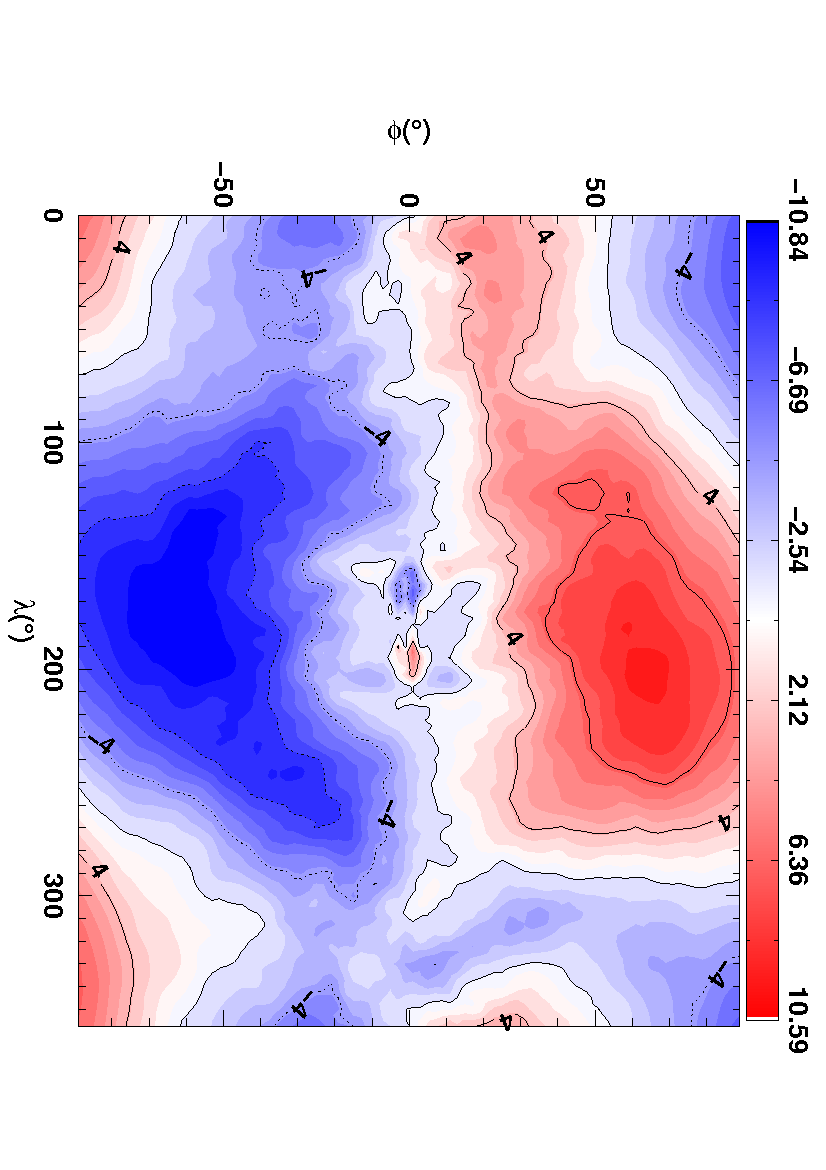}
\hspace*{-1.0cm}\includegraphics[width=6.5cm,angle=90.0]{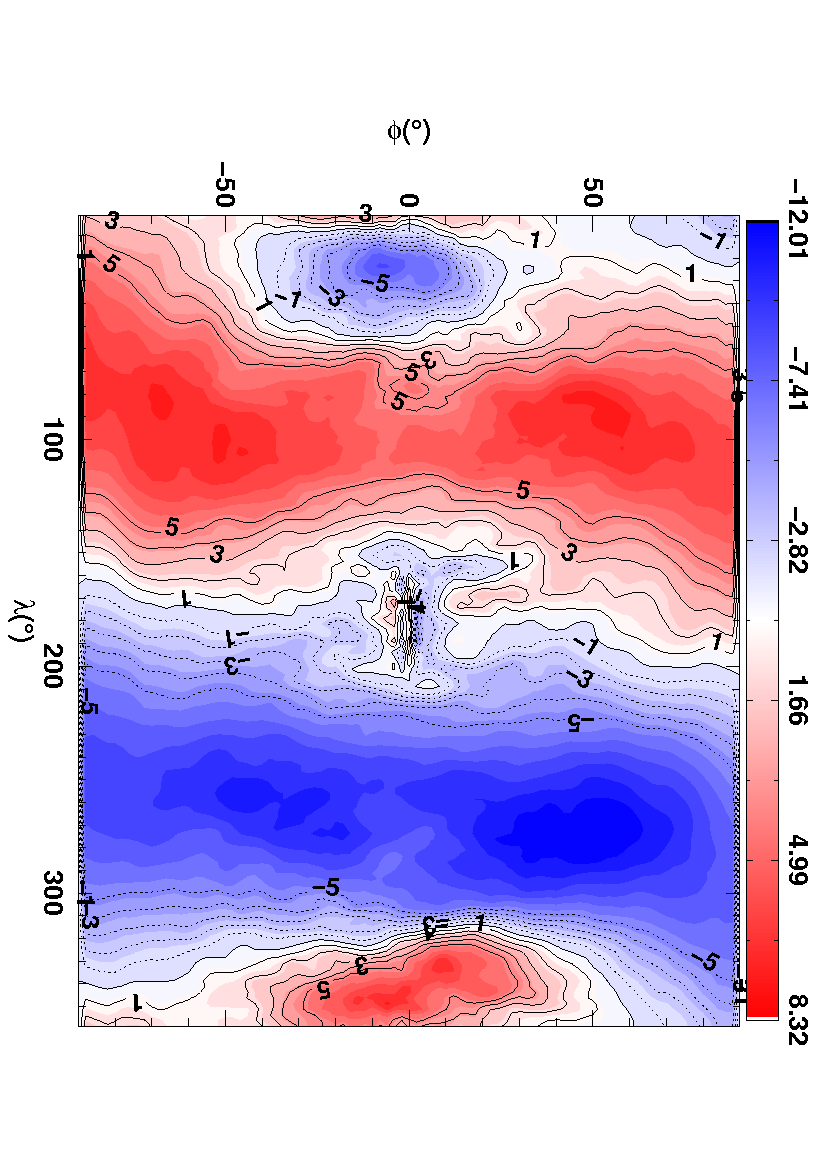}
\hspace*{-1.0cm}\includegraphics[width=6.5cm,angle=90.0]{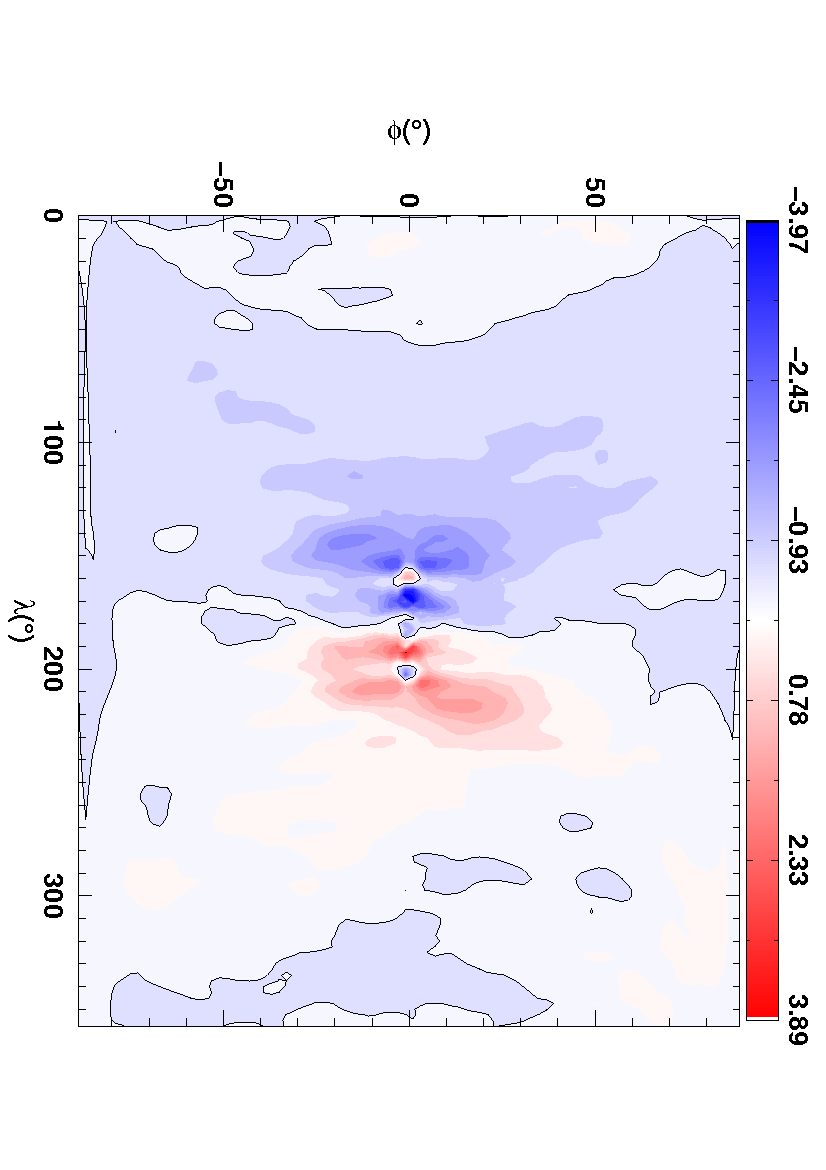}
\hspace*{-1.0cm}\includegraphics[width=6.5cm,angle=90.0]{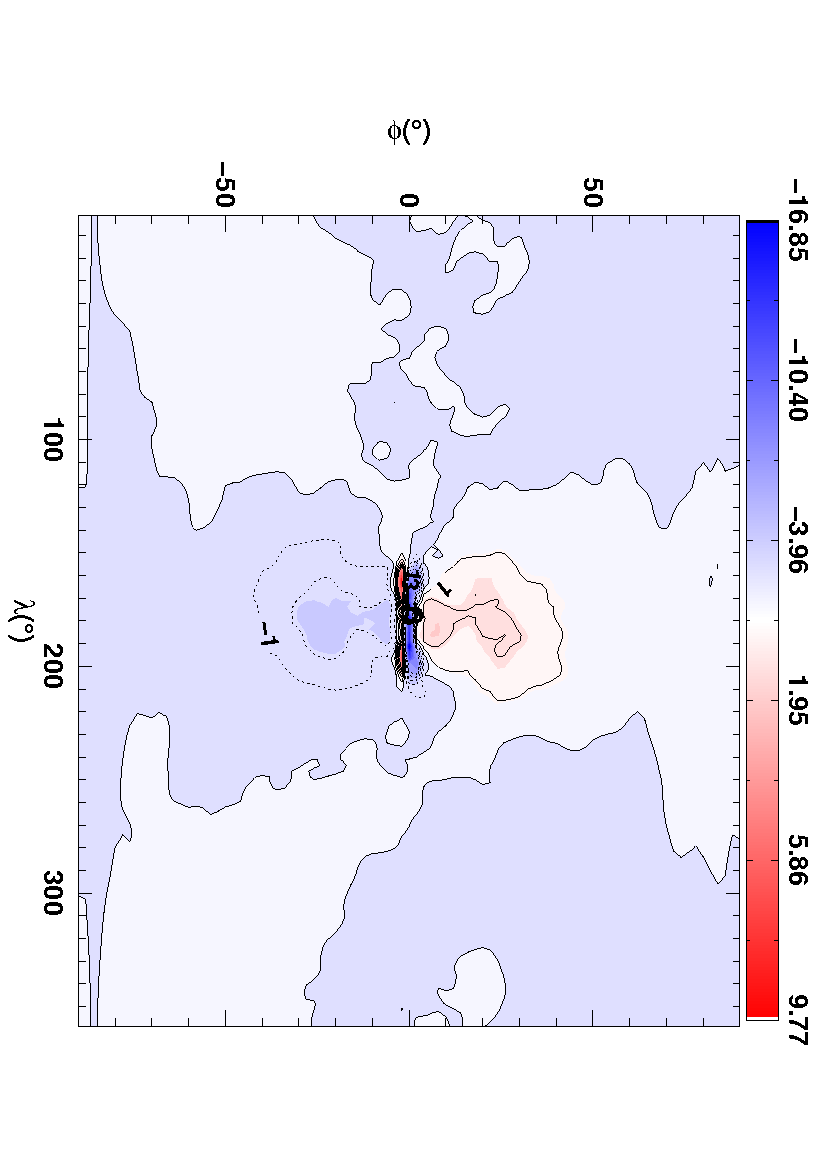}
\end{center}
\caption{Figure, Tidally Locked Earth test
  \citep{merlis_2010,heng_2011}, showing the differences EG$-$ND of
  the temporally averaged, zonal (\textit{left panels}) and meridional
  (\textit{right panels}) winds (ms$^{-1}$), at $\sigma=0.975$,
  $0.525$, and $0.225$, as the \textit{top}, \textit{middle} and
  \textit{bottom rows}, respectively (see Table \ref{model_names} for
  explanation of model types). \label{diff_TLE}}
\end{figure*}

As with the HS and EL test cases we have constructed plots of the
difference between the models. We have not produced these plots for
the instantaneous results of the temperature field, as differences in
such `snapshots' can be dominated by intrinsic temporal
variability. Additionally, as with the HS and EL test cases, the
differences between the ENDGame model results are an order of
magnitude smaller than those found between the ENDGame models and ND,
therefore only EG$-$ND is presented. Figures \ref{diff_TLE} shows the
difference, EG$-$ND, of the temporally averaged zonal and meridional
wind, as the \textit{left} and \textit{right panels} respectively, at
the surfaces presented in Figures \ref{TLE_u} and \ref{TLE_v}.

Figure \ref{diff_TLE} shows the zonal wind at $\sigma=0.225$ is faster
in the EG model, over the ND model, as the residual of EG$-$ND is
positive, for the positive flow where $\lambda>180^{\circ}$, and
negative for the negative flow where
$\lambda<180^{\circ}$. Essentially, the zonal flow (\textit{left
  panels}) away from the `hot spot' near the upper boundary is faster
in the EG model. The opposite is true for the $\sigma=0.975$ surface,
where the flow appears to be slowed in the EG, compared to the ND
model. The most intriguing difference is found at the $\sigma=0.525$
isobaric--surface where, as shown in Figure \ref{TLE_u} the flow
structure has inverted about the equator. The meridional flow is also
enhanced near the upper boundary, $\sigma=0.225$, and slowed near the
surface, in the EG model compared to the ND model (\textit{right
  panels} of Figure \ref{diff_TLE}). At the $\sigma=0.525$ surface a
systematic change either side of the equator is found, indicative of a
reversal of the flow structure one can see in the \textit{middle row}
of Figure \ref{TLE_v}. For $\lambda>180^{\circ}$ the flow is directed
towards the south pole, opposite to that found in ND, and the flow is
also reversed for $\lambda<180^{\circ}$. This reversal of flow and
difference in the diagnostic plots occurs for all ENDGame models. The
flow structure at $\sigma=0.525$ in our ENDGame models match, more
closely that found in the spectral code models of
\citet{heng_2011}. Whereas the flow for the ND model matches, more
closely that found in the finite difference model of
\citet{heng_2011}. An explicit polar filter is used in both the ND and
the \citet{heng_2011} finite difference models, but is not required in
either ENDGame or the \citet{heng_2011} spectral model. However, we
have run the TLE case using ENDGame but applying a polar filter (as
used in the ND model) and found our results still matched, more
closely the \citet{heng_2011} spectral model. This suggests that the
difference is due to improvements in the numerical scheme of ENDGame
over ND and not the polar filtering scheme.

\begin{figure}[t]
\vspace*{2mm}
\begin{center}
\hspace*{-1.0cm}\includegraphics[width=6.5cm,angle=90.0]{./Figs/TLE_EG_Deep_Temp.png}
\hspace*{-1.0cm}\includegraphics[width=6.5cm,angle=90.0]{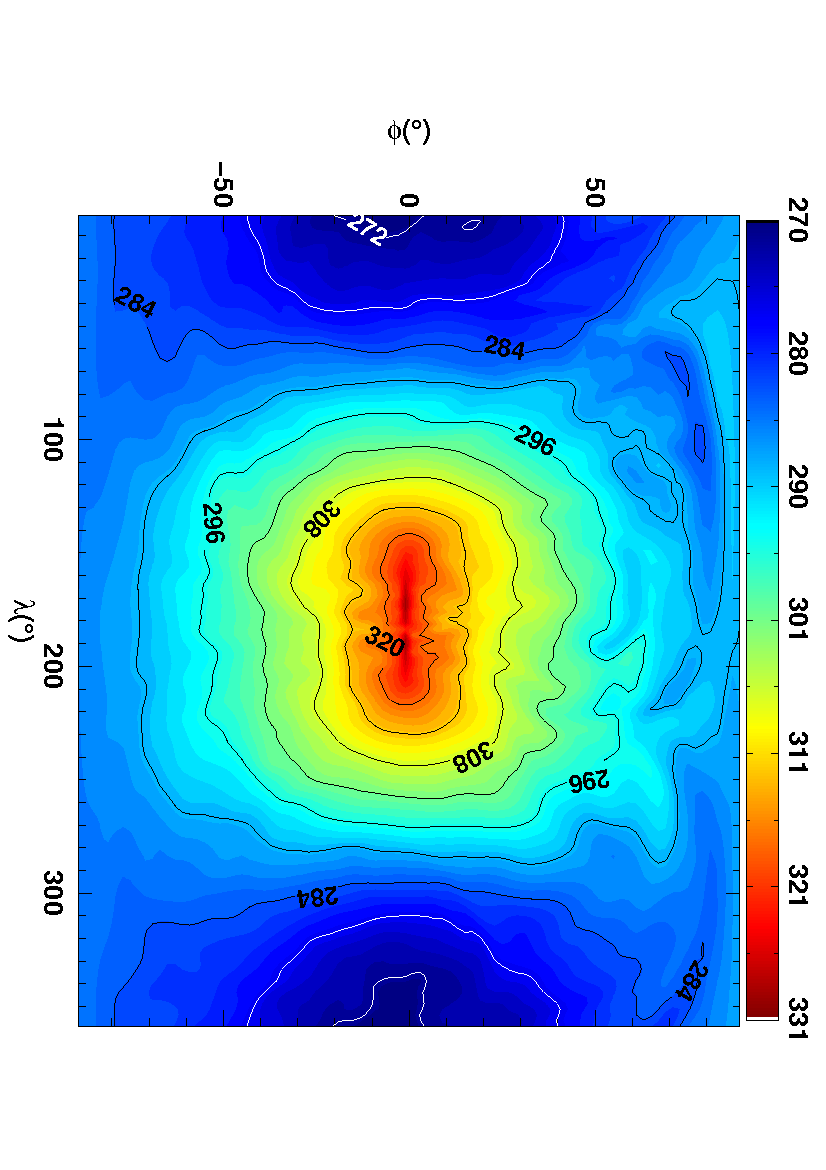}
\end{center}
\caption{Figure showing, for the Tidally Locked Earth test
  \citep{merlis_2010,heng_2011}, showing temperature at 1200 days and
  $\sigma=0.975$, for the EG models (see Table \ref{model_names} for
  explanation of model types) using $\alpha$ of $0.55$ and $1.0$
  (\textit{top} and \textit{bottom panels},
  respectively). \label{TLE_alpha}}
\end{figure}

The structure of the `hot spot' in the \textit{top panel} of Figure
\ref{TLE_u} shows the central contour is more elliptical for all the
ENDGame solutions, matching more closely (than the ND models) the
shape in Figure \ref{TLE_heng}. The structure of the `hot spot' also
seems `noisier' in the ENDGame models. The noise exhibited in the
ENDGame models is indicative of the reduced implicit damping in the
numerical scheme. This can be shown by making the ENDGame scheme more
implicit, and therefore, dissipative, by adjusting the temporal
weighting coefficient, $\alpha$. Increasing $\alpha$ leads to greater
weight being applied to the $i+1$ state and therefore a more implicit
scheme. For our ND model and all ENDGame models the $\alpha$ values
are $0.7$ and $0.55$ respectively \citep[i.e. ENDGame is more
explicit, yet is able to run stably with the same length timestep due
to the changes outlined in Section \ref{delta_num_scheme} and detailed
in][]{wood_2013}. Figure \ref{TLE_alpha} shows the temperature
structure shown in Figure \ref{TLE_u} (\textit{top panel}) for both
the EG using the standard $\alpha=0.55$ (already displayed in Figure
\ref{TLE_u}, \textit{rightmost panel}, reproduced to aid comparison)
and an EG model where $\alpha$ has been increased to 1.0. The fully
implicit model presents a smoother temperature structure.

\begin{figure}[t]
\vspace*{2mm}
\begin{center}
\hspace*{-1.0cm}\includegraphics[width=6.5cm,angle=90.0]{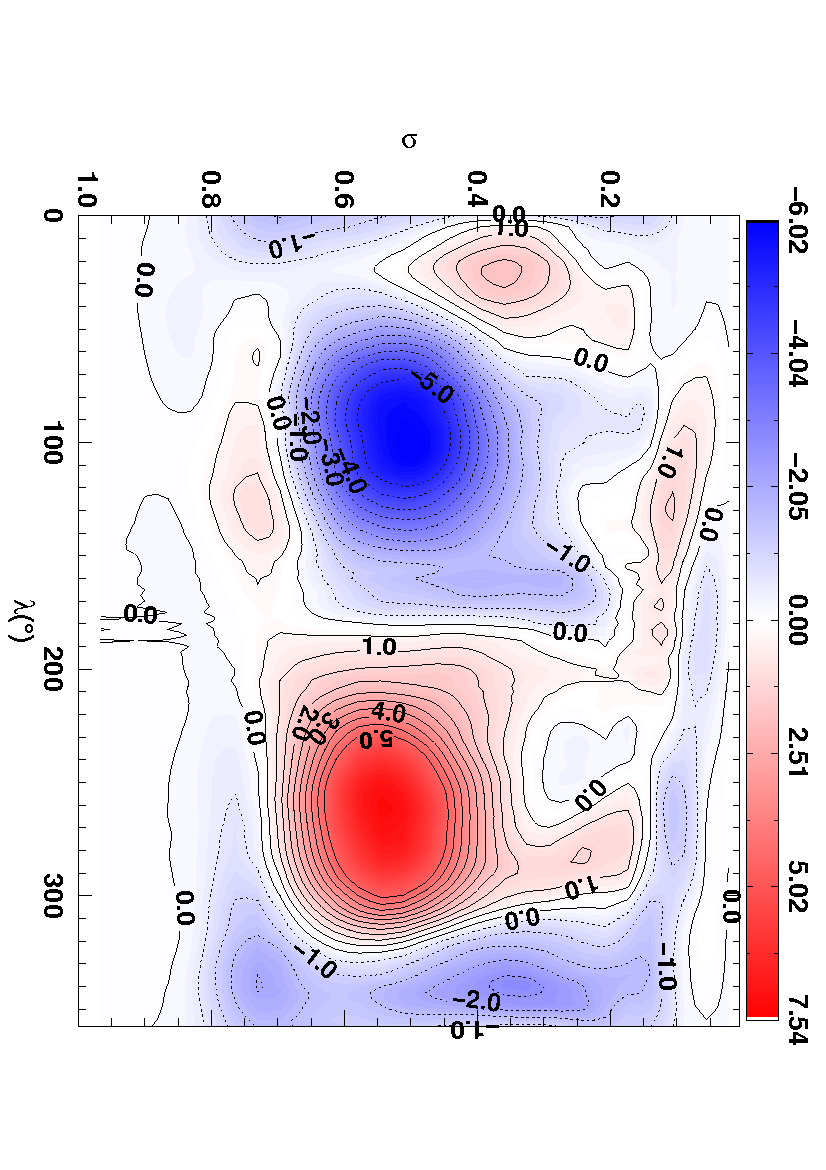}
\hspace*{-1.0cm}\includegraphics[width=6.5cm,angle=90.0]{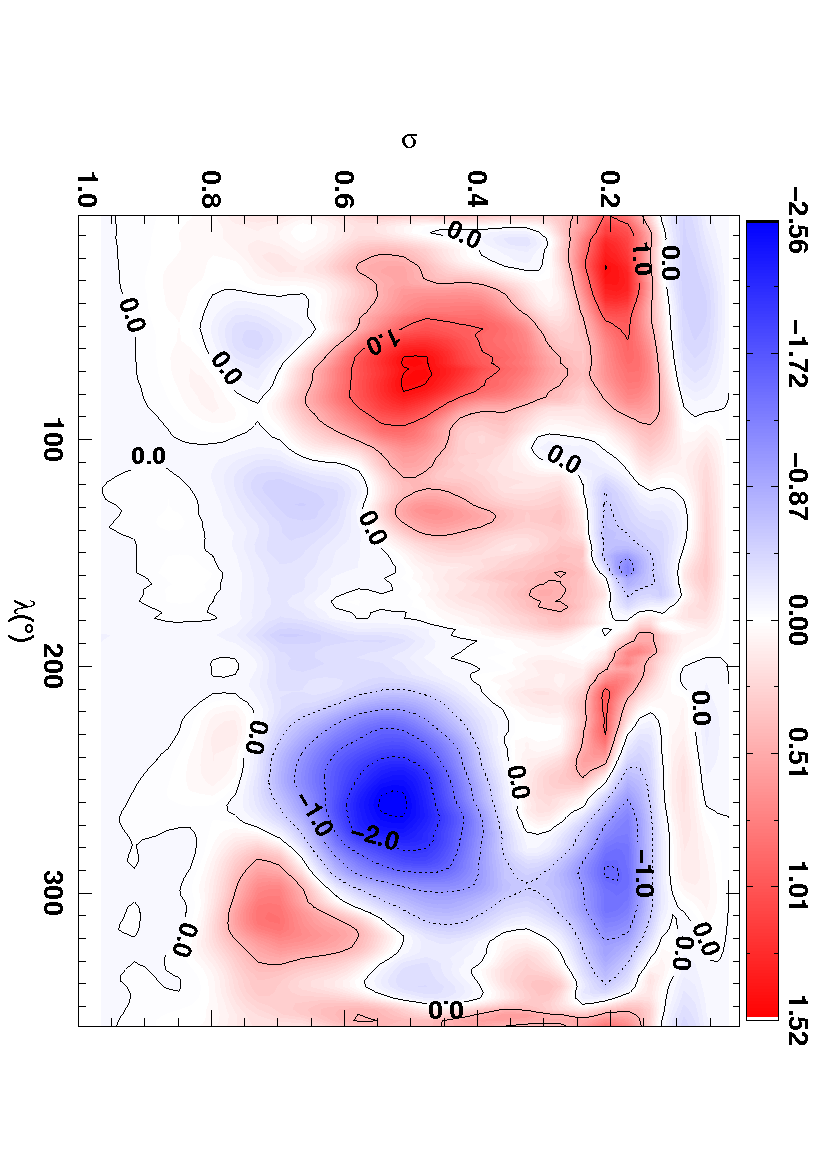}
\end{center}
\caption{Figure showing, for the Tidally Locked Earth test
  \citep{merlis_2010,heng_2011}, the temporally and meridionally
  averaged meridional flow for the ND (\textit{top panel}) and
  EG$_{\rm gc}$ (\textit{bottom panel}) models (see Table
  \ref{model_names} for explanation of model types). \label{Vvel_avg}}
\end{figure}

To attempt to isolate differences caused only by the numerical scheme
we compare the nature of the meridional circulation for the TLE models
using ND and EG$_{\rm gc}$, since the ND and EG$_{\rm gc}$ models
solve identical equations sets. Figure \ref{Vvel_avg} shows the
temporally and meridionally averaged meridional flow for the ND and
EG$_{\rm gc}$ models. The average is performed in a point--wise
fashion, i.e. $\int\,vd\,\phi$ as opposed to $\int\,\cos\phi
vd\,\phi$, to emphasise differences in flow over the pole. In a
non--rotating system, where the Coriolis force is zero, one would
expect a symmetric meridional flow, so the latitudinal average should
be close to zero. For the TLE case the rotation is slow, with a Rossby
number of, $R_o=\frac{U}{Lf}\sim\frac{30}{4\times 10^7 \times 2\times
  2\times10^{-7}}\sim 2.0$ (where $U$ is the horizontal velocity
scale, $L$ the length scale and $f=2\Omega\sin\phi$; the Coriolis
frequency or parameter), indicating negligible effects of rotation.

Figure \ref{Vvel_avg} shows that the meridional average is almost an
order of magnitude larger in the ND case, compared with the EG$_{\rm
  gc}$ model. To further examine the symmetry of meridional
circulation cells, we define a stream function ($\Psi$) as

\begin{equation}
\Psi=-2\pi\cos\phi\int_{R_{\rm p}}^rr\bar{v}(\phi,\tilde{r})d\tilde{r},\\
\label{euler_sf_eqn}
\end{equation}
where $\bar{v}$ denotes the zonally averaged meridional velocity.

\begin{figure}[t]
\vspace*{2mm}
\begin{center}
\hspace*{-1.0cm}\includegraphics[width=6.5cm,angle=90.0]{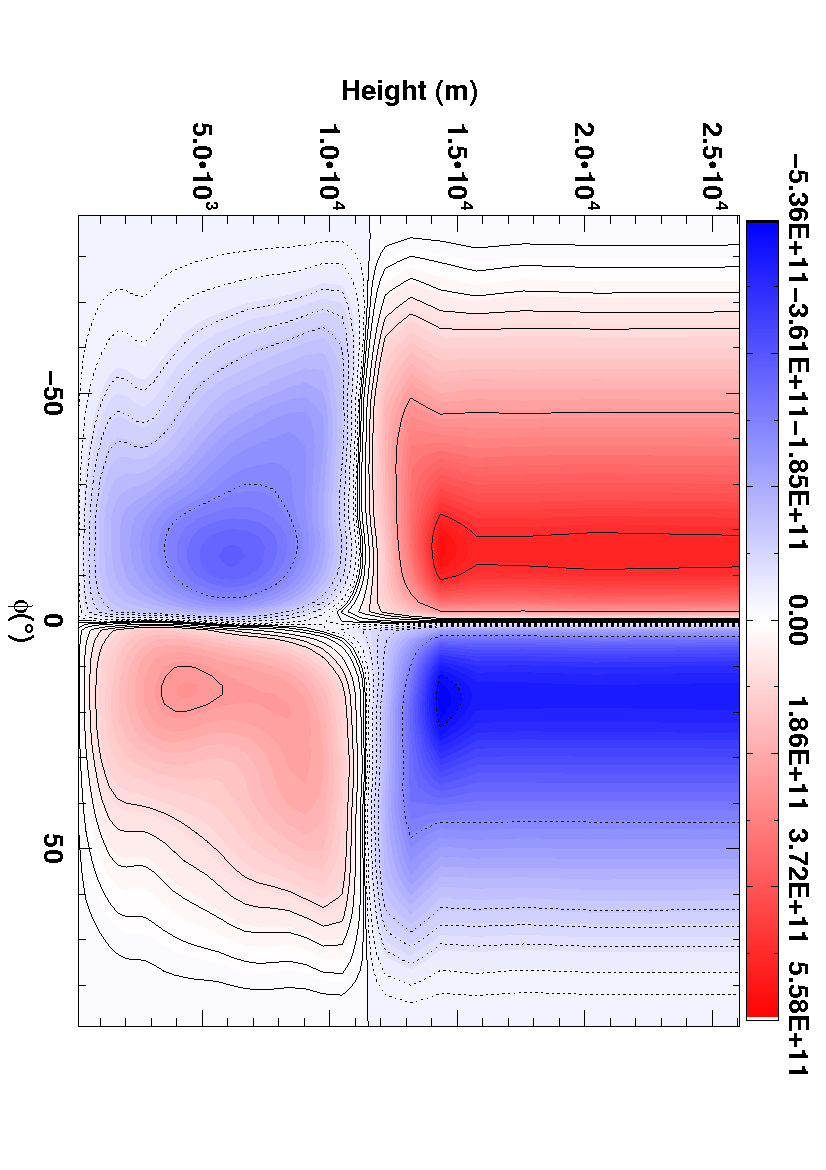}
\hspace*{-1.0cm}\includegraphics[width=6.5cm,angle=90.0]{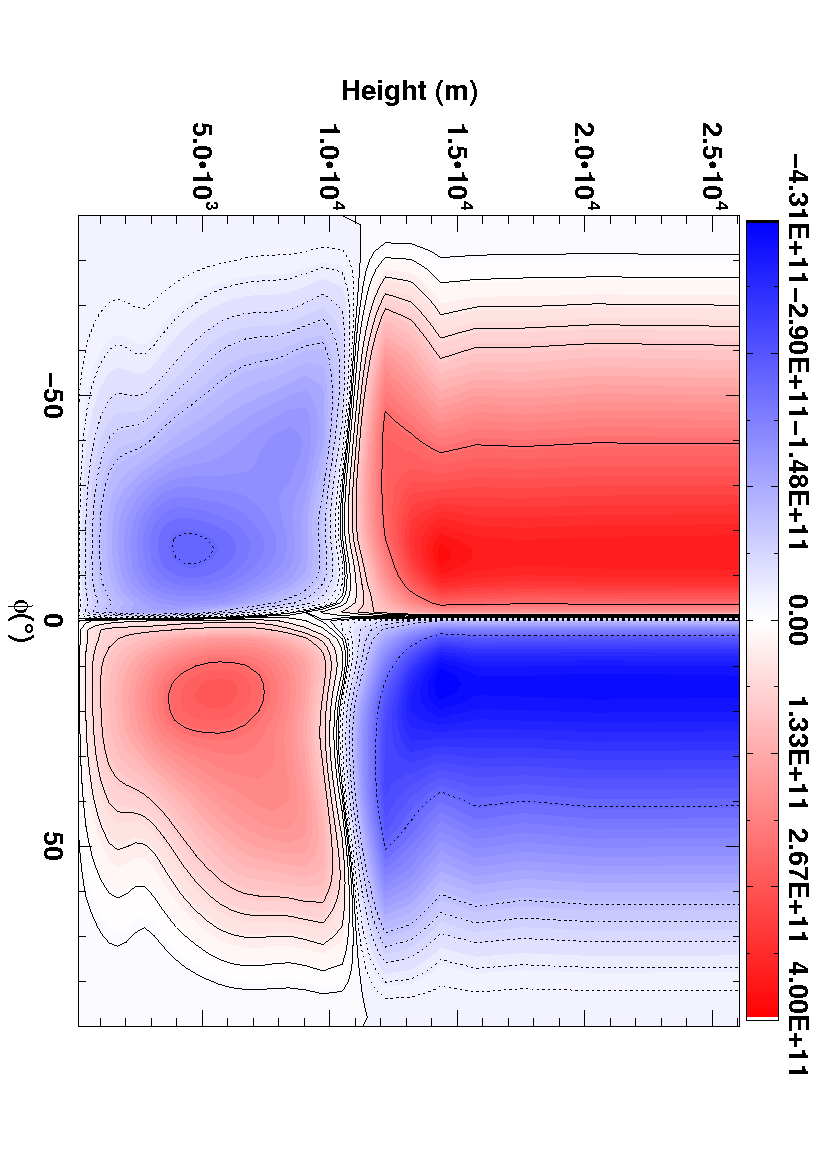}
\end{center}
\caption{Figure showing, for the Tidally Locked Earth test
  \citep{merlis_2010,heng_2011}, the streamfunction $\Psi$ (defined in
  text, see Equation \ref{euler_sf_eqn}) for the ND (\textit{top
    panel}) and EG$_{\rm gc}$ (\textit{right panel}) models (see Table
  \ref{model_names} for explanation of model types). The contours in
  both panels are the same and set at values $-5.0\times10^{11}$,
  $-2.5\times10^{11}$, $-1.0\times10^{11}$, $-7.5\times10^{10}$,
  $-5.0\times10^{10}$, $-2.5\times10^{10}$, $-1.0\times10^{10}$, 0.0,
  $1.0\times10^{10}$, $2.5\times10^{10}$, $5.0\times10^{10}$,
  $7.5\times10^{10}$, $1.0\times10^{11}$, $2.5\times10^{11}$ and
  $5.0\times10^{11}$. \label{euler_sf}}
\end{figure}

Figure \ref{euler_sf} shows this diagnostic as a function of latitude
and height for the ND and EG$_{\rm gc}$ models. The values assigned to
the contours in \textit{both panels} of Figure \ref{euler_sf} are the
same. The results are similar for both models but the circulation
cells are marignally more symmetric (especially closer to the surface)
for the EG$_{\rm gc}$ models. The lower (in altitude) circulation
cells are \textit{direct} i.e. caused by the heating of the
atmosphere, whilst the higher cells are \textit{indirect}. As shown in
\citet{heng_2011b} the circulation cells differ on the day and night
side. However, here we do not split by hemisphere as we are simply
interested in the comparison between models.

\begin{figure*}[t]
\vspace*{2mm}
\begin{center}
\hspace*{-1.0cm}\includegraphics[width=6.5cm,angle=90.0]{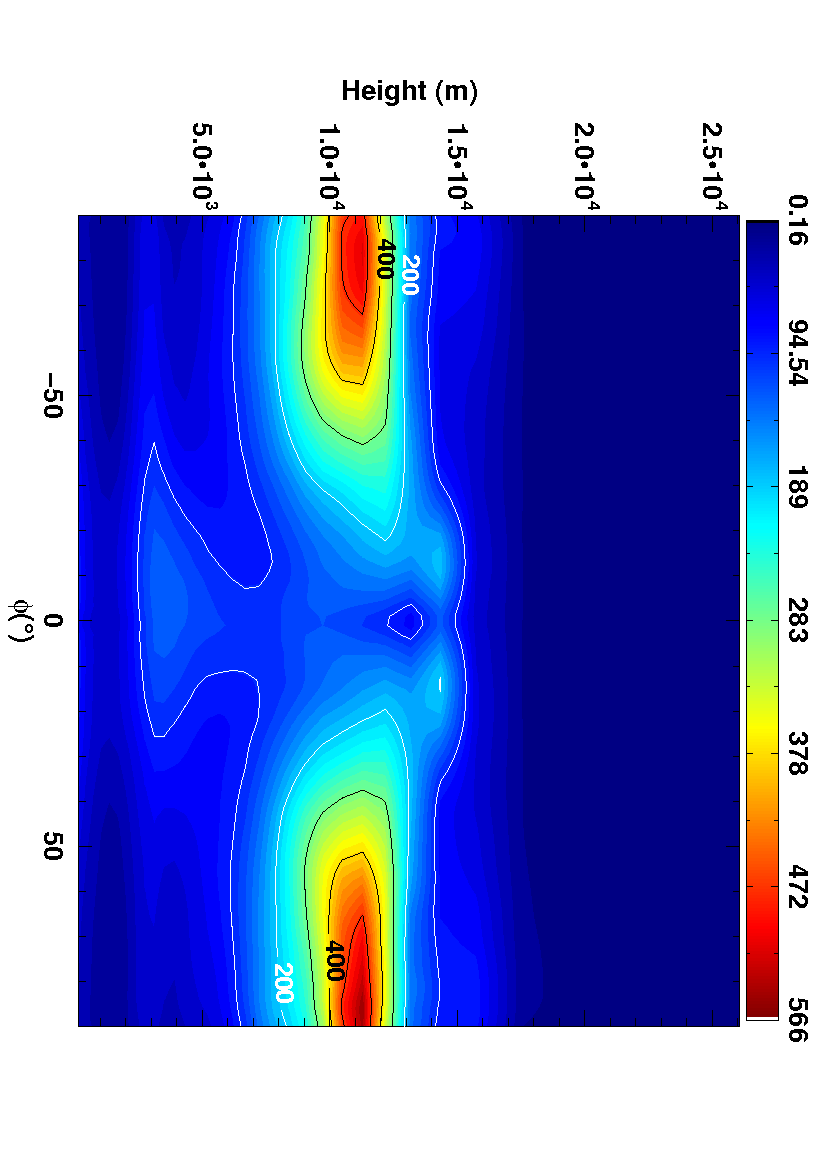}
\hspace*{-1.0cm}\includegraphics[width=6.5cm,angle=90.0]{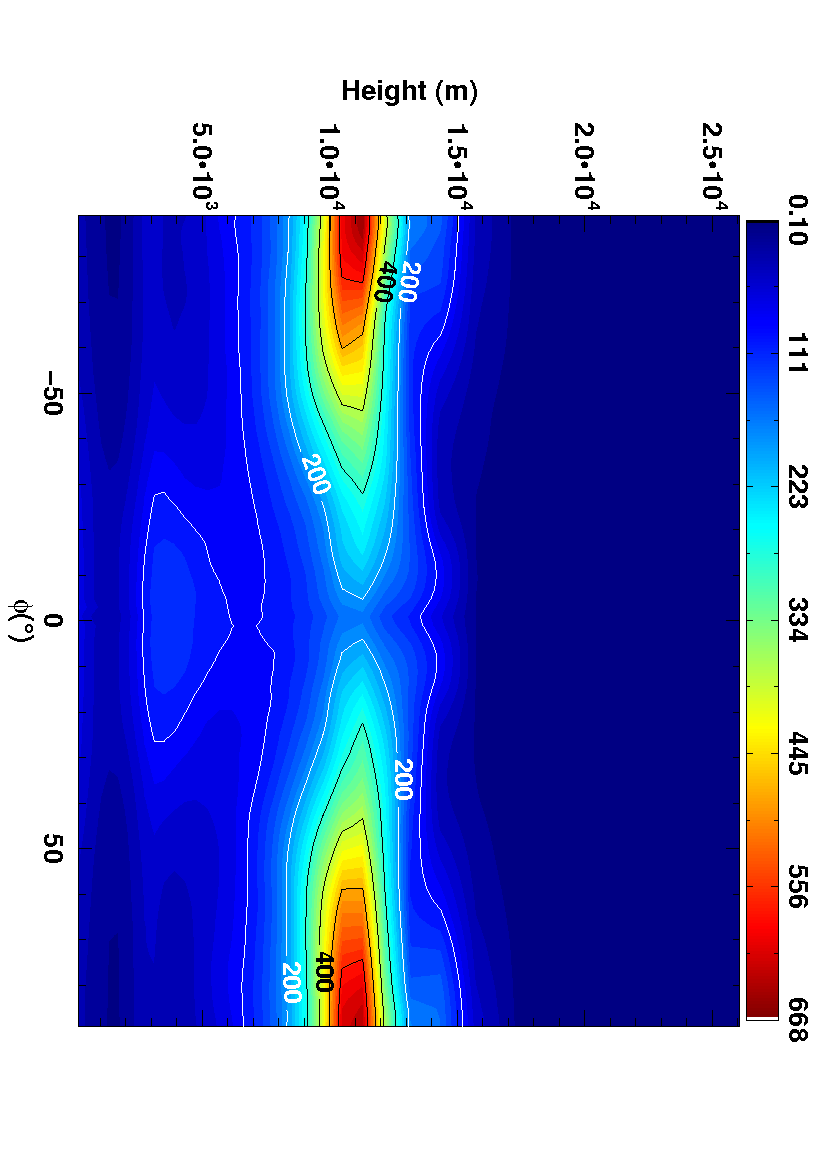}
\hspace*{-1.0cm}\includegraphics[width=6.5cm,angle=90.0]{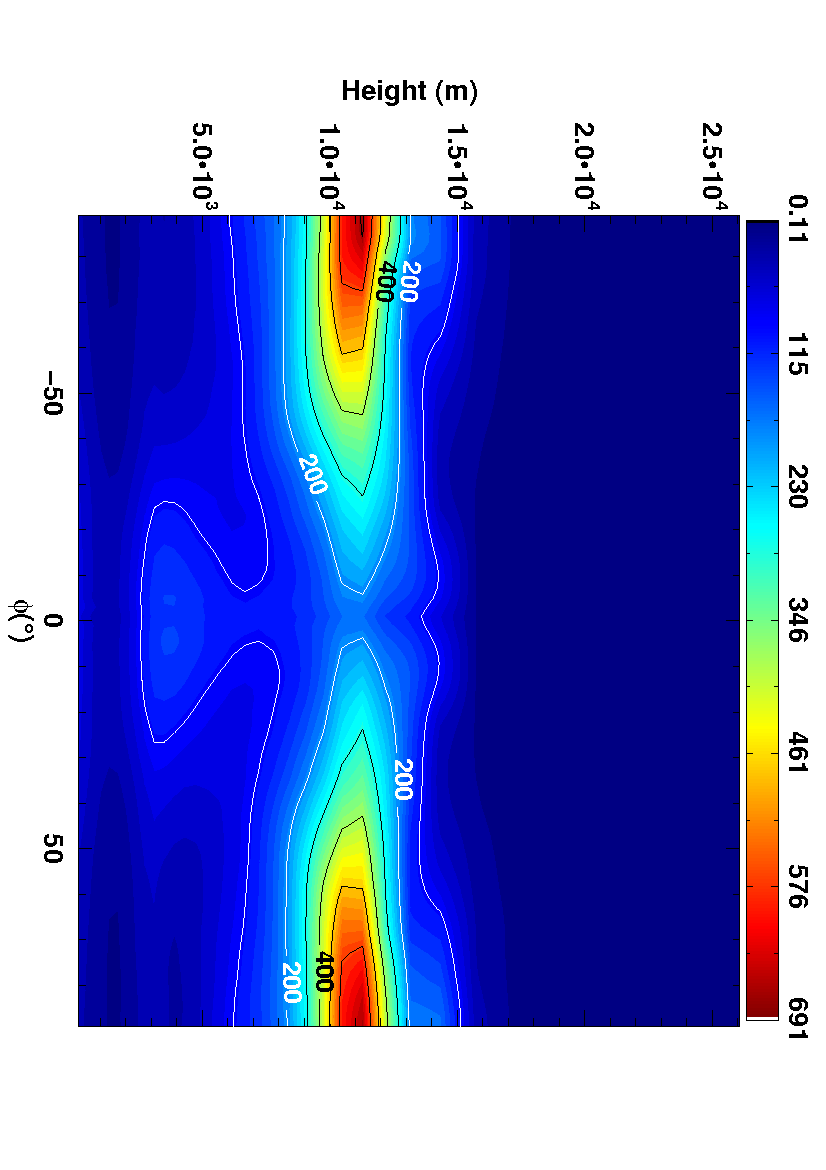}
\hspace*{-1.0cm}\includegraphics[width=6.5cm,angle=90.0]{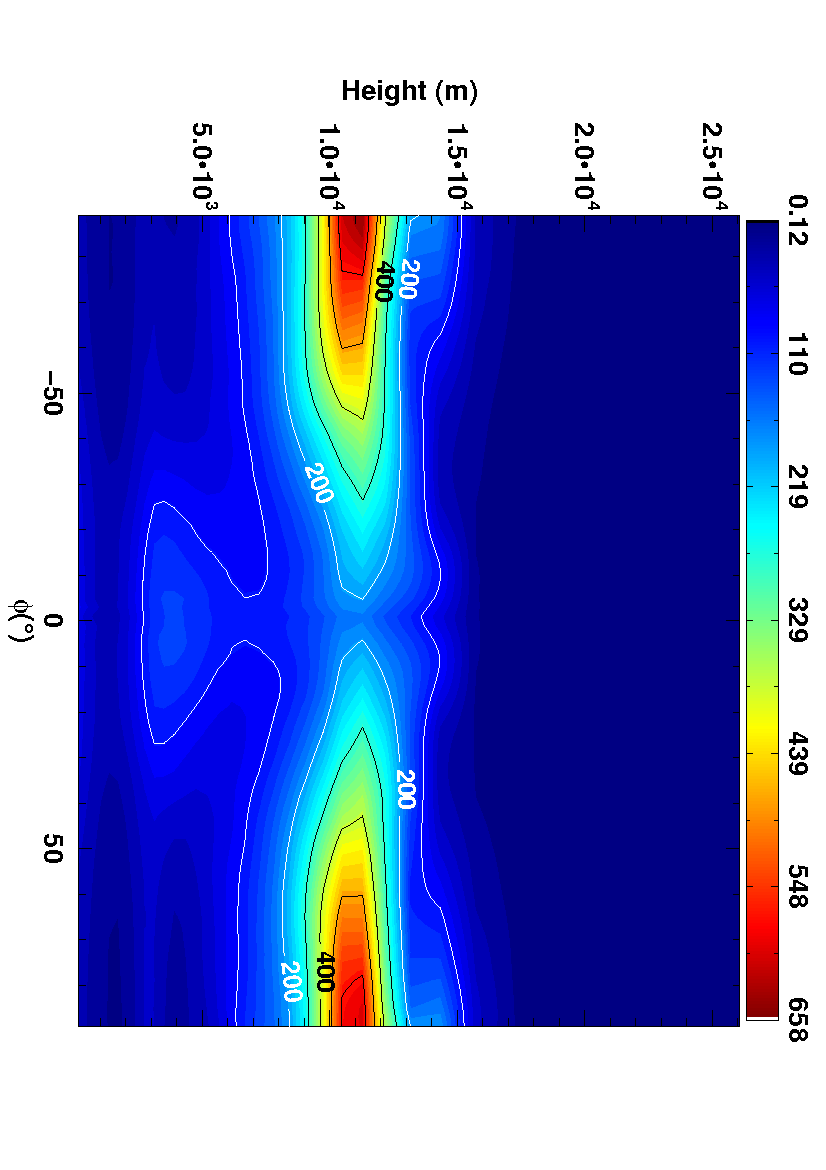}
\end{center}
\caption{Figure, for the Tidally Locked Earth test
  \citep{merlis_2010,heng_2011}, showing the zonally (in geometric
  height) and temporally averaged Eddy Kinetic Energy (EKE, see
  Section \ref{models_run}) as a function of latitude and
  height. \textit{Top left panel} ND, \textit{top right panel}
  EG$_{\rm sh}$, \textit{bottom left panel} EG$_{\rm gc}$ and
  \textit{bottom right panel} EG models (see Table \ref{model_names}
  for explanation of model types). Note the contours (solid lines) are
  the same in all plots. \label{TLE_EKE}}
\end{figure*}

\begin{figure}[t]
\vspace*{2mm}
\begin{center}
\hspace*{-1.0cm}\includegraphics[width=6.5cm,angle=90.0]{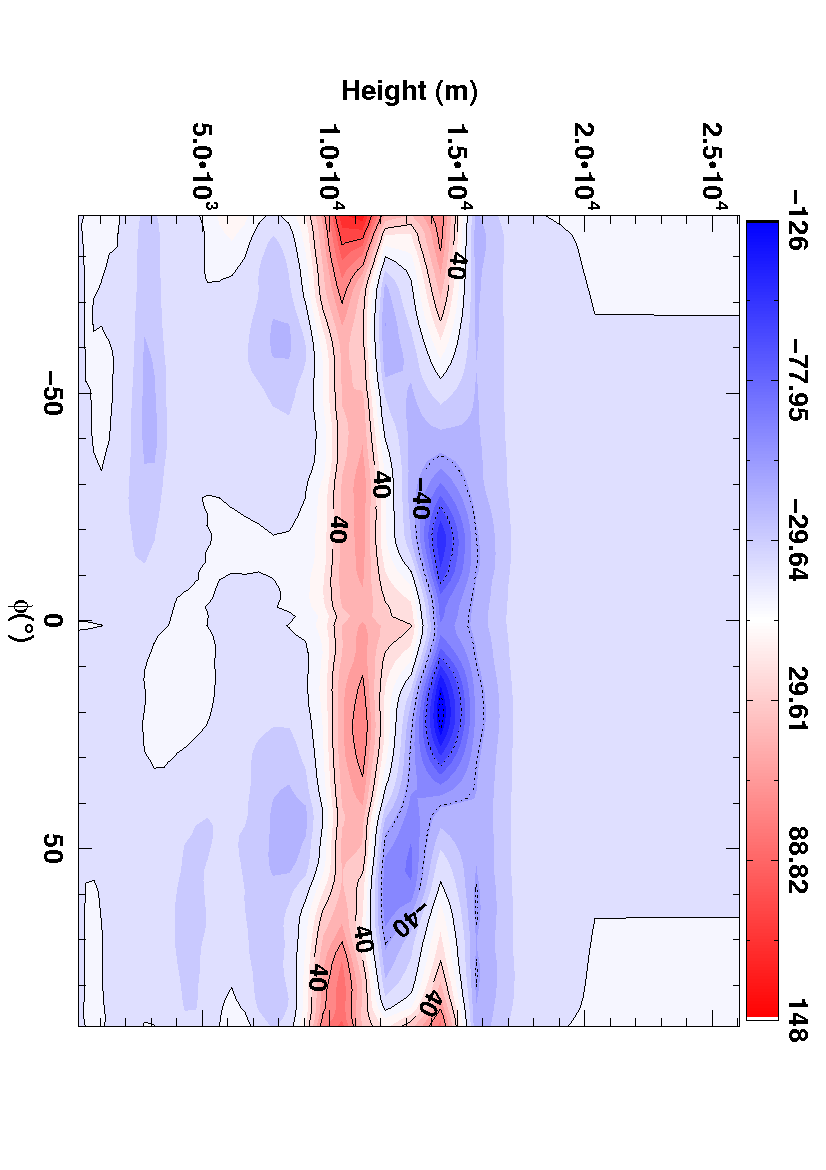}
\end{center}
\caption{Figure, for the Tidally Locked Earth test
  \citep{merlis_2010,heng_2011}, showing the differences EG$-$ND of
  the zonally and temporally averaged EKE (see Table \ref{model_names}
  for explanation of model types). \label{diff_TLE_EKE}}
\end{figure}

Figure \ref{TLE_EKE} shows the EKE, zonally (along geometric height
surfaces) and temporally averaged ($\overline{{\rm EKE}}^{\lambda_{\rm
    z}t}$), for the ND and all ENDGame models.  Figure \ref{TLE_EKE}
shows more distinct differences when comparing ND to any of the
ENDGame models, compared to the HS or EL test cases. In the TLE case
the kinetic energy associated with the eddies clearly increases when
moving from ND to ENDGame. Additionally, the structure of the peak
activity region, which extends from mid--latitudes over the poles, is
flatter (in altitude) in the ENDGame models. One can also observe a
move to increased hemispherical symmetry when moving from ND through
EG$_{\rm sh}$ and EG$_{\rm gc}$ to EG. This shows that ENDGame
produces a more spherically symmetric pattern of eddies, closer to
what one would expect in a slowly rotating system. Furthermore, it
shows that subsequent relaxation of the approximations to the
equations of motion slightly improves the symmetry of the
solution. Again, as with the EL test cases, we present the difference
in the $\overline{{\rm EKE}}^{\lambda_{\rm z},t}$, in the sense
EG$-$ND in Figure \ref{diff_TLE_EKE}, where the ENDGame model
differences are not shown as they are an order of magnitude smaller
than those between the EG and ND models.

As with the previous test cases, and evident from the prognostic
fields $T$, $u$ and $v$, all the ENDGame models show a remarkable
level of consistency in the solution. However, as in the HS and EL
test cases, significant differences in the $\overline{{\rm
    EKE}}^{\lambda_{\rm z},t}$, are found when comparing EG to ND. The
magnitude of the peak relative differences in $\overline{{\rm
    EKE}}^{\lambda_{\rm z}t}$ are $\sim$8.0, 0.40 and 0.61 for the
differences EG$-$ND, EG$-$EG$_{\rm gc}$ and EG$-$EG$_{\rm sh}$,
respectively. The relative difference for the EG$-$ND is much larger
than that found in either the HS or EL test cases. The peak
$\overline{{\rm EKE}}^{\lambda_{\rm z},t}$, is larger in the EG model
and the peak appears to shift lower in the atmosphere, when compared
to the ND model.

Whilst features such as the increased hemispherical symmetry of the
flow found in the ENDGame models, are close to what one might
physically expect, this test case (and the others) is not a definitive
test to demonstrate that the flow is handled better in
ENDGame. However, it is clear that they are at least handled
differently. The difficulty for tests such as these is that a correct,
or analytical answer, for the flow does not exist.

\conclusions  
\label{conclusions}

We have demonstrated that both the ND and ENDGame dynamical cores of
the Met Office UM produce 3D idealised large--scale and long--term
flows consistent both with previous works, and under varying
approximations to the full equations of motions. These tests are the
Held--Suarez test \citep{held_1994}, an Earth--like test
\citep{heng_2011,menou_2009} and a hypothetical tidally locked Earth
\citep{merlis_2010,heng_2011}. Qualitative agreement was found for the
results of these three idealised test cases, both between the UM
dynamical cores and when compared with literature results.
Furthermore, the consistency of the solutions was not changed when
invoking the approximations possible in the ENDGame equation set, all
of which should be applicable for our test cases, namely, the
`shallow--atmosphere' approximation, as a whole, or just the
assumption of constant gravity. We also found tentative evidence of
differences in the circulation, for the TLE case, between the ENDGame
and ND cores probably caused by changes in the temporal and spatial
discretisation.

These results should be viewed as complementary to more analytical
testing. For our project, namely adapting the UM with a
state--of--the--art dynamical core to exoplanets, this work is a
crucial first step in confirming the consistency of the code, both
with other GCMs and, under different approximations to the full
equations of motion. We have also tested the code in flow regimes with
features in common with the subset of exoplanets termed hot Jupiters
(which our project aims to characterise), i.e. a hypothetical tidally
locked earth. For the flow regimes of hot Jupiters the solutions to
the equations of motion are expected to differ under the different
approximations featured in this work. Furthermore, these objects are
severely observationally under--constrained, so rigorous testing is
required. We will present the next step of this project, involving
adaptation of the code and simplified giant planet test cases in a
future work (Mayne et al, submitted).

\appendix
\section{\\ \\ \hspace*{-7mm} Appendix}    

\subsection{A note on comparison with the work of \citet{heng_2011}.}
\label{heng_comp}
\citet{heng_2011} perform both finite--difference and spectral models
of the test cases using the same GCM (the Princeton Flexible Modeling
System, {\sc FMS}). In this work we concentrate our comparison with
the results of the finite--difference versions of the test, as the UM
also adopts a finite--difference method. Additionally, it is not clear
which $\sigma$ surface \citet{heng_2011} select when producing plots
of the atmosphere as a function of latitude and longitude, in the
spectral case. The spectral version of the {\sc FMS} dynamical core
performs vertical finite--differencing using a Simmons-Burridge
scheme. \citet{heng_2011} state, the prognostic variable output is not
exactly at the mid--point of the vertical half--levels, and when
presenting results they usually quote the $\sigma$ of the bottom pair
of half--levels. Therefore, some uncertainty exists over which
$\sigma$ surface the resulting plots are produced from. For the
finite--difference results \citet{heng_2011} state that the labeling
of the model layers adopts the same system as the spectral version,
i.e. each layer is actually labeled with the value of the larger
$\sigma$ half--level. This may result in a slight translation, or
vertical shift, when we present plots with $\sigma$ as the vertical
axis. As comparison of our results and those of \citet{heng_2011}
show, in Section \ref{el} this effect is negligible. However, for
horizontal slices at a prescribed $\sigma$ this will result in the
flow being presented at a different pressure surface. In effect,
therefore, we assume that if a figure from \citet{heng_2011} is
presented as representative of the flow at a given $\sigma$, that
actually the flow is that present at $\sigma-1.0/(2\times20)$
(i.e. $\sigma-0.025$), as \citet{heng_2011} use 20 uniformly
distributed vertical levels (with associated half--levels) spaced
evenly in $\sigma$. Therefore, our Figures will be presented using the
\emph{actual} $\sigma$ value of the model, where we have interpolated
our prognostic variables onto this $\sigma$ surface.

\subsection{Vertical Level Placements}                              
\label{vert_levs}

Table \ref{vert_levels} shows the positions of the vertical
($\theta$)\footnote{In a Charney--Phillips grid, $\rho$ levels are
  placed halfway between $\theta$ levels.}, levels in non--dimensional
height units ($\eta$), alongside the size of the domain $H$ and the
approximate $\sigma$ value (see Section \ref{models_run} for
explanation).

\begin{table*}[t]
  \caption{Table showing the dimensionless vertical coordinate for the $\theta$ levels of the three model setups, $\eta_{\theta}$ ($\eta=z/H$) alongside the approximate $\sigma$ levels and the model domain height ($H$).\label{vert_levels}}
  \vskip4mm
  \centering
\begin{tabular}{lc|ccc}
  \tophline
  \multicolumn{2}{c}{Test case:}&Held-Suarez (HS)&Earth-Like (EL)&Tidally Locked Earth (TLE)\\
  \multicolumn{2}{c}{$H$ (m)}&30975.0&30964.0&30056.0\\
  \middlehline
  Level&$\sim$$\sigma$&\multicolumn{3}{c}{$\eta_{\theta}$}\\
  \middlehline
  0&1.00&0.000000&0.000000&0.000000\\
  1&0.97&0.009072&0.004521&0.009915\\
  2&0.94&0.018111&0.009010&0.019763\\
  3&0.91&0.027506&0.026967&0.029977\\
  4&0.88&0.036901&0.036203&0.040192\\
  5&0.84&0.046295&0.045408&0.050472\\
  6&0.81&0.056433&0.055290&0.061951\\
  7&0.78&0.066764&0.065495&0.073463\\
  8&0.75&0.077094&0.075701&0.085108\\
  9&0.72&0.088103&0.086423&0.097651\\
  10&0.69&0.099467&0.097694&0.110194\\
  11&0.66&0.110896&0.109030&0.123303\\
  12&0.63&0.123099&0.121011&0.137011\\
  13&0.60&0.135626&0.133510&0.150852\\
  14&0.57&0.148539&0.146331&0.165824\\
  15&0.53&0.162260&0.159928&0.180829\\
  16&0.50&0.176303&0.174009&0.197065\\
  17&0.47&0.191251&0.188897&0.213501\\
  18&0.44&0.206780&0.204560&0.231302\\
  19&0.41&0.223245&0.221128&0.249468\\
  20&0.38&0.240613&0.238826&0.269331\\
  21&0.35&0.259112&0.257654&0.289959\\
  22&0.32&0.278935&0.278000&0.312018\\
  23&0.28&0.300371&0.300026&0.336039\\
  24&0.26&0.323584&0.324021&0.361791\\
  25&0.22&0.349379&0.350698&0.389839\\
  26&0.19&0.378563&0.380668&0.421047\\
  27&0.16&0.412365&0.415321&0.456614\\
  28&0.13&0.453010&0.457338&0.498336\\
  29&0.10&0.504310&0.510690&0.549607\\
  30&0.07&0.574851&0.583419&0.621540\\
  31&0.04&0.687780&0.698908&0.736126\\
  32&0.01&1.000000&1.000000&1.000000\\
  \bottomhline
\end{tabular}
\end{table*}

\begin{acknowledgements}
  We would like to thank Paul Ullrich and Kevin Heng for their
  valuable comments, when reviewing this manuscript. We would also
  like to thank Tom Melvin for his expert advice, and both Charline
  Marzin and Douglas Boyd for technical help. This work is supported
  by the European Research Council under the European Communitys
  Seventh Framework Programme (FP7/2007-2013 Grant Agreement
  No. 247060) and by the Consolidated STFC grant ST/J001627/1. This
  work is also partly supported by the Royal Society award
  WM090065. The calculations for this paper were performed on the
  DiRAC Facility jointly funded by STFC, the Large Facilities Capital
  Fund of BIS, and the University of Exeter.
\end{acknowledgements}

\bibliographystyle{copernicus}
\bibliography{references.bib}









\vskip4mm











\end{document}